\documentclass[aps,rmp,reprint,amsmath,amssymb,graphicx,longbibliography]{revtex4-1}

\usepackage{bm}
\usepackage{amssymb,amsfonts,amsmath,rotating}
\newcommand{\mc}{\mathcal}
\newcommand{\map}[3]{#1: #2 \rightarrow #3}
\newcommand{\real}{\mathbb{R}}
\newcommand{\transpose}{\mathsf{T}}

\newcommand{\A}{\mathbf{A}}
\newcommand{\B}{\mathbf{B}}

\newcommand{\bx}{\mathbf{x}}
\newcommand{\bu}{\mathbf{u}}

\usepackage{color}

\begin{document}

\title{Control of Dynamics in Brain Networks}

\author{Evelyn Tang}
\affiliation{Department of Bioengineering, University of Pennsylvania, PA 19104}
\author{Danielle S. Bassett}
\affiliation{Department of Bioengineering,}
\affiliation{Department of Physics \& Astronomy,}
\affiliation{Department of Neurology,}
\affiliation{Department of Electrical \& Systems Engineering, University of Pennsylvania, PA 19104}

\date{\today{}}

\begin{abstract}
The ability to effectively control brain dynamics holds great promise for the enhancement of cognitive function in humans, and the betterment of their quality of life. Yet, successfully controlling dynamics in neural systems is challenging, in part due to the immense complexity of the brain and the large set of interactions that can drive any single change. While we have gained some understanding of the control of single neurons, the control of large-scale neural systems---networks of multiply interacting components---remains poorly understood. Efforts to address this gap include the construction of tools for the control of brain networks, mostly adapted from control and dynamical systems theory. Informed by current opportunities for practical intervention, these theoretical contributions provide models that draw from a wide array of mathematical approaches. We present recent developments for effective strategies of control in dynamic brain networks, and we also describe potential mechanisms that underlie such processes. We review efforts in the control of general neurophysiological processes with implications for brain development and cognitive function, as well as the control of altered neurophysiological processes in medical contexts such as anesthesia administration, seizure suppression, and deep-brain stimulation for Parkinson's disease. We conclude with a forward-looking discussion regarding how emerging results from network control---especially approaches that deal with nonlinear dynamics or more realistic trajectories for control transitions---could be used to directly address pressing questions in neuroscience. 
\end{abstract}

\maketitle

\tableofcontents{}

\newpage
\section{Introduction}
\label{intro}

The brain displays a wealth of complex dynamics across various spatial and temporal scales \cite{betzel2016multi,kopell2014beyond}. From 302 neurons in the nematode worm \emph{C. elegans} \cite{varier2011neural,bentley2016multilayer} to some 86 billion neurons in the adult human \cite{herculano2007cellular,bartheld2016search}, the units that drive brain function are large in their number but even more complicated in their interactions. Far from the canonical models in statistical mechanics stemming from either crystalline or random structure, the brain displays a heterogeneous pattern of interconnections \cite{fraiman2009ising,castellana2014inverse,bassett2016small} that fundamentally constrains the propagation of activity. Understanding these dynamics remains of primary interest in the field of neurophysics \cite{scott1977neurophysics,gao2015simplicity}. An underlying assumption of these investigations is that such dynamics or observed neural activity can contain structure that forms representations about incoming stimuli or underlying neural processes. An emerging and increasingly tractable avenue for understanding the mechanisms of these dynamics lies in the notion of control, or how to effectively guide neural dynamics. How are brain dynamics controlled intrinsically in the awake, behaving animal? Can we harness natural principles of control in neural systems to better guide therapeutic interventions? 

The increase in available experimental neurotechnologies \cite{nag2016implantable,patil2016implantable,chang2015towards}, as well as more sophisticated computational tools \cite{marblestone2016toward,glaser2016development} and theoretical models \cite{giusti2016twos}, has recently made it possible to tackle these questions from fundamentally new angles.  While at present there is no comprehensive theory of control in the brain that we can refer to, the pursuit of such a theory remains critically important, having implications for our understanding of healthy neurophysiological processes, and our ability to intervene when those healthy processes go awry in neurological disease and psychiatric disorders \cite{johnson2013neuromodulation,chen2014harnessing,bassett2017network}. Several recent models propose new ways to control neural activity and neural rhythms, and further provide mechanistic insights into the rules by which brain dynamics are (and can be) guided. Hence, it is timely to discuss these emerging developments, and to seek to tie them together into a meaningful theoretical field that can be used to tackle current open questions in neuroscience and medicine.

Motivated by recent progress in understanding brain function from the perspective of interacting networks \cite{bassett2006small,Bassett2009,kaiser2011tutorial,Bullmore2012}, we focus on systems-level control of either local neural circuits or whole-brain connectomes \cite{Sporns2005,fornito2015connectomics}. Here we use the term ``network'' in the sense that is common in network science \cite{newman2010networks}. A \emph{brain network} is a graph whose nodes represent units of the brain that perform a specific function, like vision or audition \cite{bullmore2011brain}. At the large-scale, these units may be several centimeters of tissue, while at the small scale, these units may be individual neurons. In structural brain graphs, the edges can represent structural links such as fiber bundles at the large scale \cite{Hagmann2008,Bassett2011} or synapses at the small scale. In functional brain graphs, the edges represent synchronized dynamics that form functional links \cite{achard2006resilient,stam2004functional} between these units. While both structural and functional links can be measured directly from structural and functional data, respectively, extensive efforts have also sought to address the questions of (i) whether structural topology can be inferred from functional traces (using, for example, structural equation modeling), and (ii) whether functional traces can be inferred from structural linkage patterns (using, for example, neural mass models). Throughout this exposition, we will assume that structural links have been directly measured, rather than inferred. 

The use of the network formalism to probe brain dynamics has a rich and pervasive heritage in seminal work at the intersection between physics and neuroscience. One particularly impactful contribution was that of Hopfield, who successfully connected dynamical processes to neural representations in an Ising model \cite{Hopfield01041982}. States that minimized the energy function formed dynamical attractors and representations of memory. This early contribution was extended and formalized by \textcite{PhysRevA.32.1007,0295-5075-4-4-016}, clearly demonstrating the power of interacting networks in the modeling of complex neural processes. Here we expand the link between physics and neuroscience in the context of the network formalism by focusing on the control of brain networks, enabling us to build a theoretical understanding regarding biological processes and associated dynamics that occur across spatially distributed neural systems.  In addition, strategies for intervention and control targets can be designed through modeling dynamics in networks of neurons or brain regions. Should the reader instead be searching for an excellent treatment of various control methods for single neurons or for ensembles of neurons, we direct them to the recent textbook by \textcite{Schiff2012}, and to references therein. For further details on emerging control technologies in the brain---especially invasive electrical and optical stimulation at rapid timescales (milliseconds or below)---and associated modelling approaches, please see the recent review by \textcite{7171915}.

The remainder of this Colloquium is organized as follows. In Sec.~\ref{s:cognitive_control} we draw inspiration for understanding control of brain networks by considering how the brain itself enacts intrinsic control. In particular, we briefly discuss important computational paradigms of cognitive control, a basic ability that each of us has to control our neural activity and by extension our behavior. This discussion motivates the introduction of network control theory in Sec.~\ref{s:network_control}, which offers a useful theoretical framework in which to probe control in brain networks constructed from neuroimaging data. We next turn in Sec.~\ref{s:understanding_healthy} to detailing a few examples of how we can use network control theory, or its extensions, to understand healthy brain function. In Sec.~\ref{s:understanding_disease}, we describe the utility of network control in targeting interventions when healthy brain function goes awry. We next turn in Sec.~\ref{s:trajectories} to modeling the controlled \emph{versus} uncontrolled trajectories of neural dynamics, and we close in Sec.~\ref{s:future} by outlining emerging frontiers at the intersection of dynamical systems theory, control theory, and complex systems. Throughout, we keep neuroscience jargon to a minimum, although some terminology specific to the technique or context remains unavoidable. Our goal is to stimulate discussion through reviewing existing work (rather than presenting new data), in order to encourage further work from physicists, control theorists, practitioners, and others in this exciting and rapidly developing field.

\section{How does the brain control itself?}
\label{s:cognitive_control}

While there may be many ways of tackling the question of how to control brain dynamics, arguably one of the simplest is to ask how the brain controls itself. Perhaps by understanding intrinsic mechanisms of control in the brain, we could harness that knowledge to inform therapeutic interventions for people with mental illness. In considering this idea, it is useful to distinguish between external control, which is enacted on the system from the outside, and internal control, which is a feature of the system itself. In the brain, internal control processes include phenomena as conceptually diverse as homeostasis, which refers to processes that maintain equilibrium of dynamics \cite{nelson2008strength,nelson2015excitatory}, and cognitive control, which refers to processes that exert top-down influence to drive the system between various dynamical states \cite{botvinick2015motivation,heatherton2011cognitive}.

Here we focus on cognitive control because it is conceptually akin to the idea of extrinsic control: driving dynamics from one type to another. What can we learn from cognitive control that might help us to develop a theory for external control? To answer that question, we begin by turning to history. An early computational model that explained the production of decisions based on a given set of inputs was the perceptron \cite{rosenblatt1958perceptron,freund1999large}, a simple artificial neural network \cite{mcculloch1943logical,bishop1995neural}. The perceptron and associated notions were developed by proponents of connectionism \cite{medler1998brief}, which suggests that cognition is an emergent process of interconnected networks. The complexity of the connection architecture in these models was thought to support a complexity of brain dynamics, such as the separation of parallel neural processes and distributed neural representations propounded by the parallel distributed processing (PDP) model \cite{rumelhart1986parallel}.  The PDP model holds that cognitive processes can be explained by activation flowing through networks that link nodes together. Every new event changes the strength of connections among relevant units by altering the connection weights.

\begin{figure}[t]
\includegraphics[width=0.98\linewidth]{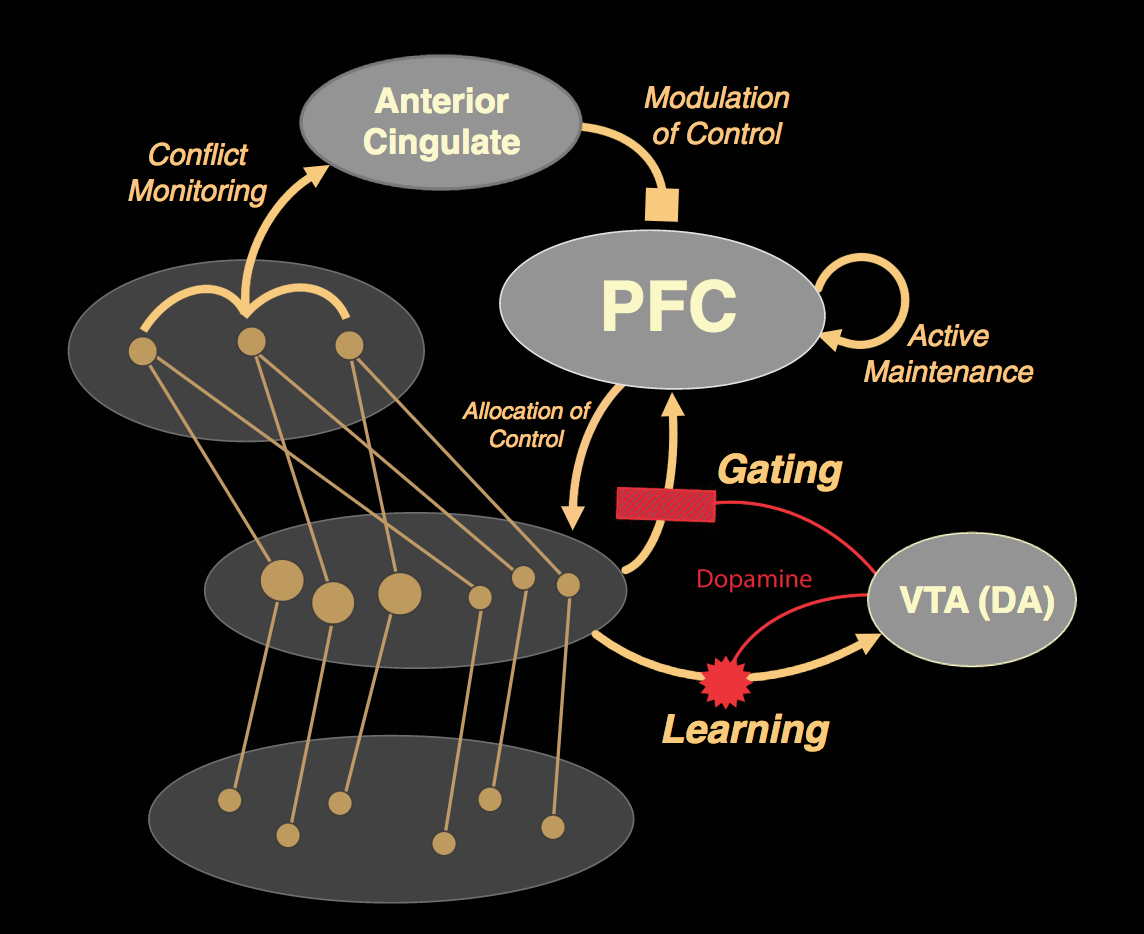}
\caption{\textbf{Model for adaptive cognitive control showing distinct mechanisms between different brain regions.} Schematic of a neural network connecting the prefrontal cortex, which executes much of the ``top-down'' control actions, to other brain regions. Another part of the brain -- the anterior cingulate cortex -- serves as a conflict monitoring mechanism that modulates the activity of control representations. Meanwhile, an adaptive gating mechanism regulates the updating of control representations in prefrontal cortex through dopaminergic (DA) projections from the ventral tegmental area (VTA), that can also be facilitated through reinforcement learning (red asterisk). From \textcite{COGS:COGS12126}.}\label{fig:cohen}
\end{figure}

Notably, the PDP model offers conceptual explanations for the processes characteristic of cognitive control \cite{COGS:COGS12126}. These ideas are built on the notion that the development of control systems in the brain \cite{chai2017evolution} can be seen as responding to the structure of naturalistic tasks, and therefore that control can be defined as the optimal parameterization of task processing. Within such a parameterization, two specific features of cognitive control appear particularly critical: (i) its remarkable flexibility, which supports diverse behaviors, and (ii) its clear constraints, which limit the number of control-demanding behaviors that can be executed simultaneously. Addressing these two features, models inspired by the PDP approach allow for cognitive control as instantiated in processes of selection from competing inputs or adaptation based on reward (Fig. \ref{fig:cohen}). 

These and related computational models emphasize the role of specific brain areas in cognitive control, including prefrontal cortex, anterior cingulate, parietal cortex, and brainstem. Yet, studying any of these areas in isolation will likely provide an impoverished undestanding of the system's function. Indeed, \textcite{Eisenreich077685} argue that control in the brain is not localized to small regions or modules, but is instead very broadly distributed, enabling versatility in both information transfer and executive control. Such a distributed -- and even perhaps overlapping -- network architecture can also offer usefully fuzzy boundaries between controllers and processors \cite{Eisenreich077685,6777299}. How exactly information is processed on these distributed systems remains an open question, but some promising modeling approaches include those that use Bayesian inference, sparse-coding, and information entropy to characterize control \cite{6777299}. Specifically, a few recent efforts draw heavily from the idea of probabilistic reasoning to formulate a model for risk control -- posited to be an overarching function of the prefrontal cortex -- characterized by a closed-loop feedback structure describing executive attention. 
 
To briefly summarize, previous computational models of cognitive control have included the eclectic notions of neural networks, regional localization, distributed processing, and information theory. Collectively, these notions motivate the construction of a model or theory that explicitly builds on the emerging capability to measure the brain's true network structure to better understand control. In the next section, we will describe recent developments in dynamical systems and control theory as applied to complex networks, whose application to the brain may offer explanatory mechanisms of neural dynamics and provide insights into the distributed nature of cognitive control.

\section{Network control theory}
\label{s:network_control}

Conceptually, it is interesting to ask the question whether and to what degree cognitive control (as defined by neuroscientists) is similar to network control (as defined by physicists, mathematicians, and engineers). To address this interesting question, we must first define what it is that we mean by network control. Controllability of a dynamical system refers to the possibility of driving the current state of the system to a specific target state by means of an external control input, see \textcite{REK-YCH-SKN:63}. Developments in engineering and physics have recently extended these ideas to the control of networks, as we describe in more detail below.  

\subsection{Control of linear dynamics}
\label{s:network}

\begin{figure}[t]
\includegraphics[width=\linewidth]{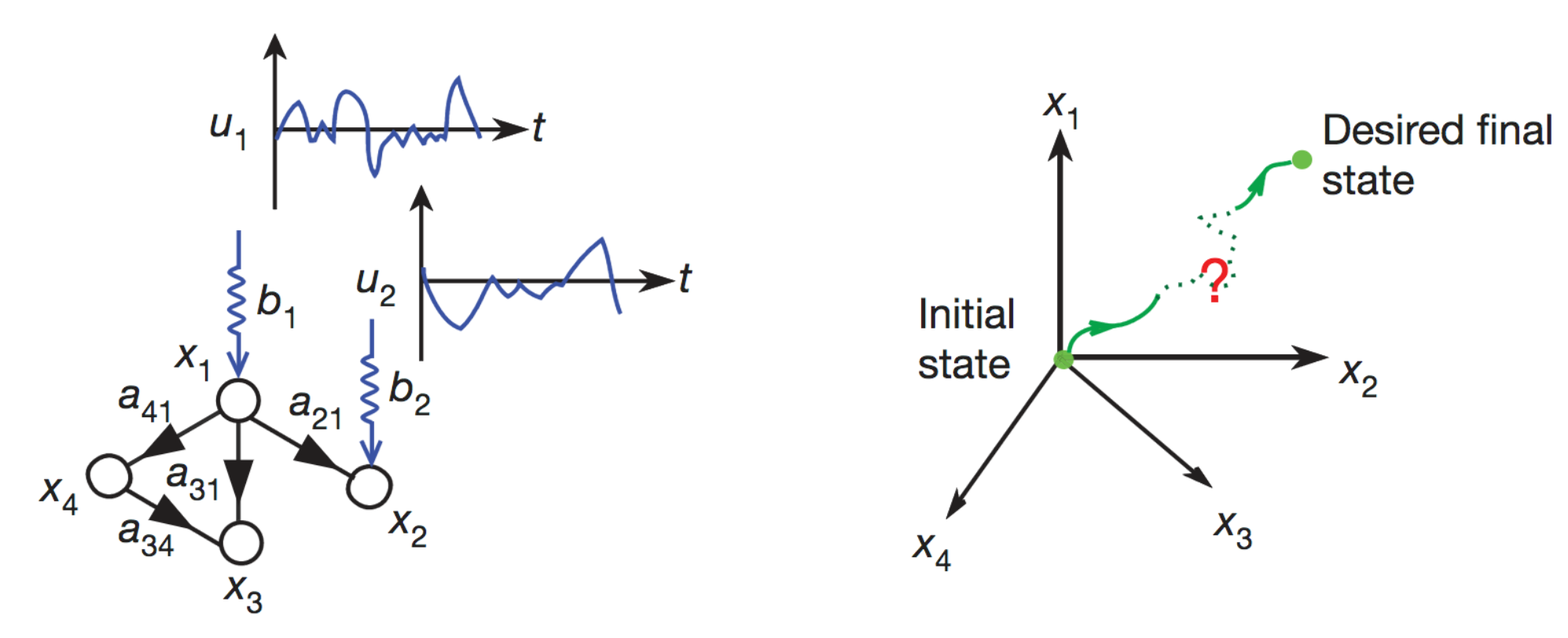}
\caption{\textbf{Controlling a simple network.} This small network can be controlled by an input vector $ \mathbf{u}_{\mc K}=(u_1(t),u_2(t))^T$ (left), allowing us to move the network within the state space, from its initial state to some desired final state (right). From \textcite{YYL-JJS-ALB:11}.} \label{fig:liubarabasi}
\end{figure}

We begin by describing a general framework for the control of linear dynamics on a complex network. Consider a network represented by the directed graph $\mc G = (\mc V, \mc E)$, where $\mc V$ and $\mc E$ are the vertex and edge sets, respectively. Let $a_{ij}$ be the weight associated with the edge $(i,j) \in \mc E$, and define the \emph{weighted adjacency matrix} of $\mc G$ as $\mathbf{A} = [a_{ij}]$, where $a_{ij} = 0$ whenever $(i,j) \not\in \mc E$. We associate a real valued (\emph{state}) with each node, collect the node states into a vector (\emph{network state}), and define the map $\map{\mathbf{x}}{\mathbb{N}_{\ge 0}}{\mathbb{R}^n}$ to describe the evolution (\emph{network dynamics}) of the network state over time. A simple way to begin is to describe the network dynamics by a discrete time, linear,
and time-invariant recursion
\begin{align}\label{eq:system}
\mathbf{x} (t+1) = \mathbf{A} \mathbf{x}(t).
\end{align}
Let a subset of nodes $\mc K = \{k_1, \dots, k_m \}$ be independently controlled, and let
\begin{align}\label{eq: B}
 \mathbf{B}_{\mc K} :=
  \begin{bmatrix}
    e_{k_1} & \cdots & e_{k_m}
  \end{bmatrix}
\end{align}
be the \emph{input matrix}, where $e_i$ denotes the $i$-th canonical vector of dimension $n$. The network with control nodes $\mc K$ reads as
\begin{align}\label{eq:controlled}
  \mathbf{x} (t+1) = \mathbf{A} \mathbf{x}(t) + \mathbf{B}_{\mc K} \mathbf{u}_{\mc K} (t) ,
\end{align}
where $\map{ \mathbf{u}_{\mc K}}{\mathbb{N}_{\ge 0}}{\mathbb{R}}$ is the control
signal injected into the network via the nodes $\mc K$ (see Fig. \ref{fig:liubarabasi}). The network
\eqref{eq:controlled} is controllable in $T$ steps by the nodes $\mc K$ if, for every state
$ \mathbf{x}_f$, there exists a control input $ \mathbf{u}_{\mc K}$ such that
$ \mathbf{x}(T) =  \mathbf{x}_f$ with $ \mathbf{x}(0) =  \mathbf{0}$ \cite{kailath1980linear}. 

Controllability of this type of system can be ensured by different structural conditions \cite{kailath1980linear,KJR:88}. For instance, let $\mathbf{ C}_{\mc K,T}$ be the \emph{controllability matrix} defined as
\begin{align*}
 \mathbf{ C}_{\mc K,T} :=
  \begin{bmatrix}
   \mathbf{ B}_{\mc K} & \mathbf{A} \mathbf{B}_{\mc K} & \cdots & \mathbf{A}^{T-1}\mathbf{ B}_{\mc K}
  \end{bmatrix}
  .
\end{align*}
The network \eqref{eq:controlled} is controllable in $T$ steps by the nodes $\mc K$ if and only if $\mathbf{ C}_{\mc K,T}$ is of full row rank, where $T$ is typically taken to be at least as large as the system size $n$.

\subsection{Key driver nodes}

Recent work from \textcite{YYL-JJS-ALB:11} demonstrated that the analytical framework described in the previous section could be used to study large, complex networks. In that study, the authors explored common patterns in a wide variety of networks from technology, biological, and social systems. Under certain conditions in these weighted and directed networks, the set of driver nodes capable of guiding the dynamics of the entire system could be directly estimated from the degree distribution. Since that study, others have shown that under other conditions and in other networks, the degree distribution alone may not provide enough information to adequately determine the set of driver nodes. Instead, that knowledge regarding the network's structure must be complemented with considerations of the network's dynamics, or reasonable approximations of those dynamics at each node \cite{10.1371/journal.pone.0038398}. 

In these studies, networks are allowed to contain real-valued weights on each edge. However, for some real-world networks, knowledge of the edge weights is uncertain. For such scenarios, a complementary framework is provided by structural controllability which evaluates the controllability of binary networks \cite{kailath1980linear,KJR:88}. By studying the underlying ``structure'', i.e. distinguishing merely between which edges are absent (zero) \emph{versus} present (non-zero), these methods allow the identification of minimal structures or control points that allow for full controllability of the network. Recent efforts have extended these ideas to large-scale systems, and to the problem of identifying the minimum number of nodes that need to be driven in order to achieve structural controllability \cite{PequitoJ1}.

In recent work, \textcite{7526572} extended the notion of structural controllability to situations in which edges evolve dynamically, and they identified the minimum number of driven nodes for full controllability of the system. Their methods would appear particularly relevant in situations like those observed in \textcite{khambhati2015dynamic}, where dynamic functional connectivity in epileptic patients was characterized by edges within seizure-generating areas that were almost constant over time, whereas edges outside these areas in healthy tissue exhibited higher variability over time. An important potential goal of control would then be to steer function on these edges away from pathological regimes \cite{7526572}, i.e. towards dynamics that demonstrate more edge weight variability.

While network control and structural controllability are particularly relevant concepts for brain network control, many other key contributions have been made to the study of control in complex networks, which lie outside the scope of this article. We wish to point interested readers to the following reviews that focus entirely on network control tools. For a review of methods to identify control points to affect particular dynamics such as synchrony, see \textcite{Chen2014}. For more general background and detail on network control in complex systems, the recent review by \textcite{RevModPhys.88.035006} provides an excellent summary of the latest developments. 

\subsection{Control energy and metrics\label{controlmetrics}}

Another important area of work lies in the development of metrics that characterize different control strategies for real networks. We define the controllability $\mathbf{W}_{\mc K,T}$  as
\begin{eqnarray}
  \mathbf{W}_{\mathcal{K},T} &=& \sum_{\tau =0}^{T-1}\mathbf{A}^\tau
  \mathbf{B}_{\mathcal{K}}\mathbf{B}_{\mathcal{K}}^\transpose(\mathbf{A}^\transpose)^\tau\\ 
  &=&\mathbf{C}_{\mathcal{K},T}\mathbf{C}_{\mathcal{K},T}^\transpose
\label{eq:Gramian}
\end{eqnarray}
which has to be full rank for the network \eqref{eq:controlled} with the set of network nodes $\mc K$ to be controllable, equivalent to the condition for the controllability matrix in Section \ref{s:network}.

In practical applications, controllable networks featuring small Gramian eigenvalues cannot be steered to certain states because the control energy is limited. This fact motivated \textcite{pasqualetti2014controllability} to propose certain control strategies and associated metrics based on minimizing the control energy; these include average, modal, and boundary controllability. 

To define these control metrics, we first let the network be controllable in $T$ steps, and let $\mathbf{x}_f=\mathbf{x}(T) $ be the desired final state in time $T$, with $||\mathbf{x}_f||_2 = 1$, where the subscript denotes the Euclidean norm, i.e. $||\mathbf{v}||_2 := \sqrt{\mathbf{v}^\transpose\mathbf{v}}$. Following from Eq.~\ref
{eq:controlled}, where  $\mathbf{u}_{\mc K}$ is the injected control
signal, we can define the energy of the control input $\mathbf{u}_\mathcal{K}$ as
\begin{equation}
E(\mathbf{u}_\mathcal{K},T) = ||\mathbf{u}_\mathcal{K} ||^2_{2,T} = \sum_{\tau=0}^{T-1}||\mathbf{ u}_\mathcal{K}(\tau)||_2^2,
\end{equation}
where $T$ is the control horizon. The unique control input that steers the network state from $\mathbf{x}(0)=\mathbf{0}$ to $\mathbf{x}(T)=\mathbf{x}_f$ with minimum energy is \cite{kailath1980linear}
\begin{equation}
\mathbf{u}_{\mathcal{K}}^*(t) =\mathbf{ B}^\transpose_{\mathcal{K}}(\mathbf{A}^\transpose)^{T-t-1}\mathbf{W}_{\mathcal{K},T}^{-1}\mathbf{x}_f
\end{equation}
with $t\in\{0,\dots,T-1\}$. Then it can be seen that
\begin{equation}
E(\mathbf{u}_{\mathcal{K}^*}, T) = \sum_{\tau=0}^{T-1} ||\mathbf{u}_{\mathcal{K}}^*(\tau)||_2^2 = \mathbf{x}_f^\transpose\mathbf{W}_{\mathcal{K},T}^{-1}\mathbf{x}_f\leq\lambda_{\min}^{-1}(\mathbf{W}_{\mathcal{K},T}),\label{eq:energy}
\end{equation}
where $\lambda_{\min}$ is the smallest eigenvalue. Note that
equality is achieved whenever $\mathbf{x}_f$ is an eigenvector of $\mathbf{W}_{\mathcal{K},T}$ associated with $\lambda_{\min}(\mathbf{W}_{\mathcal{K},T})$ \cite{pasqualetti2014controllability}. 

\begin{figure}[t]
\includegraphics[width=0.65\linewidth]{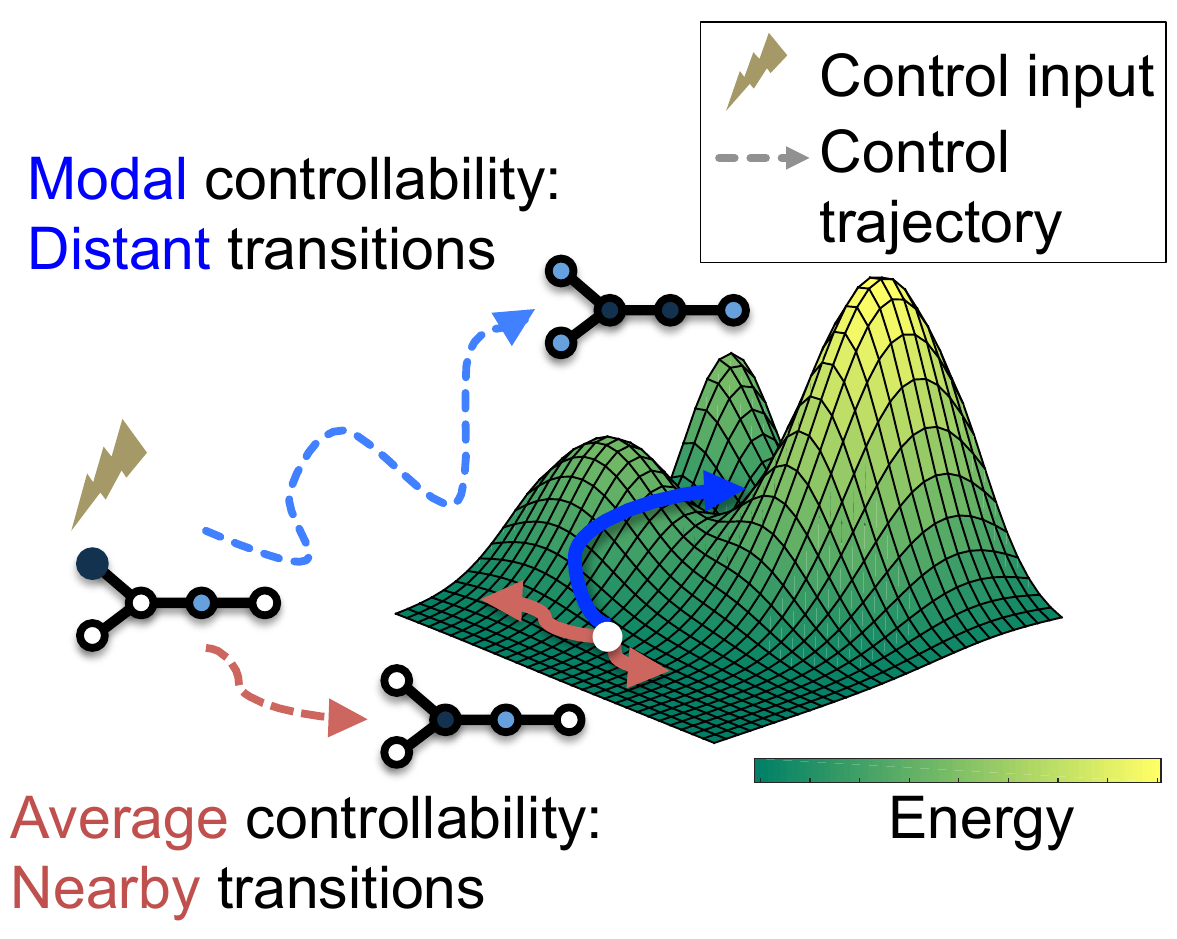}
\caption{\textbf{Energetic costs of controllabilitry metrics.}  \textcite{pasqualetti2014controllability} propose realistic control strategies that include the energetic costs of control \eqref{eq:energy}. Average controllability describes transitions nearby on an energy landscape, while modal controllability describes transitions distant on this landscape.} \label{fig:energy}
\end{figure}

\textit{Average controllability} identifies network nodes that, on average, can steer the system into different states with little effort (i.e., input energy), see Fig. \ref{fig:energy}. The average controllability in a network---formally defined as $\text{Trace} (\mathbf{W}_{\mathcal{K},T}^{-1})$---equals the average input energy from a set of control nodes and over all possible target states \cite{BM-DK-DG:04,hrs-mt:12}. Instead, $\text{Trace}( \mathbf{W}_{\mathcal{K},T} )$ is often adopted as a measure of average controllability, motivated by the relation $\text{Trace} (\mathbf{W}_{\mathcal{K},T}^{-1}) \ge N^2 / \text{Trace} (\mathbf{W}_{\mathcal{K},T})$ \cite{Summers20143784}, and the fact that $\mathbf{W}_{\mathcal{K},T}$ is close to singularity even for networks of small cardinality. Note that the maximization of $\text{Trace}(\mathbf{W}_{\mathcal{K},T})$ does not automatically ensure controllability. However, independent tests to verify the controllability can be made using Eq. (4) and were done for individual regions in brain networks \cite{gu2015controllability} (and more generally in \textcite{menara2017structural}). It should be noticed that $\text{Trace}(\mathbf{W}_{\mathcal{K},T})$ encodes a well-defined control metric, namely the energy of the network impulse response or, equivalently, the network $H_2$ norm \cite{kailath1980linear}. For practical computations,  the limit of $T\to\infty$ and $\mathbf{A}$ satisfying Schur stability is used, as this permits a closed-form solution and easier analysis. Intuitively, network nodes with high average controllability are most influential in the control of network dynamics over all possible target states.

\textit{Modal controllability} identifies network nodes that can push the network activity into difficult-to-reach states, which are those that require substantial input energy. To quantify modal controllability, we first note that the behavior of a dynamical system is fully determined by the eigenvalues (modes) and eigenvectors of its system matrix. Regarding controllability, the Popov-Belovich-Hautus test ensures that a system with matrix $\mathbf{A}$ is controllable by an input matrix $\mathbf{B}$ if and only if all its modes are controllable or, equivalently, if and only if there exists no left eigenvector of $\mathbf{A}$ orthogonal to the columns of $\mathbf{B}$ \cite{kailath1980linear}. By extension from this PBH test, if the entry $v_{ij}$ is small, then the $j$-th mode is poorly controllable from node $i$. Hence \textcite{pasqualetti2014controllability} define $\phi_i = \sum_{j} (1 - \xi_j^2 (\mathbf{A})) v_{ij}^2$ as a scaled measure of the modal controllability of all $N$ modes $\xi_0 (\mathbf{A}),\dots, \xi_{N-1}
  (\mathbf{A})$ from the brain region $i$. Intuitively, network nodes with high modal controllability are able to control all of the dynamic modes of the network, and hence to drive the dynamics towards hard-to-reach configurations. 

\textit{Boundary controllability} identifies network nodes that lie at the boundaries between network communities, beginning from communities at the largest scale and moving down across consecutive hierarchical levels of community structure -- and thus intuitively measures the ability to control the integration and segregation of network modules. This metric depends on the choice of a method for detecting boundary control points, for which an algorithm is proposed in \textcite{pasqualetti2014controllability}. This algorithm can be altered as needed for the physical system under study, e.g., to enhance the accuracy in estimating an initial partition of the network into communities, and to sharpen or loosen the boundary point criteria. Intuitively, network nodes with high boundary controllability are able to gate information between different communities, across topological scales in the network.

Overall, these three metrics provide useful estimates for real systems especially when considering dynamics over the whole network \cite{wuyan2018benchmarking}. Further work could be done to investigate other scenarios such as dynamics in just parts of the network, or how different patterns of community structure change the resulting controllability. These and more general questions about the relationship between network topology and the resulting dynamics remain open areas of study, which we discuss in more detail at the end of this article.

\subsection{Application to brain networks}

\begin{figure}[t]
\includegraphics[width=0.85\linewidth]{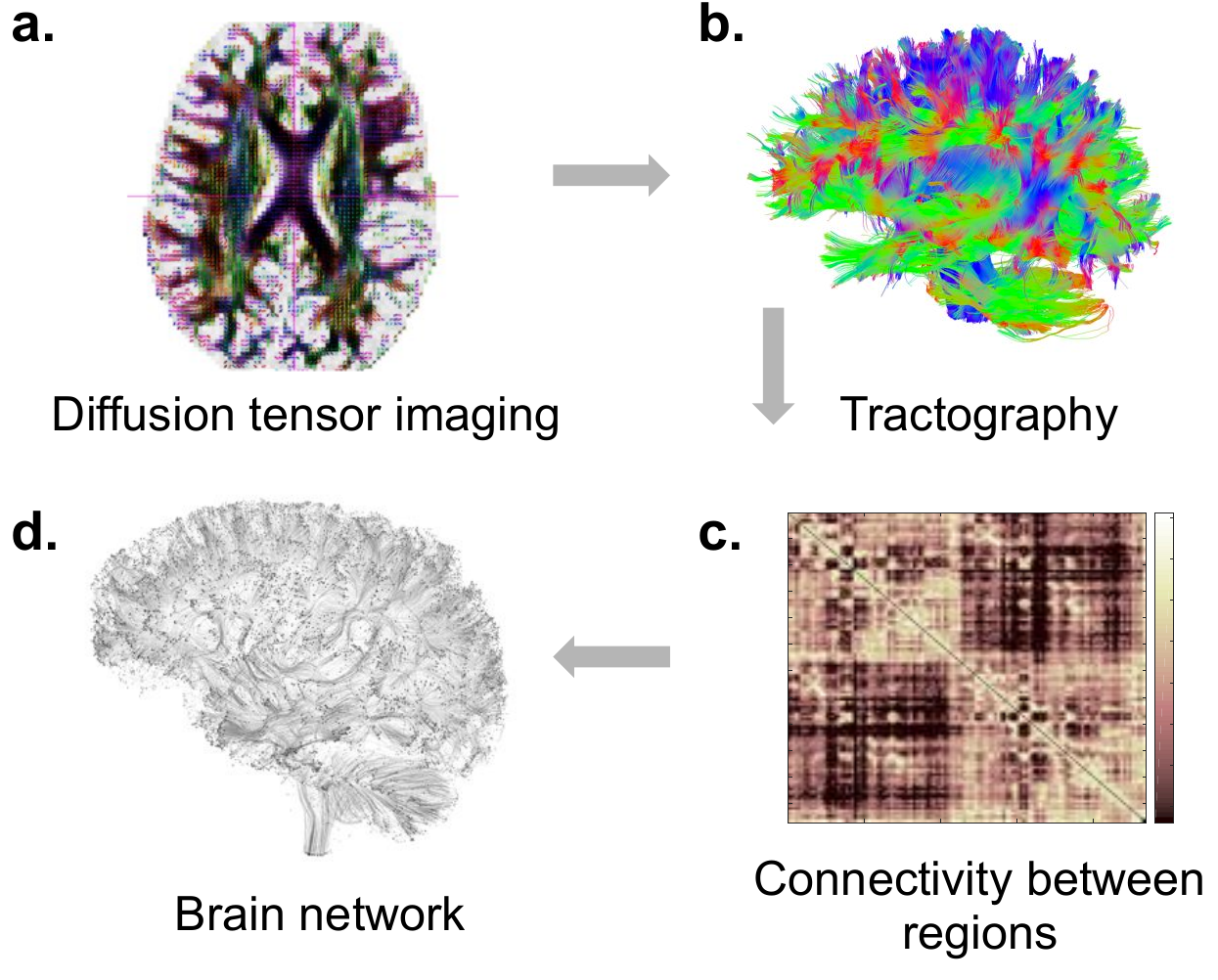}
\caption{\textbf{Construction of a human brain structural network. (a)} Diffusion imaging measures the direction of water diffusion in the human brain. \textbf{(b)} From these data, white matter streamlines can be reconstructed that connect brain regions. \textbf{(c)} An adjacency matrix representation of the structural connectivity: entries denote the estimated strength of white matter connectivity between brain regions. \textbf{(d)} The resulting brain network where nodes are brain regions, and where edges are the connection strengths between them.} \label{fig:brainnetworks}
\end{figure}

To use these methods to answer questions in neuroscience, we must begin by constructing networks based on our knowledge of brain connectivity.  At the large scale, network nodes in the brain are often defined based on regional differences in cellular architecture \cite{brodmann1909verg,glasser2016multi} or local gradients in fine-scale functional connectivity \cite{yeo2011organization,power2011functional}. Connectivity between these nodes can be estimated with emerging neurotechnologies, which we illustrate with the following examples. In humans, one particularly powerful non-invasive probe of connectivity uses magnetic resonance imaging (MRI) to infer structural pathways in the brain \cite{wedeen2012geometric} by exploiting molecular resonances of water molecules as they diffuse along white-matter tracts \cite{makris1997morphometry,basser1994MR}, see Fig. \ref{fig:brainnetworks}. By reconstructing the pathways that exist between brain regions and by estimating the strengths of those pathways, a brain network (weighted, symmetric graph) is obtained where the network edges are given by the inter-regional connection strengths \cite{Hagmann2008,Bassett2011}. Similar techniques can be used in rodents, cats, dogs, and non-human primates by way of a small-bore magnet \cite{duong2010diffusion}. Of course, tract-tracing techniques and other invasive methods are also a powerful way to image structural pathways in non-human animals \cite{okano2015brain,markov2011weight}.

Recently, \textcite {gu2015controllability} applied network control theory to such whole-brain structural networks in humans. Using networks composed of between 83 and 1015 nodes, the authors study the three controllability metrics of average, modal, and boundary controllability \cite{pasqualetti2014controllability} discussed in the previous section. Their work and others will be discussed in detail in the next section on understanding healthy brain function. While these techniques have not yet been ubiquitously applied to non-human imaging \cite{badhwar2015control,tang2012identifying}, the mathematics is generalizable to any estimate of structural connectivity in a neural system. Conceptually, this approach supports the general study of the kinds of dynamics predicted by the constraints of structural connectivity, particularly for the scenario in which a given brain region is acting as a control point for the rest of the network. On a methodological note, the results were verified across a range of network sizes. Although the connectivity studied is at a relatively coarse scale, it would be interesting to complement these observations with studies at cellular resolutions.

An integral aspect of control theory is that of system observability, which examines how measurable the system is to an observer. It is dual to system controllability; hence limits on the observability of the system will naturally impair efforts to control the system. This fact has important implications in neuroscience, where the lack of complete and constant detection, especially in living, behaving systems, introduces nontrivial uncertainty in both data and models. In non-invasive neuroimaging, systematic biases in data acquisition and processing may hamper accurate predictions built from individual measurements, e.g. that arise from the physical embedding of the brain \cite{MORRIS20081329,YamadaE14}. Common attempts to combat this possibility include verifying the reproducibility of results under a variety of choices made in the estimation of anatomical connectivity and in the construction of brain networks, for instance by comparing the results from multiple brain parcellations or tractography procedures. In time-varying networks, it should be verified that any conclusions hold over several time window lengths, and a minimum length of window should be chosen to ensure statistical signficance. Still, further work should be done to quantify how systematic biases in data acquisition or system observability, such as the effects of the physical embedding of the brain, result in bounds on the possible control predictions.

\section{Understanding healthy brain function through control theory}
\label{s:understanding_healthy}

In this section, we explore the utility of network control theory for offering mechanisms of cognitive control, providing explanations for individual differences in cognitive control across people, and capturing the evolution of control as we grow from children to adults. We close this section by discussing open questions in cognitive neuroscience that appear particularly amenable to extensions of network control theory.

\subsection{Network control as a partial mechanism for cognitive control}

\begin{figure*}[t]
\includegraphics[width=0.72\linewidth]{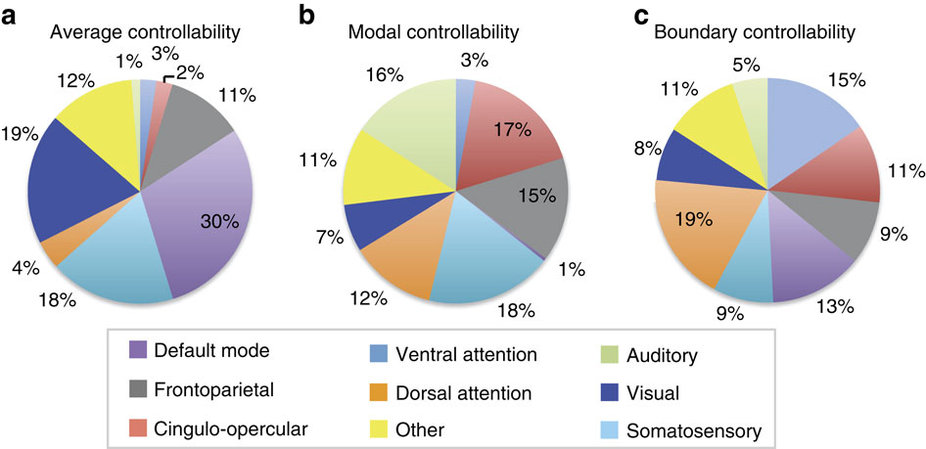}
\caption{\textbf{Cognitive control hubs are differentially located across cognitive systems. (a)} Hubs of average controllability are preferentially located in the default mode system. \textbf{(b)} Hubs of modal controllability are predominantly located in cognitive control systems, including both the frontoparietal and cingulo-opercular systems. \textbf{(c)} Hubs of boundary controllability are distributed throughout all systems, with the two predominant systems being ventral and dorsal attention systems  \cite{gu2015controllability}. } \label{fig:gu}
\end{figure*}  

A simple question to ask about any theory is whether or not it offers predictions of observed processes. One particularly straightforward and testable hypothesis is that the common control strategies studied in control and dynamical systems theory are strategies that the brain uses to control its own intrinsic dynamics. In a recent study, \textcite{gu2015controllability} addressed this hypothesis by first calculating the controllability strengths for each brain region, and then by identifying the preferences of each brain region for different types of control. The authors found that strong average controllers, strong modal controllers, and strong boundary controllers were located in quite different areas of the brain, see Fig. \ref{fig:gu}. 

Notably, the different sorts of controllers appeared to map on to the types of function that each brain region is thought to perform. For example, strong average controllers were disproportionately located in the default mode system, which is a spatially distributed set of brain regions that are markedly active when a person is simply resting \cite{raichle2015brains}. This is particularly interesting because it suggests that areas of the brain that are active in the ``ground state'' are also areas that are structurally predicted to optimally push the system into many local easily-reachable states, close by on the underlying energy landscape. Furthermore, strong modal controllers were disproportionately located in cognitive control systems, including both the frontoparietal and cingulo-opercular systems. This is particularly interesting because it suggests that the areas of the brain that are active during tasks that demand high levels of cognitive control or task switching \cite{botvinick2015motivation} are structurally predicted to optimally push the system into distant states, far away on the underlying energy landscape. Lastly, strong boundary controllers were disproportionately located in regions implicated in attention \cite{corbetta2002control}, supporting their predicted role in gating \cite{womelsdorf2015long,eldar2013effects} information between network communities. 

This study offers a possible mechanistic explanation for how the brain might move between cognitive states that depends fundamentally on white matter microstructure. The work suggests that structural network differences between the default mode, cognitive control, and attentional control systems dictate their distinct roles in brain network function. While the results need to be validated in other species and data sets, the broad trends indicate the relevance of control theory for capturing canonical concepts in cognitive control.

\subsection{Network control and cognitive performance}

In the previous section, we reviewed evidence that notions from network control applied to neuroimaging data can provide insight into the roles that brain regions may play in the control of neural dynamics. Here we ask the more specific question of whether the brain in one person (or animal) might be optimized for a different type of control than the brain in another person \cite{kim2018role}. That is, can controllability metrics explain why cognitive performance differs across individuals \cite{cornblath2018sex}? 

While still a very open question, two recent studies suggest that indeed each brain displays a different profile of control, and differences across people are correlated with differences in their cognitive capacities. In one study in healthy adult humans, \textcite{1606.09185} compare the predictions from network control theory applied to individual brain images to the performance of these same individuals on traditional  cognitive control tasks. More specifically, the authors calculate modal and boundary controllability (see \ref{controlmetrics}) on brain networks obtained from diffusion imaging, and they also test the performance of subjects in cognitive control tasks that measure the inhibition of behavior, the shifting of attention, vigilance, and working memory capacity.  The study reports key regional controllers in the brain whose controllability strength is correlated with task performance measures across individuals, thus providing a second line of evidence that network control may be a partial mechanism for cognitive control in humans.

Turning from adults to children, \textcite{Tang2017} evaluated the controllability strength of brain regions as well as more general cognitive performance (not specific to cognitive control) in a community-based sample of healthy youth. The authors found that the relative strength of average controllers in subcortical \textit{versus} cortical regions (which are the earliest evolving and latest evolving brain areas, respectively) is an important predictor of improved cognitive performance. This relationship held true even when accounting for differences in age across the cohort, suggesting that it is a fundamental characteristic of human brain structure and dynamics. A follow-up study further tied these differences to individual differences in cognitive control specifically \cite{cornblath2018sex}.

\subsection{Evolution of network control in development\label{dev}}

The identification of age-invariant relationships between controllability metrics and cognitive function begs the question of whether controllability metrics of brain networks change with age, either in their magnitude or in their spatial distribution. To address this question, \textcite{Tang2017} studied the controllability metrics of average controllability and modal controllability in 882 healthy youth from 8 to 22 years of age, and quantified a single value of controllability for a person as given by the average of controllability strengths across all brain regions. This coarse-graining of the data enabled the authors to study how brain networks facilitate energetically easy transitions (average controllability) as well as energetically costly ones (modal controllability). 

\begin{figure}[t]
\includegraphics[width=0.98\linewidth]{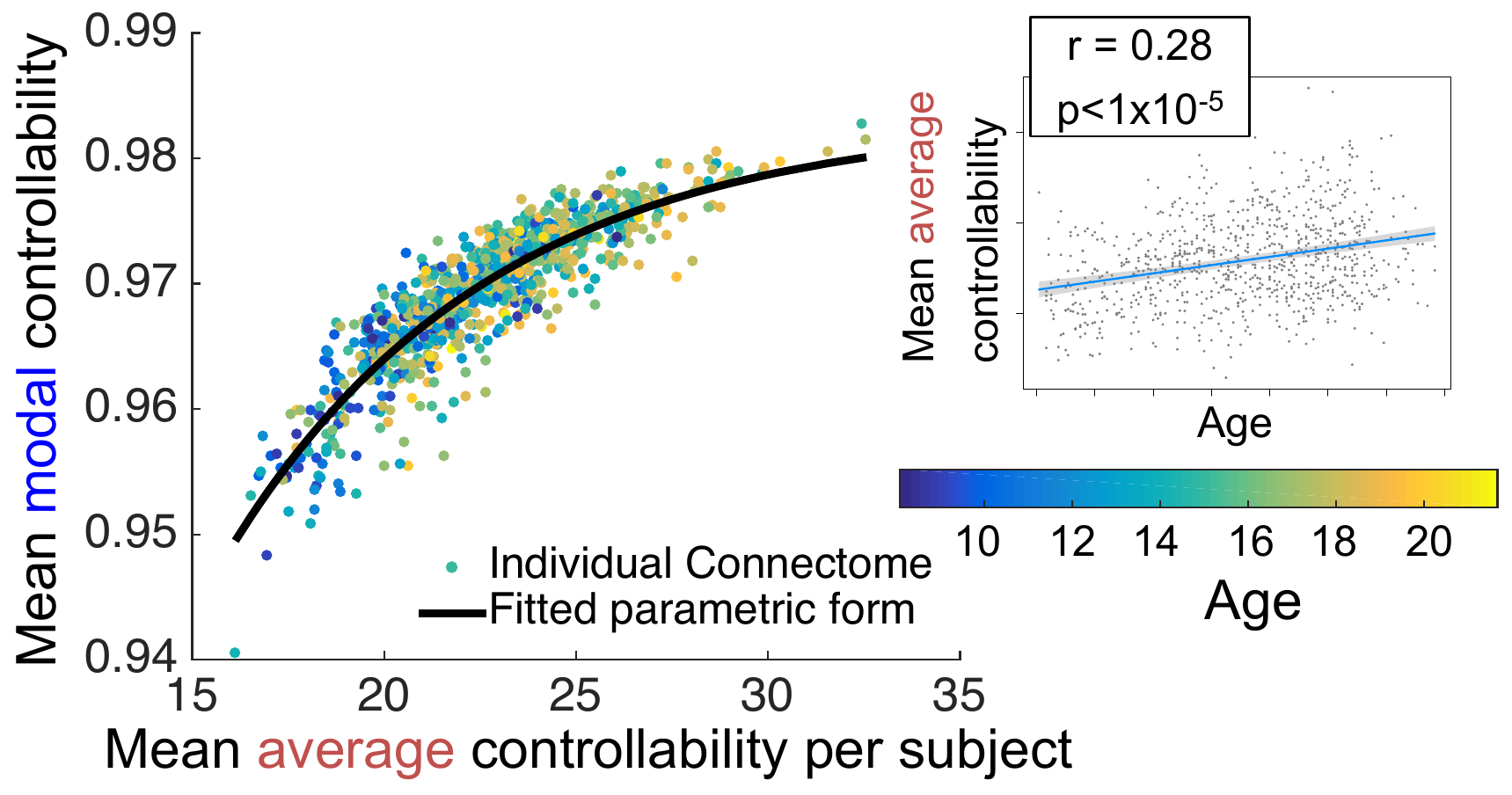}
\caption{\textbf{Controllability metrics are positively correlated with age, with older youth displaying greater average and modal controllability than younger youth.} Each data point represents the average strength of controllability metrics calculated on the brain network of a single individual, in a cohort of 882 healthy youth from ages 8 to 22 years. Brain networks were found to be optimized to support energetically easy transitions (average controllability) as well as energetically costly ones (modal controllability). There is a significant correlation between age and the ability to support this diverse range of dynamics: see inset or color (online) that denotes the age of the subjects. Note that modal controllability being a weighted sum of normalized eigenvectors is always capped at 1, hence its smaller range as compared to average controllability is not meaningful; rather, the relative differences between the values are meaningful here. From \cite{Tang2017}.} \label{fig:dev}
\end{figure}

They found that brain networks are highly optimized to support a diverse range of possible dynamics (as compared with randomized versions of the networks) and that this range of supported dynamics increases with age, see Fig. \ref{fig:dev}. Seeking to investigate structural mechanisms that support these changes, the authors simulate network evolution with a set of growth rules, to find that all brain networks -- from child to adult -- become increasingly structured in a manner highly optimized for network control. These results suggest key neurophysiological changes that may be occuring during development, driving the system towards an increasing capability to traverse a larger surface of the energy landscape. It would be interesting in the future to assess whether these metrics are altered in youth with neuropsychiatric disorders, or whether they could be used to predict transition to psychosis. 
 
\subsection{Open questions in control and cognition}

It is important to note that linear models of neural dynamics \cite{galan2008how,honey2009predicting} for use in network control theory have both advantages and disadvantages. Their advantage is that one has access to a wide array of theoretical observations that can offer intuition about the system's (controlled) dynamics, particularly around an operating point \cite{gu2015controllability}. The disadvantage is that they cannot speak to neural processes that transition from one dynamical regime (limit cycles, fixed points, attractors) to another \cite{deco2012ongoing,golos2015multistability,muldoon2016stimulation}. In these cases, developing additional methods for control of nonlinear systems may be necessary.

One simple scenario in which limit cycles -- or transitions between them -- may be particularly important for the processes of cognitive control is that of human decision-making \cite{chand2016salience,chand2016face}. For example, oscillatory activity in specific brain regions has been linked to rational \emph{versus} irrational decision-making in a task that requires financial judgements (akin to gambling). \textcite{7591459} studied a group of human subjects in which multiple depth electrodes were implanted in deep brain structures as a part of routine presurgical evaluation for medically refractory epilepsy. By recording the local field potentials at each of these electrodes, the authors were able to monitor the activity of neuronal ensembles in the precuneus and show that high-frequency activity (70-100 Hz) increased when irrational decisions were made. Further, transitions between various mental states such as rational or irrational decision making could be described using a state space model of activity from these electrodes, illustrating the network aspect of concerted activity between regions. This and similar studies in other areas of higher-order cognitive function that depend upon synchronized oscillatory activity in neuronal ensembles \cite{kopell2000gamma,bassett2009cognitive} suggest the possibility that control strategies could be devised that use brain stimulation to alter the frequency of neuronal synchrony to modulate cognitive processes. Such a possibility will depend on accurately extending linear control models to nonlinear ones, isolating the dynamics relevant for the cognitive process of interest, and localizing the region that is most impacted.

These studies cover a range of experimental probes from non-invasive neuroimaging to implanted electrodes, and computational models from linear models to nonlinear models. Together, they illustrate the breadth of scenarios in healthy cognitive function available for further investigation, and invite further work that identifies connections or common themes within these studies.

\section{Targeting therapeutic interventions to maximize beneficial outcomes to patients}
\label{s:understanding_disease}

In this section, we broaden our focus from linear models of network control in order to more generally discuss emerging engineering approaches for the control of brain dynamics in the context of clinical medicine. We separate our discussion into methods for modulating consciousness via anesthesia administration, methods for ongoing monitoring and treatment of Parkinson's disease, methods for non-invasive stimulation, and methods for the control of transient epileptic seizures. These topics are in no way meant to be comprehensive of the field, but simply to highlight important directions of clinical relevance. Examples are chosen based on their focus on distributed control and analysis over many brain regions, in view of the system as an interacting whole, where network models are often explicitly employed.

\subsection{Anesthesia titration}

Anesthesia is used in medical institutions to modulate consciousness through drugs during surgery, potentially by altering distributed circuitry \cite{crone2016testing}. Accurately titrating the levels of anesthetic for each person, and at each time point during the surgery, is critically important for the comfort, health, and survival of the patient. Recent efforts seek to optimize this titration using a closed-loop system \cite{doi:10.1097/ALN.0b013e31829d4ab4}, where the challenge is to maintain a medically-induced coma by delivering propofol via an intravenous catheter or pump. Using a computer to control this delivery system, precise amounts of anesthetic can be chosen, administered, and adapted in a time-dependent manner, potentially reducing the incidence of propofol overdose which is accompanied by debilitating side effects.

 \begin{figure}[t]
\includegraphics[width=\linewidth]{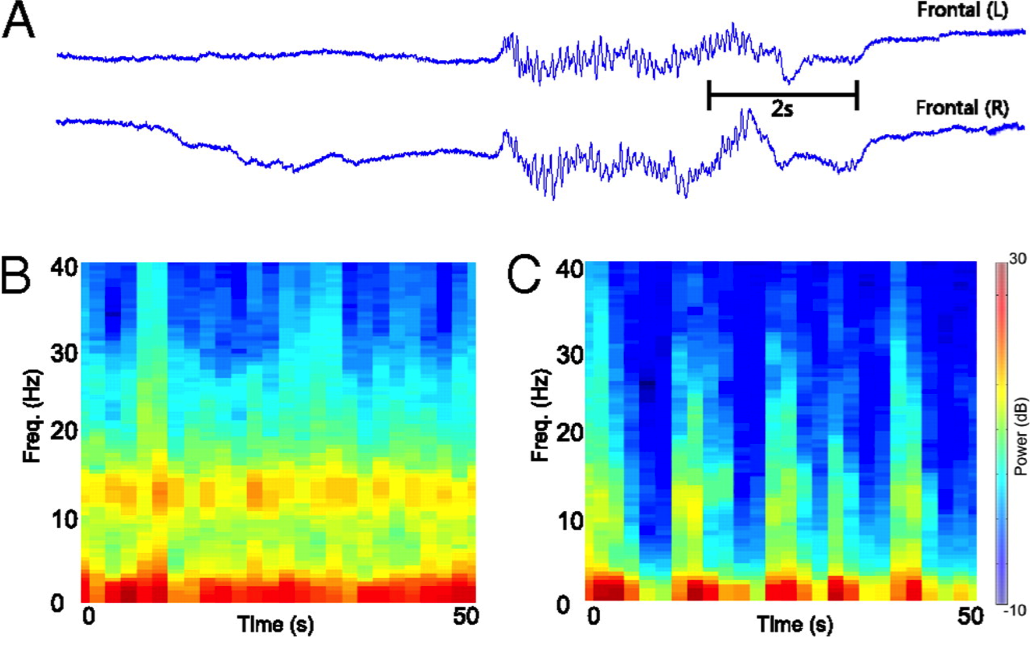}
\caption{\textbf{Burst suppression phenomenology. (a)} A typical recording of burst suppression from a human subject anesthetized with propofol -- a type of general anesthesia. The bursts manifest concurrently across the scalp (here, shown for left and right frontal electrodes). \textbf{(b)} Spectrogram for a frontal electrode during deep, but not burst-suppression, general anesthesia. \textbf{(c)} At a deeper level of general anesthesia, burst suppression is achieved (the spectrogram clearly displays epochs of quiescence). From \textcite{ching2012neurophysiological}.} \label{fig:bursts}
\end{figure}

Building on their earlier biophysical model, \textcite{doi:10.1097/ALN.0b013e31829d4ab4} demonstrate the real-time monitoring and control of the brain's burst suppression state from the electroencephalogram (see Fig. \ref{fig:bursts}), which indicates a state of highly reduced electrical and metabolic activity \cite{ching2012neurophysiological} and allows tracking of the level of consciousness. This state is illustrated via small model networks of two principal cell types (cortical pyramidal cells and inhibitory interneurons). Control of this state can then be done using an on-line parameter estimation procedure and proportional-integral controller. The technique has already been validated in rodents, where it can be used to successfully monitor and control the burst suppression state. Translating this work into humans will require more extensive computational estimation of model parameters and empirical validation over periods of several hours.

\subsection{Deep-brain stimulation for Parkinson's disease}

High-frequency deep brain stimulation (DBS), commonly used to treat Parkinson's disease, is one of the oldest examples of successful dynamical manipulation of brain function to alleviate clinical symptoms. Yet, it remains unclear exactly how and why it works so well. Control and systems theory approaches are useful for modelling the underlying circuitry to understand the mechanisms by which deep-brain stimulation affects behavioral phenotypes \cite{Santaniello10022015,PhysRevLett.81.3291,10.1371/journal.pcbi.1004673}.

Recent work has highlighted the network-level mechanisms of the diseased dynamics, and the control necessary to treat them. For example, \textcite{Santaniello10022015} move from localized functions to the relevant circuitry, positing that DBS increases the regularity of firing patterns in the basal ganglia, thereby decreasing symptoms of Parkinson's disease \cite{chiken2014disrupting}. The authors suggest that high-frequency stimulation of 130 Hz in DBS is effective because it is a resonant frequency of the overall cortico-basal ganglia-thalamo-cortical loop. The authors explore the effects of different stimulation conditions by simulating hundreds of biophysically realistic neurons from different regions of the circuitry that are thought to have very different functions. Their results suggest a loop-based reinforcement model, where DBS proximally or distally does not individually account for resulting pattern changes, but instead relies on a combined impact across the circuit. This observation could inform the choice of stimulation frequency and location when using DBS clinically \cite{johnson2013neuromodulation}. 

While identifying the resonant frequency of a critical circuit may provide a useful target for control, other mechanisms may also exist, and it is possible that interventions targeting more than one mechanism could be more effective than targeting one mechanism alone. Other candidate mechanisms include coupling between peripheral tremor rhythms, and the phase locking of the activity of primary and secondary motor areas. For example, \textcite{PhysRevLett.81.3291} propose two techniques to identify the relative phase locking between two MEG signals, thereby detecting synchronization of neuronal activity and mapping its relationship to peripheral tremors. Other attempts to uncover mechanisms include the investigation of entrainment and desynchronization dynamics, both seen in populations of neurons, as a result of DBS. \textcite{10.1371/journal.pcbi.1004673} study a population of model neurons and the effects of stimulation, to observe underlying low-dimensional patterns that can illuminate collective processes in spiking neurons. The simplicity of that particular model affords theoretical insight into a potential mechanism that governs DBS. 

Once the optimal mechanism(s) have been identified, a key goal is the use of control theory to create a closed-loop system for more effective treament. \textcite{Holt2014} identify their goal for DBS as the suppression of pathological frequencies that occur during Parkinson's disease. They simulate the physiology of the basal ganglia using a network model to create a mean-field description of the closed-loop system, which allows for the tuning of stimulation parameters based on patient physiology. This setup provides significant advantages over the current method of trial-and-error tuning, which is based on the clinician's past experience. If such a model can be empirically validated, it would be an important step towards improving the efficacy of DBS for patients with Parkinson's disease.

\subsection{Non-invasive transcranial stimulation}

While such invasive monitoring and stimulation paradigms are not accessible to most humans, other non-invasive methods of brain stimulation are becoming increasingly feasible. The most common is that of transcranial magnetic (electric) stimulation, which is the application of a magnetic (electric) field through the scalp for a short period of time \cite{bikson2016safety}. While the effects of transcranial stimulation tend to be diffuse, they have demonstrated utility in treating depression and other neurological and psychiatric disorders \cite{kedzior2016cognitive}. In healthy subjects, transcranial electric stimulation has been shown to differentially affect endogenous versus exogenous attention in human subjects \cite{flavio}. These and similar effects can be understood to some degree by employing computational models of oscillatory and state-dependent dynamics \cite{10.1371/journal.pbio.1002424}. Computational work has also begun to directly bridge mathematical models of nonlinear neural dynamics with the predictions of network control theory in the context of such exogeneous stimulation \cite{muldoon2016stimulation}. The tractability of computational studies and the pervasive empirical use of non-invasive stimulation opens the possibility of building mechanistic models that provide a deeper understanding of stimulation's effects on the brain \cite{johnson2013neuromodulation}, and of the rules by which stimulation parameters and location can be optimized to enhance brain function.

One study directly bridges mathematical models of nonlinear neural dynamics and the predictions of network control theory in the context of such exogeneous stimulation. \textcite{muldoon2016stimulation} consider the effects of electrical stimulation to a specific brain region using a model of nonlinear oscillators connected by a coupling matrix estimated from measured diffusion imaging data (Fig. \ref{fig:brainnetworks}). By simulating dynamics in this network of Wilson-Cowan oscillators, they can test for different regimes of desired functional outcomes supported by the network---if the effects of stimulation remain focal or spread globally---and compare these with the predictions from network control theory using the controllability metrics described in \ref{controlmetrics}. Importantly, their results validate linear network control predictions over eight subjects and more generally provide a model that can be used or tested in clinical settings, in order to strengthen the connection between theory and clinical practice.

\subsection{Seizure suppression in epilepsy}

\begin{figure}[t]
\includegraphics[width=\linewidth]{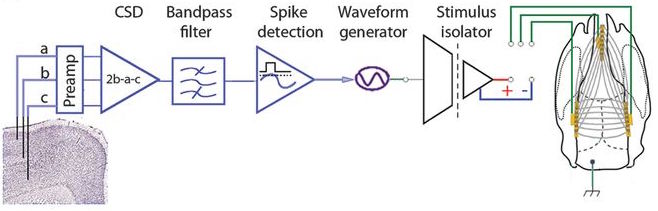}
\caption{\textbf{Closed-loop stimulation for seizure suppression in a rat.} Recordings from channels $a, b$ and $c$ in the cortex are filtered for spike detection, where signals exceeding the predetermined amplitude threshold are detected. These thresholded signals are used to trigger transcranial electric stimulation, which is applied through the scalp. From \textcite{Berenyi735}.} \label{fig:berenyi}
\end{figure}

Both invasive and non-invasive stimulation methods have been considered for the treatment of medically intractable epilepsy. This multiplicity of methods is due in part to the difficulties inherent in localizing the regions involved in seizures: different brain regions can play diverse roles in the production and propagation of epileptiform dynamics \cite{EPI:EPI13791}. Both types of interventions would seem to be preferable to the current clinical practice of resecting large sections of neural tissue thought to cause the seizure, although of course this statement is speculative \cite{stacey2008technology}. Instead, stimulation may have the potential to suppress seizures \cite{Ching2012,Berenyi735}, particularly if tailored to the underlying brain connectivity \cite{taylorruths}, and/or its associated dynamics \cite{Khambhati20161170}. In a recent practical demonstration, work from the group of \textcite{Berenyi735} shows the efficacy of brain stimulation in seizure suppression, in a rat model for epilepsy (see Fig. \ref{fig:berenyi}). Their application of transcranial electrical stimulation using a closed-loop system reduces seizure duration, on average, by 60\%. These results show great promise for the development of closed-loop stimulation that leaves other aspects of brain function unaffected, and paves the way for the use of such therapies in humans. 

For seizure suppression, some techniques appear to be effective for distributed control and others appear to be effective for local control. Theoretical modelling of the former case was done by \textcite{Ching2012}, who employ a grid of stimulating electrodes that act as actuators to help stem and direct the propagation of electrical activity. To model mesoscale cortical dynamics, they use a network of Wilson-Cowan oscillators, with both diffusive and synaptic coupling. By modelling the placement of several actuators, they demonstrate the ability to limit pathological activity (the spreading of electrical activity across a patch). By slowing the spread of activity, their method can be used in conjuction with pharmacological agents, or allow time for other self-correcting mechanisms in the brain. Naturally, their method would depend on how well the actuators contact and target the underlying tissue, as well as on accurate monitoring of seizure activity and the ability to control the system in real time. An alternative approach is put forth by \textcite{taylorruths}, whose model covers a larger spatial area and uses connectivity derived from patient MRI to facilitate personalization of stimulation. A simple dynamical model describes regional activity including epileptic spike wave dynamics, and a pseudospectral method  generates time-varying stimuli to halt simulated seizures.

\begin{figure}[t]
\includegraphics[width=\linewidth]{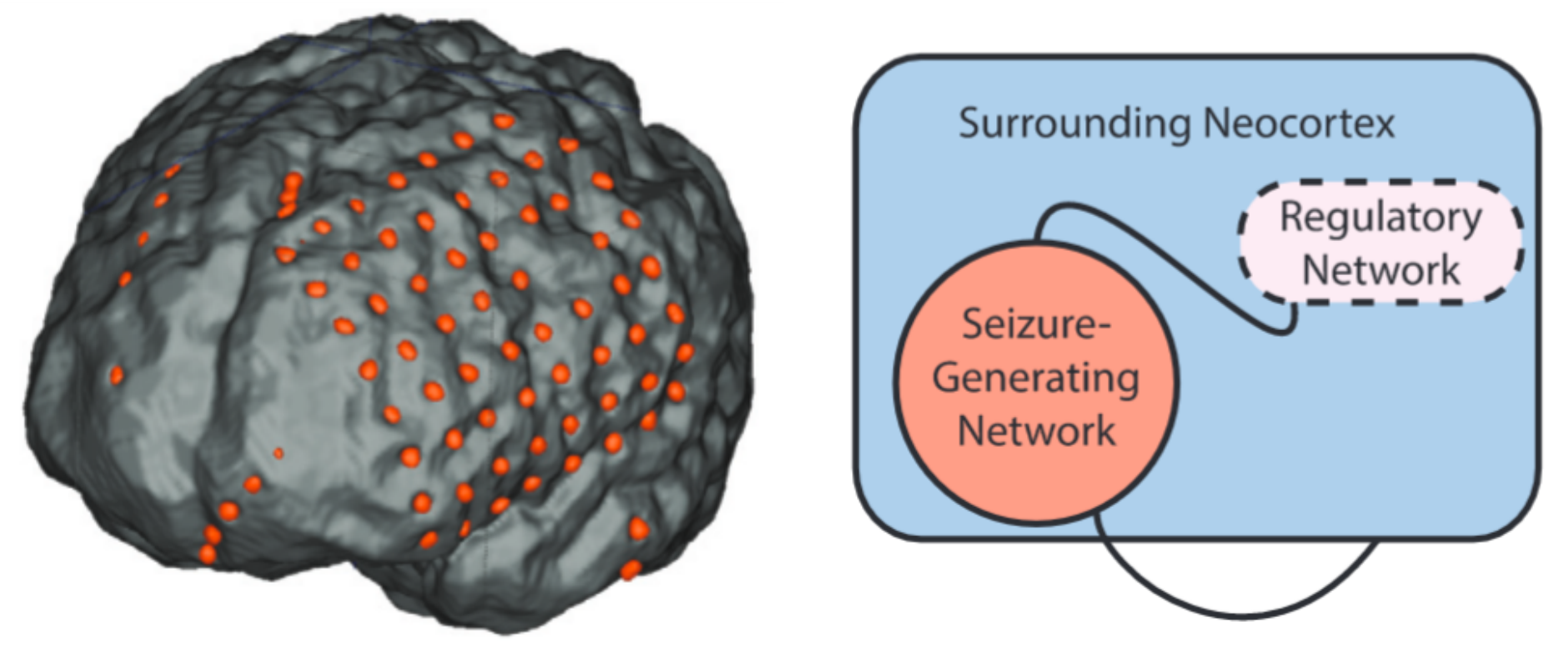}
\caption{\textbf{Schematics of patient electrophysiology and epileptic model.} \textit{Left:} Intracranial electrophysiology of patients with neocortical epilepsy. Each sensor (red dot) can be treated as a node within a functional network that uses magnitude squared coherence between sensors as network edges. \textit{Right:} A model of the epileptic network, comprised of a seizure-generating system and a hypothesized regulatory system that controls the spread of pathologic seizure activity. From \cite{Khambhati20161170}.} \label{fig:ankit}
\end{figure}

When considering translating some of these techniques to the clinic, it is useful to contrast them with existing clinical procedures. Generally speaking, clinical interventions for epilepsy can come in the form of (i) carefully modifying neural structure and dynamics, (ii) entirely quieting dynamics over short periods of time, or (iii) removing tissue to ensure silence over a lifetime. \textcite{Khambhati20161170} study methods to treat epilepsy via either short term ``lesioning'' (meaning quieting dynamics using stimulation) or long term ``resection'' (actually surgically removing the tissue). They develop methods for the identification of suitable lesion points, that affect the ability of the network to sustain synchronous activity associated with the occurence of a seizure (see Fig. \ref{fig:ankit}). These inferences are based on a measure of synchronizability of the network -- the ratio of the largest and smallest eigenvalues of the graph Laplacian \cite{barahona2002synchronization}. Virtual resection of individual brain regions \emph{in silico} can pinpoint control regions that strongly synchronize or desynchronize network dynamics, while revealing a principle of push-pull antagonism that provides a possible explanation for why seizures spread. Still, fully synchronized states only occur in a subset of seizure types, and it is therefore very likely that different sorts of control will be required for different sorts of seizure etiologies. Hence, the mapping from control type to seizure type will need to be validated experimentally, and further work is needed to clarify the translational applicability of this approach. 

Considering the large variability of epileptic synods and seizures (focal and generalized), these methods could add to the suite of possible interventions that include local control. The range of models in this section illustrates many possible direct applications of control theory to important medical questions, and the potential gains that could be made through the successful control of aberrant dynamics. This possibility for clinical impact is perhaps the most immediate motivation to study the control of brain dynamics, and we hope these examples will encourage new efforts in these areas.

\section{Control of specific neural dynamics or pathways}
\label{s:trajectories}

The example contexts in clinical medicine that we discuss in the previous sections highlight the great diversity of neural network dynamics in both health and disease. In this section, we focus on two specific types of network dynamics for which simple mathematical models can be studied, and for which control strategies can be examined analytically. The first context is that of neural synchrony, or rhythmic oscillations of neural ensembles. The second context is state transitions, where the activation profile of the brain moves from one pattern to another. We conclude the section by describing a few empirical tools that can be used to modulate these dynamics, and to test predictions from network control theory.

\subsection{Synchrony of neural populations}

\begin{figure*}[t]
\includegraphics[width=0.71\linewidth]{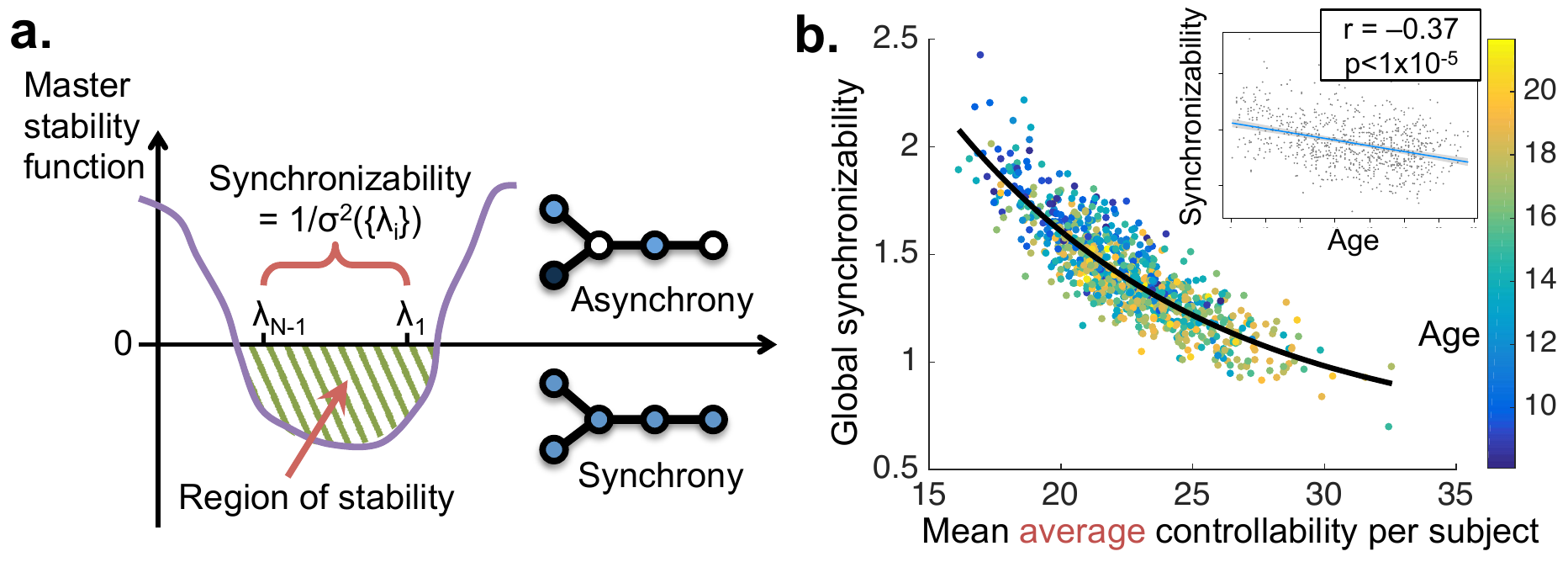}
\caption{\textbf{Synchronizability of structural brain networks and a negative correlation with age. (a)} Schematic of a master stability function (MSF) for a generic network of oscillators, which gives the perturbative stability of a globally synchronous state \cite{PhysRevLett.80.2109}. Such a state is stable when the MSF is negative for all positive eigenvalues of the graph Laplacian, hence the inverse spread of the Laplacian eigenvalues $1/\sigma^2(\{\lambda_{i}\})$ provides an estimate of synchronizability (or stability under synchrony), see \textcite{Nishikawa08062010}.  \textbf{(b)} Synchronizability in structural brain networks estimated from diffusion imaging in a large cohort of 882 youth is found to be anti-correlated with mean average controllability, as well as with age (see inset, or color online). From \cite{Tang2017}.} \label{fig:sync}
\end{figure*}

\subsubsection{Dynamical characteristics and clinical relevance}

When considering the control of specific dynamics, a natural place to start in neural systems is synchrony, which occurs when populations of neurons or brain regions exhibit the same dynamics $\mathbf{s} (t)$, i.e. $x_1 (t) = ... = x_n (t) =s (t)$ (see Fig. \ref{fig:sync}a). In many organisms, synchrony manifests as strong time-locked patterns, such as circadian rhythms and gait regularity. Moreover, the transition between synchrony and desynchrony has implications for treating epilepsy \cite{jirsa2014nature}, Parkinson's disease, or other pathological conditions. Hence the propensity towards synchrony or the ease of transitioning in and out of a synchronous state is of great interest -- both in local neuronal ensembles \cite{1741-2552-8-6-065008,lizdavison} and in distributed whole-brain networks \cite{Tang2017}. 

While this field is too large to do justice to in this small space, we highlight the work of \textcite{1741-2552-8-6-065008} as an excellent example describing the process of desynchronization in two models of coupled units (Kuramoto and a reduced phase Hodgkin-Huxley with electrotonic coupling), through the dynamic programming of inputs to a single neuron in the population. This work offers a possible mechanism for deep-brain stimulation in Parkinson's disease, where stimulation represents a single input that can affect desynchronization. Importantly, the model includes global (all-to-all) coupling between neurons, and therefore the use of more heterogeneous couplings that are characteristic of empirically measured brain networks could be an interesting future direction.

While understanding desynchronization processes is critically important, another relevant question pertains to the conditions under which synchrony can occur. While some efforts seek to address this question through the analysis of Lyapunov functions \cite{lizdavison}, the bounds are often of limited value as they are far from the regime in which we expect neural dynamics to take place. Alternatively, transient regimes toward synchrony and perturbative methods on synchronizability can be used to describe more realistic regimes. 

\subsubsection{Structural drivers of synchrony: Graph architecture and symmetries}

One framework to study the perturbative stability of a synchronous state or transients toward synchrony takes an explicitly structural approach. For instance, \textcite{PhysRevLett.80.2109} proposed the master stability function (MSF) to analyze the stability of this state on a network of oscillators. A schematic of this function for a generic network of identical oscillators is given in Fig. \ref{fig:sync}a. Within this framework, linear stability depends on the positive eigenvalues  $\{\lambda_{i}\}, i=1, ... ,N-1$ of the graph Laplacian $\mathbf{L}$ defined by $L_{ij}=\delta_{ij}\sum_{k}A_{ik}-A_{ij}$, where $\mathbf{A}$ is the network adjacency matrix defined in \ref{s:network}. More specifically, stability under perturbations exists when this function is negative for all positive eigenvalues of the Laplacian matrix.

Without a detailed specification of the properties of the dynamical units, a larger spread of Laplacian eigenvalues will typically make the system more difficult to synchronize than a smaller spread. Therefore, one natural measure of global synchronizability is the inverse variance $1/\sigma^2{(\{\lambda_{i}\})}$, as proposed by \textcite{Nishikawa08062010}:
 \begin{equation}
 \sigma^2=\frac{\sum_{i=1}^{N-1}|\lambda_i-\bar{\lambda}|^2}{d^2(N-1)}\textrm{,\quad where } \bar{\lambda}:=\frac{1}{N-1}\sum_{i=1}^{N-1}\lambda_i
 \end{equation}
and $d:=\frac{1}{N}\sum_{i}\sum_{j\neq i}A_{ij}$, the average coupling strength per node, which effectively normalizes the overall network strength.

\textcite{Tang2017} used this metric of global synchronizability to study the brain networks of 882 typically developing youth from the ages of 8 to 22 years. They found that brain networks that are more synchronizable tend to display lower average controllability (Fig.~\ref{fig:sync}b) as well as lower modal controllability. While no known relationship between synchronizability and controllability exists, the correlation is intuitive in that it suggests that individuals who are theoretically predicted to more easily transition into a variety of dynamical states are less susceptible to having many regions locked in synchrony. Interestingly, the relationship between synchronizability and controllability is partially explained by age: synchronizability decreases as children age (inset of Fig.~\ref{fig:sync}b). These results suggest that as the brain matures, its network architecture supports a larger range of dynamics (from nearby to distant states) perhaps necessary for the adult repertoire of cognitive functions, and is less able to support globally synchronized states which are instead characteristic of pathological conditions such as epilepsy. 

\begin{figure}[t]
\includegraphics[width=0.7\linewidth]{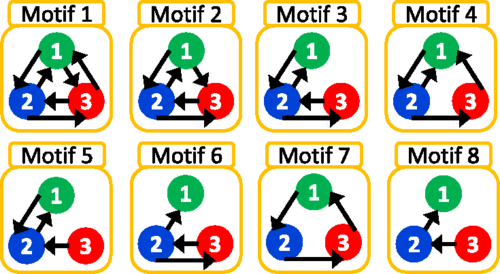}
\caption{\textbf{Motif structures that occur within networks.} The motif structures studied by \textcite{PhysRevX.5.011005}, through simulations of nonlinear biophysical neuronal models and their resulting controllability.} \label{fig:motifs}
\end{figure}

The emergence of local patterns of synchronization can follow different paths depending on the graph architecture, and hence suggest the existence of particular control strategies that may enact the desired path. \textcite{PhysRevLett.98.034101} probe this dependence on the network coupling strength and topology, as well as patterns in the transition to synchrony in a network representing structural measurements from cat cerebral cortex \cite{10.1371/journal.pone.0012313}. Such considerations that move beyond the linear stability of the synchronized state can provide insights into the design of real-world networks that often display small-world topologies. The concept of basin stability that can describe nonlocal and nonlinear systems is a powerful example, successfully describing features of neural networks such as the macaque or cat cortex \cite{Menck2013}. The control of synchrony hence has strong connections with nonlinear control, also exemplified when considering the role of structural symmetries. Indeed, critical work from \textcite{PhysRevX.5.011005} demonstrates that symmetries and motifs in the network structure have a nontrivial impact on the potential to control the system's dynamics. Their work addressing three-node motifs (see Fig. \ref{fig:motifs}) explores the possibility of introducing a group-theoretic component to the existing algebra of control theory. They conduct simulations of the motifs using biophysical neuronal models characterized by nonlinear dynamics as described by the Fitzhugh-Nagumo equations, which comprise a general representation of excitable neuronal membranes. They explore several dynamical regimes including chaotic, pulsed limit-cycle, and constant input limit-cycle, to see how different types of symmetries (such as rotational or mirror) affect the resulting controllability. Further work is needed to determine whether these effects on controllability generalize to scenarios in which the same 3-node motifs are embedded in a larger network, or in which the model of dynamics is changed from a cellular-level model to a macro-scale model of neuronal activity. In addition, other factors besides anatomical connectivity or network coupling strength (such as local dynamics or neurotransmitter levels) could also contribute to synchony and dynamics, and provide interesting directions for future investigation.

\subsection{The cost of controlling specific trajectories}

While the control metrics defined earlier (\ref{controlmetrics}) consider the cost of control, they necessarily coarse-grain over many different state transitions: average controllability measures the ability to move the system to (all) local states on the energy landscape, while modal controllability measures the ability to move the system to (all) distant states on the energy landscape. However, there are circumstances in real world networks -- and particularly in brain networks -- in which we would like to understand how to move the system from a specified initial state to a specified target state. In this general scenario, we might like to be able to compare the shape of different trajectories within state space, thereby providing intuitions regarding the feasibility of a specific transition and the accessibility of certain final states. 

\begin{figure}[b]
\includegraphics[width=0.7\linewidth]{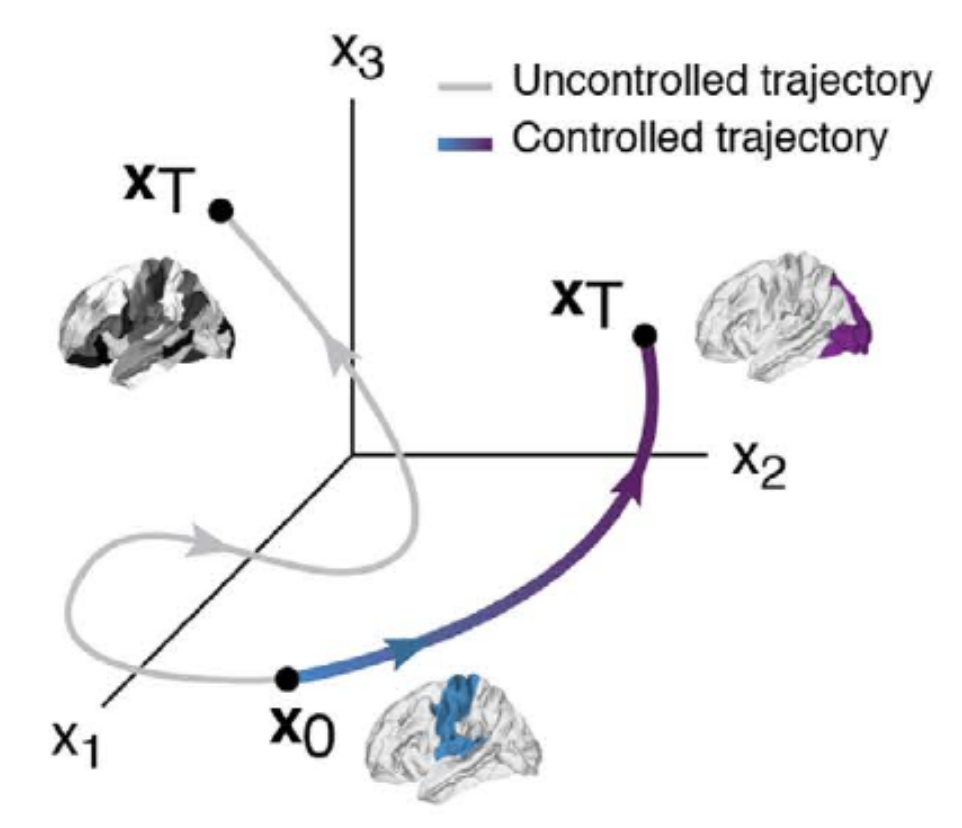}
\caption{\textbf{Example trajectory through state space.} With external input (control signals), the system at state $\bx_0$ is driven into the desired target state $\bx_T$; without input the system's passive 
dynamics leads to another state $\bx_T$ where random brain regions are more active than others. From \cite{Betzel2016}.} \label{fig:trajectories}
\end{figure}

In the context of the linear network system described earlier (Eq.~\ref{eq:controlled}), one proposed solution to this problem considers the trajectory from an initial state $\bx_0$ (one pattern of regional activation) to a target state $\bx_T$ (another pattern of regional activation), see Fig. \ref{fig:trajectories}. Our goal is to infer a control input function $\mathbf{u(t)}$ that minimizes the energy of the transition and the distance of the current state from the target (final) state:
\begin{equation}
\label{eqn_opt1}
\begin{aligned}
\min_\bu & & \int_{0}^T\left((\bx_T-\bx(t))^T(\bx_T-\bx(t)) + \rho \bu(t)^T \bu(t)\right) dt,\\
s.t. & & \dot{\bx}(t) = \A \bx(t) + \B \bu(t),\\
& & \bx(0) = \bx_0,\\
& & \bx(T) = \bx_T,
\end{aligned}
\end{equation}
where $T$ is the control horizon, $\rho\in\real_{> 0}$, and $(x_T - x(t))$ is the distance between the state at time $t$ and the target state.

Using this formulation, \textcite{1607.01706} study the energy landscape of finite-time control trajectories from the brain's baseline activation state to states with heightened activity in cortex devoted to vision, audition, and motor function. Interestingly, they observe that the most efficient drivers of these transitions were nodes in the network (or regions of the brain) with high communicability to the target state. Communicability examines the weighted sum of walks of all lengths, i.e. 
$G_{ij} = \sum_{k=0}^\infty (\frac{\mathbf{A}^k}{k!})_{ij} = (e^\mathbf{A})_{ij}$ in a binary network. The generalization to weighted networks is $G_{ij}^{w} = e^{\mathbf{A}'}_{ij}$, where $\mathbf{A}' = \mathbf{D}^{-\frac{1}{2}} \mathbf{A} \mathbf{D}^{-\frac{1}{2}}$ and $\mathbf{D}$ is the diagonal matrix with $D_{ii} = \sum_j A_{ij}$. Their results indicate the importance of long-distance walks on the network for efficient control. Moreover, by studying changes in the energetic impact of nodes on certain control actions, they also find that patients with mild traumatic brain injury show a loss of specificity in the putative control processes that their brain networks support. This work sheds light on the mechanisms that drive brain state transitions in healthy cognition and their alteration following injury. 

Similarly, \textcite{Betzel2016} simulate control trajectories among states characterized by the activation of various cognitive systems in the brain: systems devoted to visual, auditory, motor, baseline, cognitive control, salience, and attention-related functions. The goal was to compare energetic costs of these transition and to determine how this cost depends on the number of controllers used. The authors identify the brain regions that contribute most strongly to changes in energetic cost, and compare these with predictions from network control theory. In particular, they identify a group of control regions that are located in the rich club: a set of high-degree nodes that tend to also connect to one another \cite{colizza2006detecting}. Notably, these rich-club hubs acting as control regions most altered energetic outcomes when the brain's rich club organization was destroyed by simulated lesioning, an increasingly common model of neurodegenerative disease \cite{alstott2009modeling}.

Within this modeling framework, a choice of which trajectories to be simulated has to be made. Further work remains to identify the most useful trajectories for simulation that can reveal actual brain dynamics, thereby increasing biophysical relevance.

\subsection{Empirical tools for control of specific neural dynamics or pathways}

In the previous few subsections, we outlined theoretical frameworks and computational methods to model and interrogate the control of neural synchrony and brain state transitions. In each of these cases, it is and will remain important to inform and validate theories and models with empirical data, using experimental tools for control. Earlier in this report, we highlighted several of these tools in the form of brain stimulation, which have proven especially relevant for therapeutic interventions. However, in addition to these relatively large-scale tools, that are already being linked to control theory, there also exist fine-scale tools for the manipulation of single neuronal cell types \cite{lee2010global}, which could benefit from additional theoretical work. 

\begin{figure}[t]
\includegraphics[width=\linewidth]{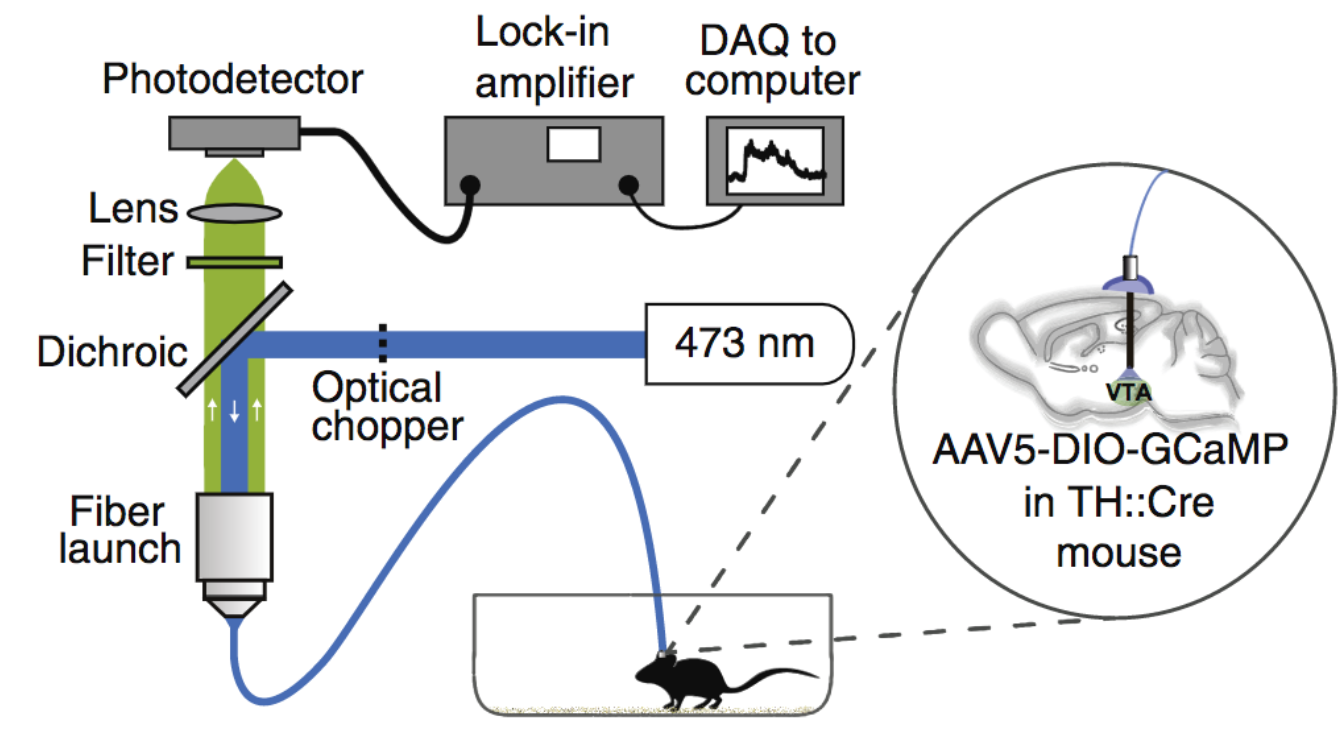}
\caption{\textbf{Setup for optogenetic control in a rat.} \textit{Left}: Fiber photometry setup showing light path for fluorescence excitation and emission through a single 400 micron fiber optic implanted in the ventral tegmental area (VTA). \textit{Right}: Recombinase-dependent viral targeting of GCaMP5 to VTA dopamine neurons. From \textcite{ref1}.}\label{fig:opto}
\end{figure}

Arguably one of the most powerful recently-developed tools for the manipulation of single cell types is optogenetics. Optogenetics offers millisecond-scale optical control of neural activity in defined cell types during animal behavior \cite{ref1}. Its marked precision, in some cases at single-cell resolution, allows the possibility to guide activity in awake animals and provide a causal investigation of neural circuitry, see Fig. \ref{fig:opto}. While mostly used in rodents, these techniques are increasingly being used in primates as well to probe basic principles of neural function, and to test strategies for therapeutic interventions such as the interruption of seizures; for further details we point readers to the recent review by \textcite{ref1}.

Meanwhile, technologies for simultaneously recording cell activity and targeting stimulation are constantly improving, and hence now allow the possibility for closed-loop control in animals. The very specificity of the stimulation and the targeted cells, means that at present specific design choices about intended outcomes have to be made. For instance, the same stimulation that evokes gamma oscillations ($>60$ Hz) at the circuit level using a relatively slow opsin variant ChR2(H134R) cannot always reliably drive individual pyramidal cells at such frequencies. Still, the ability to use such stimulation to direct behavior in animals, suggests tremendous potential for closed-loop optogenetics to reveal mechanisms for cognition.

These examples demonstrate new insights obtained through the modelling and probing of specific pathways and circuits in brain networks, and provide a controlled study of their role and contribution to the overall function of the brain. Further work could investigate how these pathways and circuits work in a concerted manner to affect cognitive function, as well as underlying principles in the design and use of these circuits.

\section{Emerging control methods with potential utility in neuroscience}
\label{s:future}
\begin{figure*}[t]
\includegraphics[width=0.98\linewidth]{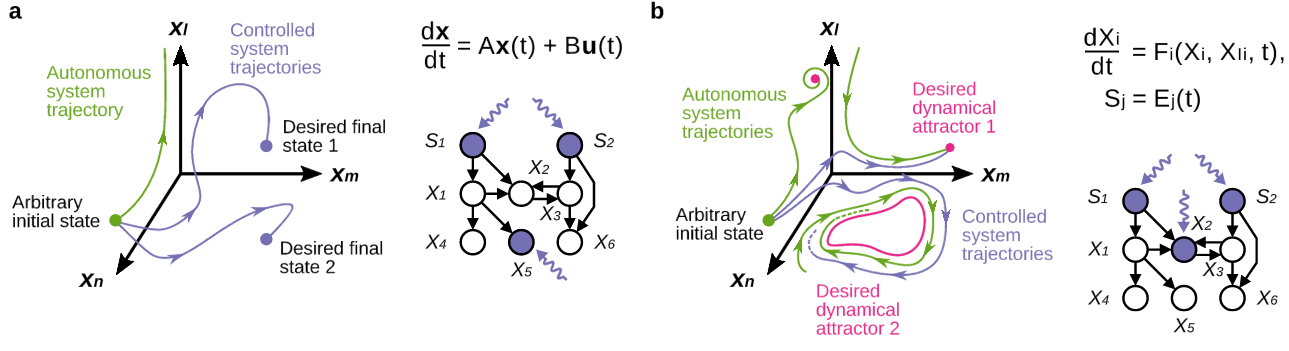}
\caption{\textbf{Comparison between structural controllability and control using feedback vertex sets. (a)} In structural controllability, the objective is to drive the network from an arbitrary initial state to any desired  final state by acting on the network with an external signal $\mathbf{u} (t)$. The dynamics are considered to be well-approximated by linear dynamics.  \textbf{(b)} In feedback vertex set control the objective is to drive the network from an arbitrary initial state to any desired dynamical attractor (e.g., a fixed point) by overriding the state of certain nodes. From \textcite{1605.08415}.}\label{fvs}
\end{figure*}

Many of these recently introduced theoretical frameworks to model the control of brain activity rest on linear or simplified models of dynamics. While they already provide useful conceptual insights and analytical descriptions for controlling neural activity, the large repertoire of dynamics in the brain requires more flexible models to capture its complexity. To close this review, we focus on two broad directions of advances in network control theory that appear particularly relevant for addressing this gap. The first is the extension of network control theory to describe a broader range of dynamical regimes -- such as nonlinear dynamics or time-dependent control -- or the study of control metrics to estimate the feasibility of control trajectories. The second examines new approaches in network control theory that exploit specific properties of the problem to better achieve desired targets, which may well differ based on the problem at hand. These include the use of perturbations, stochasticity in the system, or aspects of the network topology, to design control strategies.

\subsection{Broader control regimes}
\subsubsection{Nonlinear dynamics}

Brain activity is highly nonlinear, which can be seen especially at the level of single neurons or small groups of neurons. A recent analytical development that is mathematically exact for a broad range of nonlinear dynamics is that of feedback vertex sets (FVS) \cite{Fiedler2013}. It only requires a few conditions (e.g. continuous, dissipative, and decaying) that are typically satisfied by many real systems. This formalism identifies the set of nodes in a directed network that can control all the dynamics of the network and can steer it to the desired trajectories. Open-loop control applied to the nodes of an FVS allow for switching the dynamics of the whole system from one attractor to some other attractor. 

\textcite{1605.08415} provide an instructive discussion of the differences between structural controllability and control using FVS, as illustrated in Fig. \ref{fvs}. The authors use the FVS formalism to study several real networks. By comparing its predictions to those of classical structural controllability, they identify the topological characteristics that underlie the observed differences. In addition, they apply the FVS formalism to study dynamic models of gene regulation, in which directed networks can be used to model gene interactions. 

In cases where both the function and structure of the network are known, one can use simplified dynamical models such as logical dynamics (on/off states similar to the Ising model) to identify stable motifs that can control the dynamics of the network. Indeed, \textcite{10.1371/journal.pcbi.1004193} demonstrate that such an approach need only be applied transiently for the network to reach and remain in the desired state. The authors illustrate this method using a leukemia signaling network and a network for cell differentiation, giving rise to several predicted interventions that are supported by experiments.

\subsubsection{Time-dependent control} 

Given a possible lack of full information about the network, which is usually the case when one is estimating a brain network from empirical data, it is possible to identify strategies based on available data to define an uncertainty set containing all networks that are coherent with empirical observations. Indeed, \textcite{victorconical} propose a method to control the spread of a viral epidemic, taking place in a directed contact network with unknown contact rates. They assume that they have access to time series data describing the evolution of the spreading process, and propose a data-driven optimization framework to find the optimal allocation of protection resources. This method is illustrated using partial data about the dynamics of a hypothetical epidemic outbreak over a finite period of time---paving the way for inferring control strategies based on limited observational data over finite periods of time. These or similar methods may be particularly relevant for the control of seizure spread in the human brain given that the ``resource'' of brain stimulation is limited by the fact that too much stimulation causes heating of the tissue and eventual cell death.

Indeed, the question of cost and limited resources is futher investigated by \textcite{1607.06168}, who point out the possibility to take advantage of dynamically changing edges in a network to inform time-dependent control strategies, that may actually reach controllability faster than time-independent control strategies. This idea is based on the premise of energy savings in such strategies, by exploiting the changing topology to avoid energetically costly directions. For instance, they exert control towards the desired final state when the topology renders the energy cost acceptable, and pause when the topology makes the cost prohibitive. While suggestive of new designs for time-dependent control strategies that may prove more effective than static strategies, further work is needed to examine their relevance and feasibility in real neural systems. 

\subsubsection{Realistic control trajectories} 

\begin{figure}[t]
\includegraphics[width=0.7\linewidth]{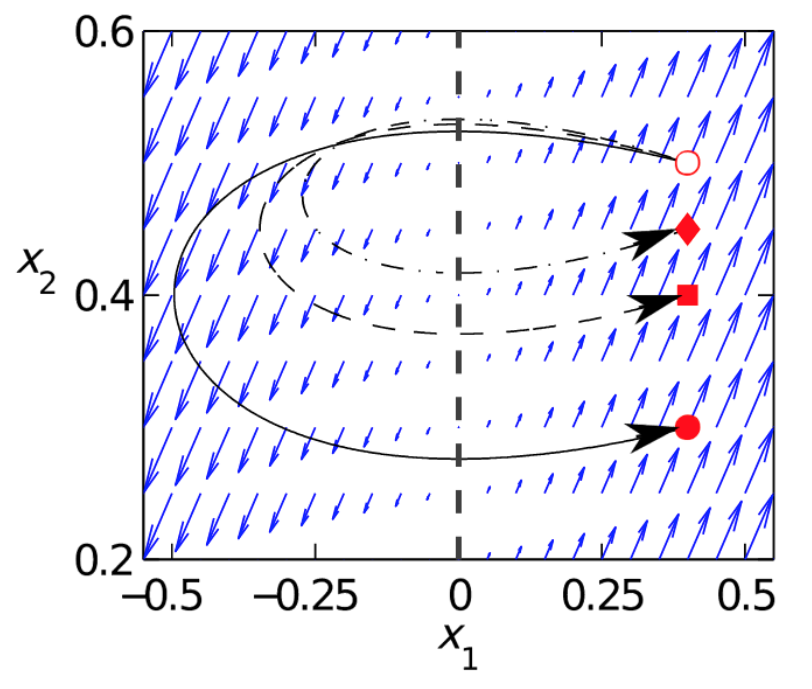}
\caption{\textbf{Two-dimensional example of nonlocal trajectories.} Example system $\dot{x}_1=x_1 + u_1(t),\dot{x}_2=x_1$, where the curves indicate minimal-energy control trajectories for the given initial state (open symbol) and target states (solid symbols). Background arrows indicate the vector field in the absence of control.  From \textcite{PhysRevLett.110.208701}.}\label{fig:nonlocal}
\end{figure}

\textcite{PhysRevLett.110.208701} investigate the control of dynamical trajectories in practice and what determines their energetics or feasibility. In particular, they point to the condition number of the controllability Gramian \eqref{eq:Gramian} as crucial for understanding control in practice, even if the corresponding Kalman's controllability matrix is well conditioned. Furthermore, they point out that numerical control fails even for linear systems if the Gramian is ill conditioned, and that control trajectories are generally nonlocal in the phase space (see Fig. \ref{fig:nonlocal}). Futher, they provide a condition for the numerical success rate of control strategies that depends on the number of control inputs, which they term the numerical controllability transition. Their work points towards additional criteria that would be relevant when considering the practicality of various control strategies in real systems.

\subsection{Exploiting system properties}
 
\subsubsection{Compensatory perturbations or noise} 

It is important to note that the study of control of brain network dynamics could also benefit from other methods that target neither nodes nor edges but instead identify effective parameters to design new strategies for control. The advantage of such approaches is their applicability for realistic regimes including nonlinear dynamics or stochastic systems. One such method proposed by \textcite{Cornelius2013} uses compensatory perturbations to steer the system to desired states: that is, perturbations to state variables that bring the system to the basin of attraction of the desired target state. The authors present methods to iteratively identify such compensatory perturbations, through consideration of the physically admissible perturbations, and through nonlinear optimization on this space of possible changes. Their approach is effective in bringing the system to a desired target state even when this state is not directly accessible, as they demonstrated through the mitigation of cascading failures in a power grid and the identification of drug targets in a cancer signaling network. 

Another such method identifies interventions that can reshape the topography of the underlying quasipotential in a desired way \cite{PhysRevX.5.031036}. This is achieved by determining the minimum action paths---those followed by the likely noise-induced transition trajectories---and the corresponding transition rates between all pairs of stable states. By optimizing these transition rates, the authors effectively alter quasipotential barriers between different stable states, which could be achieved biologically through, for example, a genome editing approach. This proposal exploits the response of biological systems to noise to induce a desired cell state, and thereby to predict and control noise-induced switching in genetic networks. While this method is demonstrated on models of cell differentiation, it is potentially useful for control in other classes of noisy complex networks.  

\subsubsection{Network topology}

Finally, understanding control in brain networks could benefit greatly from a better understanding of which  topological features and symmetries determine the controllability of a network. Recent work on this front has been pioneered by \textcite{GB-FP-SZ:15}, who study the controllability degree of complex networks as a function of the network diameter and the weights. By examining the energy required by a group of nodes to control the network to a desired state, the authors find that networks with a long diameter and anisotropic weights are easier to control than networks with a short diameter or isotropic weights. Here weights are defined to be isotropic if they allow a (control) signal to propagate equally in all directions, and to be anisotropic otherwise. 

Separately, \textcite{Ruths1373} discuss control profiles in real networks, by identifying topological features of the network (such as sources and sinks) that correlate with control properties. Building on these ideas, \textcite{Campbell2015} show that the number of source and sink nodes, the form of the in- and out-degree distributions, and local complexity (e.g., cycles) shape the control profile in empirical networks. Other work by \textcite{Posfai2013} examines the effects of clustering, modularity, and degree correlations on the minimal number of driver nodes required to control a network (similar to the problem posed by  \textcite{YYL-JJS-ALB:11}). They find that under certain conditions, only degree correlations have a discernible effect. 

Lastly, \textcite{1303.3907} investigate analytical relationships between network modularity or symmetries, and the resulting dynamics. They show that continuous time network dynamics can be decomposed into collections of interacting local control systems --- and that a class of maps called \textit{graph fibrations} give rise to conjugate dynamical systems. Their work provides a robust mathematical formalism to generalize existing understanding such as the relationship between symmetries and synchrony, through the broad notion of modularity.

\section{Conclusion}

We have discussed many new developments in the exciting field of controlling brain network dynamics and more importantly, attempted to highlight some of the many remaining open questions. This is an exciting time that has seen rapid theoretical and technological progress in methods of brain network control, or innovations that could be useful for brain network control. By outlining the potential in this young and emerging field, we hope to entice new practioners and further efforts towards this important goal of controlling brain network dynamics, that has great implications for the bettering of our health and cognitive function.

\begin{acknowledgments}

We thank Sergio Pequito, Ankit N. Khambhati, and Richard Betzel for helpful comments on earlier versions of this manuscript. D.S.B. and E.T. would also like to acknowledge support from the John D. and
Catherine T. MacArthur Foundation, the Alfred P. Sloan Foundation, the Army Research Laboratory and the Army Research Office through contract numbers W911NF-10-2-0022 and W911NF-14-1-0679, the National Institute of Health (2-R01-DC-009209-11, 1R01HD086888-01, R01-MH107235, R01-MH107703, R01MH109520, 1R01NS099348 and R21-M MH-106799), the Office of Naval Research, and the National Science Foundation (BCS-1441502, CAREER PHY-1554488, BCS-1631550, and CNS-1626008).The content is solely the responsibility of the authors and does not necessarily represent the official views of any of the funding agencies.
\end{acknowledgments}

\bibliography{braincontrol,bibfile_2,bibfile_new,bibfile_original}

\begin{thebibliography}{147}%
\makeatletter
\providecommand \@ifxundefined [1]{%
 \@ifx{#1\undefined}
}%
\providecommand \@ifnum [1]{%
 \ifnum #1\expandafter \@firstoftwo
 \else \expandafter \@secondoftwo
 \fi
}%
\providecommand \@ifx [1]{%
 \ifx #1\expandafter \@firstoftwo
 \else \expandafter \@secondoftwo
 \fi
}%
\providecommand \natexlab [1]{#1}%
\providecommand \enquote  [1]{``#1''}%
\providecommand \bibnamefont  [1]{#1}%
\providecommand \bibfnamefont [1]{#1}%
\providecommand \citenamefont [1]{#1}%
\providecommand \href@noop [0]{\@secondoftwo}%
\providecommand \href [0]{\begingroup \@sanitize@url \@href}%
\providecommand \@href[1]{\@@startlink{#1}\@@href}%
\providecommand \@@href[1]{\endgroup#1\@@endlink}%
\providecommand \@sanitize@url [0]{\catcode `\\12\catcode `\$12\catcode
  `\&12\catcode `\#12\catcode `\^12\catcode `\_12\catcode `\%12\relax}%
\providecommand \@@startlink[1]{}%
\providecommand \@@endlink[0]{}%
\providecommand \url  [0]{\begingroup\@sanitize@url \@url }%
\providecommand \@url [1]{\endgroup\@href {#1}{\urlprefix }}%
\providecommand \urlprefix  [0]{URL }%
\providecommand \Eprint [0]{\href }%
\providecommand \doibase [0]{http://dx.doi.org/}%
\providecommand \selectlanguage [0]{\@gobble}%
\providecommand \bibinfo  [0]{\@secondoftwo}%
\providecommand \bibfield  [0]{\@secondoftwo}%
\providecommand \translation [1]{[#1]}%
\providecommand \BibitemOpen [0]{}%
\providecommand \bibitemStop [0]{}%
\providecommand \bibitemNoStop [0]{.\EOS\space}%
\providecommand \EOS [0]{\spacefactor3000\relax}%
\providecommand \BibitemShut  [1]{\csname bibitem#1\endcsname}%
\let\auto@bib@innerbib\@empty
\bibitem [{\citenamefont {Achard}\ \emph {et~al.}(2006)\citenamefont {Achard},
  \citenamefont {Salvador}, \citenamefont {Whitcher}, \citenamefont
  {Suckling},\ and\ \citenamefont {Bullmore}}]{achard2006resilient}%
  \BibitemOpen
  \bibfield  {author} {\bibinfo {author} {\bibnamefont {Achard}, \bibfnamefont
  {S}}, \bibinfo {author} {\bibfnamefont {R}~\bibnamefont {Salvador}}, \bibinfo
  {author} {\bibfnamefont {B}~\bibnamefont {Whitcher}}, \bibinfo {author}
  {\bibfnamefont {J}~\bibnamefont {Suckling}}, \ and\ \bibinfo {author}
  {\bibfnamefont {E}~\bibnamefont {Bullmore}}} (\bibinfo {year} {2006}),\
  \bibfield  {title} {\enquote {\bibinfo {title} {A resilient, low-frequency,
  small-world human brain functional network with highly connected association
  cortical hubs},}\ }\href@noop {} {\bibfield  {journal} {\bibinfo  {journal}
  {J Neurosci}\ }\textbf {\bibinfo {volume} {26}}~(\bibinfo {number} {1}),\
  \bibinfo {pages} {63--72}}\BibitemShut {NoStop}%
\bibitem [{\citenamefont {Alagapan}\ \emph {et~al.}(2016)\citenamefont
  {Alagapan}, \citenamefont {Schmidt}, \citenamefont {Lefebvre}, \citenamefont
  {Hadar}, \citenamefont {Shin},\ and\ \citenamefont
  {Fr\"{o}hlich}}]{10.1371/journal.pbio.1002424}%
  \BibitemOpen
  \bibfield  {author} {\bibinfo {author} {\bibnamefont {Alagapan},
  \bibfnamefont {Sankaraleengam}}, \bibinfo {author} {\bibfnamefont
  {Stephen~L.}\ \bibnamefont {Schmidt}}, \bibinfo {author} {\bibfnamefont
  {J\'{e}r\'{e}mie}\ \bibnamefont {Lefebvre}}, \bibinfo {author} {\bibfnamefont
  {Eldad}\ \bibnamefont {Hadar}}, \bibinfo {author} {\bibfnamefont {Hae~Won}\
  \bibnamefont {Shin}}, \ and\ \bibinfo {author} {\bibfnamefont {Flavio}\
  \bibnamefont {Fr\"{o}hlich}}} (\bibinfo {year} {2016}),\ \bibfield  {title}
  {\enquote {\bibinfo {title} {Modulation of cortical oscillations by
  low-frequency direct cortical stimulation is state-dependent},}\ }\href
  {\doibase 10.1371/journal.pbio.1002424} {\bibfield  {journal} {\bibinfo
  {journal} {PLOS Biology}\ }\textbf {\bibinfo {volume} {14}}~(\bibinfo
  {number} {3}),\ \bibinfo {pages} {1--21}}\BibitemShut {NoStop}%
\bibitem [{\citenamefont {Alstott}\ \emph {et~al.}(2009)\citenamefont
  {Alstott}, \citenamefont {Breakspear}, \citenamefont {Hagmann}, \citenamefont
  {Cammoun},\ and\ \citenamefont {Sporns}}]{alstott2009modeling}%
  \BibitemOpen
  \bibfield  {author} {\bibinfo {author} {\bibnamefont {Alstott}, \bibfnamefont
  {J}}, \bibinfo {author} {\bibfnamefont {M}~\bibnamefont {Breakspear}},
  \bibinfo {author} {\bibfnamefont {P}~\bibnamefont {Hagmann}}, \bibinfo
  {author} {\bibfnamefont {L}~\bibnamefont {Cammoun}}, \ and\ \bibinfo {author}
  {\bibfnamefont {O}~\bibnamefont {Sporns}}} (\bibinfo {year} {2009}),\
  \bibfield  {title} {\enquote {\bibinfo {title} {Modeling the impact of
  lesions in the human brain},}\ }\href@noop {} {\bibfield  {journal} {\bibinfo
   {journal} {PLoS Comput Biol}\ }\textbf {\bibinfo {volume} {5}}~(\bibinfo
  {number} {6}),\ \bibinfo {pages} {e1000408}}\BibitemShut {NoStop}%
\bibitem [{\citenamefont {Amit}\ \emph {et~al.}(1985)\citenamefont {Amit},
  \citenamefont {Gutfreund},\ and\ \citenamefont
  {Sompolinsky}}]{PhysRevA.32.1007}%
  \BibitemOpen
  \bibfield  {author} {\bibinfo {author} {\bibnamefont {Amit}, \bibfnamefont
  {Daniel~J}}, \bibinfo {author} {\bibfnamefont {Hanoch}\ \bibnamefont
  {Gutfreund}}, \ and\ \bibinfo {author} {\bibfnamefont {H.}~\bibnamefont
  {Sompolinsky}}} (\bibinfo {year} {1985}),\ \bibfield  {title} {\enquote
  {\bibinfo {title} {Spin-glass models of neural networks},}\ }\href {\doibase
  10.1103/PhysRevA.32.1007} {\bibfield  {journal} {\bibinfo  {journal} {Phys.
  Rev. A}\ }\textbf {\bibinfo {volume} {32}},\ \bibinfo {pages}
  {1007--1018}}\BibitemShut {NoStop}%
\bibitem [{\citenamefont {Badhwar}\ and\ \citenamefont
  {Bagler}(2015)}]{badhwar2015control}%
  \BibitemOpen
  \bibfield  {author} {\bibinfo {author} {\bibnamefont {Badhwar}, \bibfnamefont
  {R}}, \ and\ \bibinfo {author} {\bibfnamefont {G}~\bibnamefont {Bagler}}}
  (\bibinfo {year} {2015}),\ \bibfield  {title} {\enquote {\bibinfo {title}
  {Control of neuronal network in {C}aenorhabditis elegans},}\ }\href@noop {}
  {\bibfield  {journal} {\bibinfo  {journal} {PLoS One}\ }\textbf {\bibinfo
  {volume} {10}}~(\bibinfo {number} {9}),\ \bibinfo {pages}
  {e0139204}}\BibitemShut {NoStop}%
\bibitem [{\citenamefont {Barahona}\ and\ \citenamefont
  {Pecora}(2002)}]{barahona2002synchronization}%
  \BibitemOpen
  \bibfield  {author} {\bibinfo {author} {\bibnamefont {Barahona},
  \bibfnamefont {M}}, \ and\ \bibinfo {author} {\bibfnamefont {L~M}\
  \bibnamefont {Pecora}}} (\bibinfo {year} {2002}),\ \bibfield  {title}
  {\enquote {\bibinfo {title} {Synchronization in small-world systems},}\
  }\href@noop {} {\bibfield  {journal} {\bibinfo  {journal} {Phys Rev Lett}\
  }\textbf {\bibinfo {volume} {89}}~(\bibinfo {number} {5}),\ \bibinfo {pages}
  {054101}}\BibitemShut {NoStop}%
\bibitem [{\citenamefont {von Bartheld}\ \emph {et~al.}(2016)\citenamefont {von
  Bartheld}, \citenamefont {Bahney},\ and\ \citenamefont
  {Herculano-Houzel}}]{bartheld2016search}%
  \BibitemOpen
  \bibfield  {author} {\bibinfo {author} {\bibnamefont {von Bartheld},
  \bibfnamefont {C~S}}, \bibinfo {author} {\bibfnamefont {J}~\bibnamefont
  {Bahney}}, \ and\ \bibinfo {author} {\bibfnamefont {S}~\bibnamefont
  {Herculano-Houzel}}} (\bibinfo {year} {2016}),\ \bibfield  {title} {\enquote
  {\bibinfo {title} {The search for true numbers of neurons and glial cells in
  the human brain: A review of 150 years of cell counting},}\ }\href@noop {}
  {\bibfield  {journal} {\bibinfo  {journal} {J Comp Neurol}\ }\textbf
  {\bibinfo {volume} {524}}~(\bibinfo {number} {18}),\ \bibinfo {pages}
  {3865--3895}}\BibitemShut {NoStop}%
\bibitem [{\citenamefont {Bartolomei}\ \emph {et~al.}(2017)\citenamefont
  {Bartolomei}, \citenamefont {Lagarde}, \citenamefont {Wendling},
  \citenamefont {McGonigal}, \citenamefont {Jirsa}, \citenamefont {Guye},\ and\
  \citenamefont {Bénar}}]{EPI:EPI13791}%
  \BibitemOpen
  \bibfield  {author} {\bibinfo {author} {\bibnamefont {Bartolomei},
  \bibfnamefont {Fabrice}}, \bibinfo {author} {\bibfnamefont {Stanislas}\
  \bibnamefont {Lagarde}}, \bibinfo {author} {\bibfnamefont {Fabrice}\
  \bibnamefont {Wendling}}, \bibinfo {author} {\bibfnamefont {Aileen}\
  \bibnamefont {McGonigal}}, \bibinfo {author} {\bibfnamefont {Viktor}\
  \bibnamefont {Jirsa}}, \bibinfo {author} {\bibfnamefont {Maxime}\
  \bibnamefont {Guye}}, \ and\ \bibinfo {author} {\bibfnamefont {Christian}\
  \bibnamefont {Bénar}}} (\bibinfo {year} {2017}),\ \bibfield  {title}
  {\enquote {\bibinfo {title} {Defining epileptogenic networks: Contribution of
  {SEEG} and signal analysis},}\ }\href {\doibase 10.1111/epi.13791} {\bibfield
   {journal} {\bibinfo  {journal} {Epilepsia}\ }\textbf {\bibinfo {volume}
  {58}}~(\bibinfo {number} {7}),\ \bibinfo {pages} {1131--1147}}\BibitemShut
  {NoStop}%
\bibitem [{\citenamefont {Basser}\ \emph {et~al.}(1994)\citenamefont {Basser},
  \citenamefont {Mattiello},\ and\ \citenamefont {LeBihan}}]{basser1994MR}%
  \BibitemOpen
  \bibfield  {author} {\bibinfo {author} {\bibnamefont {Basser}, \bibfnamefont
  {P~J}}, \bibinfo {author} {\bibfnamefont {J}~\bibnamefont {Mattiello}}, \
  and\ \bibinfo {author} {\bibfnamefont {D}~\bibnamefont {LeBihan}}} (\bibinfo
  {year} {1994}),\ \bibfield  {title} {\enquote {\bibinfo {title} {{MR}
  diffusion tensor spectroscopy and imaging},}\ }\href@noop {} {\bibfield
  {journal} {\bibinfo  {journal} {Biophys J}\ }\textbf {\bibinfo {volume}
  {66}}~(\bibinfo {number} {1}),\ \bibinfo {pages} {259--267}}\BibitemShut
  {NoStop}%
\bibitem [{\citenamefont {Bassett}\ \emph {et~al.}(2011)\citenamefont
  {Bassett}, \citenamefont {Brown}, \citenamefont {Deshpande}, \citenamefont
  {Carlson},\ and\ \citenamefont {Grafton}}]{Bassett2011}%
  \BibitemOpen
  \bibfield  {author} {\bibinfo {author} {\bibnamefont {Bassett}, \bibfnamefont
  {D~S}}, \bibinfo {author} {\bibfnamefont {J~A}\ \bibnamefont {Brown}},
  \bibinfo {author} {\bibfnamefont {V}~\bibnamefont {Deshpande}}, \bibinfo
  {author} {\bibfnamefont {J~M}\ \bibnamefont {Carlson}}, \ and\ \bibinfo
  {author} {\bibfnamefont {S~T}\ \bibnamefont {Grafton}}} (\bibinfo {year}
  {2011}),\ \bibfield  {title} {\enquote {\bibinfo {title} {Conserved and
  variable architecture of human white matter connectivity},}\ }\href@noop {}
  {\bibfield  {journal} {\bibinfo  {journal} {Neuroimage}\ }\textbf {\bibinfo
  {volume} {54}}~(\bibinfo {number} {2}),\ \bibinfo {pages}
  {1262--1279}}\BibitemShut {NoStop}%
\bibitem [{\citenamefont {Bassett}\ and\ \citenamefont
  {Bullmore}(2006)}]{bassett2006small}%
  \BibitemOpen
  \bibfield  {author} {\bibinfo {author} {\bibnamefont {Bassett}, \bibfnamefont
  {D~S}}, \ and\ \bibinfo {author} {\bibfnamefont {E}~\bibnamefont {Bullmore}}}
  (\bibinfo {year} {2006}),\ \bibfield  {title} {\enquote {\bibinfo {title}
  {Small-world brain networks},}\ }\href@noop {} {\bibfield  {journal}
  {\bibinfo  {journal} {Neuroscientist}\ }\textbf {\bibinfo {volume}
  {12}}~(\bibinfo {number} {6}),\ \bibinfo {pages} {512--523}}\BibitemShut
  {NoStop}%
\bibitem [{\citenamefont {Bassett}\ and\ \citenamefont
  {Bullmore}(2009)}]{Bassett2009}%
  \BibitemOpen
  \bibfield  {author} {\bibinfo {author} {\bibnamefont {Bassett}, \bibfnamefont
  {D~S}}, \ and\ \bibinfo {author} {\bibfnamefont {E~T}\ \bibnamefont
  {Bullmore}}} (\bibinfo {year} {2009}),\ \bibfield  {title} {\enquote
  {\bibinfo {title} {Human brain networks in health and disease},}\ }\href@noop
  {} {\bibfield  {journal} {\bibinfo  {journal} {Curr Opin Neurol}\ }\textbf
  {\bibinfo {volume} {22}}~(\bibinfo {number} {4}),\ \bibinfo {pages}
  {340--347}}\BibitemShut {NoStop}%
\bibitem [{\citenamefont {Bassett}\ and\ \citenamefont
  {Bullmore}(2016)}]{bassett2016small}%
  \BibitemOpen
  \bibfield  {author} {\bibinfo {author} {\bibnamefont {Bassett}, \bibfnamefont
  {D~S}}, \ and\ \bibinfo {author} {\bibfnamefont {E~T}\ \bibnamefont
  {Bullmore}}} (\bibinfo {year} {2016}),\ \bibfield  {title} {\enquote
  {\bibinfo {title} {Small-world brain networks revisited},}\ }\href@noop {}
  {\bibinfo  {journal} {Neuroscientist}\ ,\ \bibinfo {pages}
  {1073858416667720}}\BibitemShut {NoStop}%
\bibitem [{\citenamefont {Bassett}\ \emph {et~al.}(2009)\citenamefont
  {Bassett}, \citenamefont {Bullmore}, \citenamefont {Meyer-Lindenberg},
  \citenamefont {Apud}, \citenamefont {Weinberger},\ and\ \citenamefont
  {Coppola}}]{bassett2009cognitive}%
  \BibitemOpen
\bibfield  {journal} {  }\bibfield  {author} {\bibinfo {author} {\bibnamefont
  {Bassett}, \bibfnamefont {D~S}}, \bibinfo {author} {\bibfnamefont {E~T}\
  \bibnamefont {Bullmore}}, \bibinfo {author} {\bibfnamefont {A}~\bibnamefont
  {Meyer-Lindenberg}}, \bibinfo {author} {\bibfnamefont {J~A}\ \bibnamefont
  {Apud}}, \bibinfo {author} {\bibfnamefont {D~R}\ \bibnamefont {Weinberger}},
  \ and\ \bibinfo {author} {\bibfnamefont {R}~\bibnamefont {Coppola}}}
  (\bibinfo {year} {2009}),\ \bibfield  {title} {\enquote {\bibinfo {title}
  {Cognitive fitness of cost-efficient brain functional networks},}\
  }\href@noop {} {\bibfield  {journal} {\bibinfo  {journal} {Proc Natl Acad Sci
  U S A}\ }\textbf {\bibinfo {volume} {106}}~(\bibinfo {number} {28}),\
  \bibinfo {pages} {11747--11752}}\BibitemShut {NoStop}%
\bibitem [{\citenamefont {Bassett}\ and\ \citenamefont
  {Khambhati}(2017)}]{bassett2017network}%
  \BibitemOpen
  \bibfield  {author} {\bibinfo {author} {\bibnamefont {Bassett}, \bibfnamefont
  {D~S}}, \ and\ \bibinfo {author} {\bibfnamefont {A~N}\ \bibnamefont
  {Khambhati}}} (\bibinfo {year} {2017}),\ \bibfield  {title} {\enquote
  {\bibinfo {title} {A network engineering perspective on probing and
  perturbing cognition with neurofeedback},}\ }\href@noop {} {\bibfield
  {journal} {\bibinfo  {journal} {Annals of the New York Academy of Sciences}\
  }\textbf {\bibinfo {volume} {In Revision}}}\BibitemShut {NoStop}%
\bibitem [{\citenamefont {Bentley}\ \emph {et~al.}(2016)\citenamefont
  {Bentley}, \citenamefont {Branicky}, \citenamefont {Barnes}, \citenamefont
  {Chew}, \citenamefont {Yemini}, \citenamefont {Bullmore}, \citenamefont
  {Vertes},\ and\ \citenamefont {Schafer}}]{bentley2016multilayer}%
  \BibitemOpen
  \bibfield  {author} {\bibinfo {author} {\bibnamefont {Bentley}, \bibfnamefont
  {B}}, \bibinfo {author} {\bibfnamefont {R}~\bibnamefont {Branicky}}, \bibinfo
  {author} {\bibfnamefont {C~L}\ \bibnamefont {Barnes}}, \bibinfo {author}
  {\bibfnamefont {Y~L}\ \bibnamefont {Chew}}, \bibinfo {author} {\bibfnamefont
  {E}~\bibnamefont {Yemini}}, \bibinfo {author} {\bibfnamefont {E~T}\
  \bibnamefont {Bullmore}}, \bibinfo {author} {\bibfnamefont {P~E}\
  \bibnamefont {Vertes}}, \ and\ \bibinfo {author} {\bibfnamefont {W~R}\
  \bibnamefont {Schafer}}} (\bibinfo {year} {2016}),\ \bibfield  {title}
  {\enquote {\bibinfo {title} {The multilayer connectome of caenorhabditis
  elegans},}\ }\href@noop {} {\bibfield  {journal} {\bibinfo  {journal} {PLoS
  Comput Biol}\ }\textbf {\bibinfo {volume} {12}}~(\bibinfo {number} {12}),\
  \bibinfo {pages} {e1005283}}\BibitemShut {NoStop}%
\bibitem [{\citenamefont {Ber{\'e}nyi}\ \emph {et~al.}(2012)\citenamefont
  {Ber{\'e}nyi}, \citenamefont {Belluscio}, \citenamefont {Mao},\ and\
  \citenamefont {Buzs{\'a}ki}}]{Berenyi735}%
  \BibitemOpen
  \bibfield  {author} {\bibinfo {author} {\bibnamefont {Ber{\'e}nyi},
  \bibfnamefont {Antal}}, \bibinfo {author} {\bibfnamefont {Mariano}\
  \bibnamefont {Belluscio}}, \bibinfo {author} {\bibfnamefont {Dun}\
  \bibnamefont {Mao}}, \ and\ \bibinfo {author} {\bibfnamefont {Gy{\"o}rgy}\
  \bibnamefont {Buzs{\'a}ki}}} (\bibinfo {year} {2012}),\ \bibfield  {title}
  {\enquote {\bibinfo {title} {Closed-loop control of epilepsy by transcranial
  electrical stimulation},}\ }\href {\doibase 10.1126/science.1223154}
  {\bibfield  {journal} {\bibinfo  {journal} {Science}\ }\textbf {\bibinfo
  {volume} {337}}~(\bibinfo {number} {6095}),\ \bibinfo {pages}
  {735--737}}\BibitemShut {NoStop}%
\bibitem [{\citenamefont {Betzel}\ and\ \citenamefont
  {Bassett}(2016)}]{betzel2016multi}%
  \BibitemOpen
  \bibfield  {author} {\bibinfo {author} {\bibnamefont {Betzel}, \bibfnamefont
  {R~F}}, \ and\ \bibinfo {author} {\bibfnamefont {D~S}\ \bibnamefont
  {Bassett}}} (\bibinfo {year} {2016}),\ \bibfield  {title} {\enquote {\bibinfo
  {title} {Multi-scale brain networks},}\ }\href@noop {} {\bibfield  {journal}
  {\bibinfo  {journal} {Neuroimage}\ }\textbf {\bibinfo {volume}
  {S1053-8119}}~(\bibinfo {number} {16}),\ \bibinfo {pages}
  {30615--30612}}\BibitemShut {NoStop}%
\bibitem [{\citenamefont {Betzel}\ \emph {et~al.}(2016)\citenamefont {Betzel},
  \citenamefont {Gu}, \citenamefont {Medaglia}, \citenamefont {Pasqualetti},\
  and\ \citenamefont {Bassett}}]{Betzel2016}%
  \BibitemOpen
  \bibfield  {author} {\bibinfo {author} {\bibnamefont {Betzel}, \bibfnamefont
  {Richard~F}}, \bibinfo {author} {\bibfnamefont {Shi}\ \bibnamefont {Gu}},
  \bibinfo {author} {\bibfnamefont {John~D.}\ \bibnamefont {Medaglia}},
  \bibinfo {author} {\bibfnamefont {Fabio}\ \bibnamefont {Pasqualetti}}, \ and\
  \bibinfo {author} {\bibfnamefont {Danielle~S.}\ \bibnamefont {Bassett}}}
  (\bibinfo {year} {2016}),\ \bibfield  {title} {\enquote {\bibinfo {title}
  {Optimally controlling the human connectome: the role of network topology},}\
  }\href {http://dx.doi.org/10.1038/srep30770} {\bibfield  {journal} {\bibinfo
  {journal} {Scientific Reports}\ }\textbf {\bibinfo {volume} {6}},\ \bibinfo
  {pages} {30770}}\BibitemShut {NoStop}%
\bibitem [{\citenamefont {Bianchin}\ \emph {et~al.}(2015)\citenamefont
  {Bianchin}, \citenamefont {Pasqualetti},\ and\ \citenamefont
  {Zampieri}}]{GB-FP-SZ:15}%
  \BibitemOpen
  \bibfield  {author} {\bibinfo {author} {\bibnamefont {Bianchin},
  \bibfnamefont {G}}, \bibinfo {author} {\bibfnamefont {F.}~\bibnamefont
  {Pasqualetti}}, \ and\ \bibinfo {author} {\bibfnamefont {S.}~\bibnamefont
  {Zampieri}}} (\bibinfo {year} {2015}),\ \bibfield  {title} {\enquote
  {\bibinfo {title} {The role of diameter in the controllability of complex
  networks},}\ }in\ \href@noop {} {\emph {\bibinfo {booktitle} {IEEE Conference
  on Decision and Control}}},\ pp.\ \bibinfo {pages} {980--985}\BibitemShut
  {NoStop}%
\bibitem [{\citenamefont {Bikson}\ \emph {et~al.}(2016)\citenamefont {Bikson},
  \citenamefont {Grossman}, \citenamefont {Thomas}, \citenamefont {Zannou},
  \citenamefont {Jiang}, \citenamefont {Adnan}, \citenamefont {Mourdoukoutas},
  \citenamefont {Kronberg}, \citenamefont {Truong}, \citenamefont {Boggio},
  \citenamefont {Brunoni}, \citenamefont {Charvet}, \citenamefont {Fregni},
  \citenamefont {Fritsch}, \citenamefont {Gillick}, \citenamefont {Hamilton},
  \citenamefont {Hampstead}, \citenamefont {Jankord}, \citenamefont {Kirton},
  \citenamefont {Knotkova}, \citenamefont {Liebetanz}, \citenamefont {Liu},
  \citenamefont {Loo}, \citenamefont {Nitsche}, \citenamefont {Reis},
  \citenamefont {Richardson}, \citenamefont {Rotenberg}, \citenamefont
  {Turkeltaub},\ and\ \citenamefont {Woods}}]{bikson2016safety}%
  \BibitemOpen
  \bibfield  {author} {\bibinfo {author} {\bibnamefont {Bikson}, \bibfnamefont
  {M}}, \bibinfo {author} {\bibfnamefont {P}~\bibnamefont {Grossman}}, \bibinfo
  {author} {\bibfnamefont {C}~\bibnamefont {Thomas}}, \bibinfo {author}
  {\bibfnamefont {A~L}\ \bibnamefont {Zannou}}, \bibinfo {author}
  {\bibfnamefont {J}~\bibnamefont {Jiang}}, \bibinfo {author} {\bibfnamefont
  {T}~\bibnamefont {Adnan}}, \bibinfo {author} {\bibfnamefont {A~P}\
  \bibnamefont {Mourdoukoutas}}, \bibinfo {author} {\bibfnamefont
  {G}~\bibnamefont {Kronberg}}, \bibinfo {author} {\bibfnamefont
  {D}~\bibnamefont {Truong}}, \bibinfo {author} {\bibfnamefont {P}~\bibnamefont
  {Boggio}}, \bibinfo {author} {\bibfnamefont {A~R}\ \bibnamefont {Brunoni}},
  \bibinfo {author} {\bibfnamefont {L}~\bibnamefont {Charvet}}, \bibinfo
  {author} {\bibfnamefont {F}~\bibnamefont {Fregni}}, \bibinfo {author}
  {\bibfnamefont {B}~\bibnamefont {Fritsch}}, \bibinfo {author} {\bibfnamefont
  {B}~\bibnamefont {Gillick}}, \bibinfo {author} {\bibfnamefont {R~H}\
  \bibnamefont {Hamilton}}, \bibinfo {author} {\bibfnamefont {B~M}\
  \bibnamefont {Hampstead}}, \bibinfo {author} {\bibfnamefont {R}~\bibnamefont
  {Jankord}}, \bibinfo {author} {\bibfnamefont {A}~\bibnamefont {Kirton}},
  \bibinfo {author} {\bibfnamefont {H}~\bibnamefont {Knotkova}}, \bibinfo
  {author} {\bibfnamefont {D}~\bibnamefont {Liebetanz}}, \bibinfo {author}
  {\bibfnamefont {A}~\bibnamefont {Liu}}, \bibinfo {author} {\bibfnamefont
  {C}~\bibnamefont {Loo}}, \bibinfo {author} {\bibfnamefont {M~A}\ \bibnamefont
  {Nitsche}}, \bibinfo {author} {\bibfnamefont {J}~\bibnamefont {Reis}},
  \bibinfo {author} {\bibfnamefont {J~D}\ \bibnamefont {Richardson}}, \bibinfo
  {author} {\bibfnamefont {A}~\bibnamefont {Rotenberg}}, \bibinfo {author}
  {\bibfnamefont {P~E}\ \bibnamefont {Turkeltaub}}, \ and\ \bibinfo {author}
  {\bibfnamefont {A~J}\ \bibnamefont {Woods}}} (\bibinfo {year} {2016}),\
  \bibfield  {title} {\enquote {\bibinfo {title} {Safety of transcranial direct
  current stimulation: Evidence based update 2016},}\ }\href@noop {} {\bibfield
   {journal} {\bibinfo  {journal} {Brain Stimul}\ }\textbf {\bibinfo {volume}
  {9}}~(\bibinfo {number} {5}),\ \bibinfo {pages} {641--661}}\BibitemShut
  {NoStop}%
\bibitem [{\citenamefont {Bishop}(1995)}]{bishop1995neural}%
  \BibitemOpen
  \bibfield  {author} {\bibinfo {author} {\bibnamefont {Bishop}, \bibfnamefont
  {C~M}}} (\bibinfo {year} {1995}),\ \href@noop {} {\emph {\bibinfo {title}
  {Neural Networks for Pattern Recognition}}}\ (\bibinfo  {publisher} {Oxford
  University Press})\BibitemShut {NoStop}%
\bibitem [{\citenamefont {Botvinick}\ and\ \citenamefont
  {Braver}(2015)}]{botvinick2015motivation}%
  \BibitemOpen
  \bibfield  {author} {\bibinfo {author} {\bibnamefont {Botvinick},
  \bibfnamefont {M}}, \ and\ \bibinfo {author} {\bibfnamefont {T}~\bibnamefont
  {Braver}}} (\bibinfo {year} {2015}),\ \bibfield  {title} {\enquote {\bibinfo
  {title} {Motivation and cognitive control: from behavior to neural
  mechanism},}\ }\href@noop {} {\bibfield  {journal} {\bibinfo  {journal} {Annu
  Rev Psychol}\ }\textbf {\bibinfo {volume} {66}},\ \bibinfo {pages}
  {83--113}}\BibitemShut {NoStop}%
\bibitem [{\citenamefont {Botvinick}\ and\ \citenamefont
  {Cohen}(2014)}]{COGS:COGS12126}%
  \BibitemOpen
  \bibfield  {author} {\bibinfo {author} {\bibnamefont {Botvinick},
  \bibfnamefont {Matthew~M}}, \ and\ \bibinfo {author} {\bibfnamefont
  {Jonathan~D.}\ \bibnamefont {Cohen}}} (\bibinfo {year} {2014}),\ \bibfield
  {title} {\enquote {\bibinfo {title} {The computational and neural basis of
  cognitive control: Charted territory and new frontiers},}\ }\href {\doibase
  10.1111/cogs.12126} {\bibfield  {journal} {\bibinfo  {journal} {Cognitive
  Science}\ }\textbf {\bibinfo {volume} {38}}~(\bibinfo {number} {6}),\
  \bibinfo {pages} {1249--1285}}\BibitemShut {NoStop}%
\bibitem [{\citenamefont {Brodmann}(1909)}]{brodmann1909verg}%
  \BibitemOpen
  \bibfield  {author} {\bibinfo {author} {\bibnamefont {Brodmann},
  \bibfnamefont {K}}} (\bibinfo {year} {1909}),\ \href@noop {} {\emph {\bibinfo
  {title} {Vergleichende Lokalisationslehre der Grosshirnrinde}}}\ (\bibinfo
  {publisher} {Johann Ambrosius Barth})\BibitemShut {NoStop}%
\bibitem [{\citenamefont {Bullmore}\ and\ \citenamefont
  {Sporns}(2012)}]{Bullmore2012}%
  \BibitemOpen
  \bibfield  {author} {\bibinfo {author} {\bibnamefont {Bullmore},
  \bibfnamefont {E}}, \ and\ \bibinfo {author} {\bibfnamefont {O}~\bibnamefont
  {Sporns}}} (\bibinfo {year} {2012}),\ \bibfield  {title} {\enquote {\bibinfo
  {title} {The economy of brain network organization},}\ }\href@noop {}
  {\bibfield  {journal} {\bibinfo  {journal} {Nat Rev Neurosci}\ }\textbf
  {\bibinfo {volume} {13}}~(\bibinfo {number} {5}),\ \bibinfo {pages}
  {336--349}}\BibitemShut {NoStop}%
\bibitem [{\citenamefont {Bullmore}\ and\ \citenamefont
  {Bassett}(2011)}]{bullmore2011brain}%
  \BibitemOpen
  \bibfield  {author} {\bibinfo {author} {\bibnamefont {Bullmore},
  \bibfnamefont {E~T}}, \ and\ \bibinfo {author} {\bibfnamefont {D~S}\
  \bibnamefont {Bassett}}} (\bibinfo {year} {2011}),\ \bibfield  {title}
  {\enquote {\bibinfo {title} {Brain graphs: graphical models of the human
  brain connectome},}\ }\href@noop {} {\bibfield  {journal} {\bibinfo
  {journal} {Annu Rev Clin Psychol}\ }\textbf {\bibinfo {volume} {7}},\
  \bibinfo {pages} {113--140}}\BibitemShut {NoStop}%
\bibitem [{\citenamefont {Campbell}\ \emph {et~al.}(2015)\citenamefont
  {Campbell}, \citenamefont {Ruths}, \citenamefont {Ruths}, \citenamefont
  {Shea},\ and\ \citenamefont {Albert}}]{Campbell2015}%
  \BibitemOpen
  \bibfield  {author} {\bibinfo {author} {\bibnamefont {Campbell},
  \bibfnamefont {Colin}}, \bibinfo {author} {\bibfnamefont {Justin}\
  \bibnamefont {Ruths}}, \bibinfo {author} {\bibfnamefont {Derek}\ \bibnamefont
  {Ruths}}, \bibinfo {author} {\bibfnamefont {Katriona}\ \bibnamefont {Shea}},
  \ and\ \bibinfo {author} {\bibfnamefont {R{\'e}ka}\ \bibnamefont {Albert}}}
  (\bibinfo {year} {2015}),\ \bibfield  {title} {\enquote {\bibinfo {title}
  {Topological constraints on network control profiles},}\ }\href
  {http://dx.doi.org/10.1038/srep18693} {\bibfield  {journal} {\bibinfo
  {journal} {Scientific Reports}\ }\textbf {\bibinfo {volume} {5}},\ \bibinfo
  {pages} {18693}}\BibitemShut {NoStop}%
\bibitem [{\citenamefont {Castellana}\ and\ \citenamefont
  {Bialek}(2014)}]{castellana2014inverse}%
  \BibitemOpen
  \bibfield  {author} {\bibinfo {author} {\bibnamefont {Castellana},
  \bibfnamefont {M}}, \ and\ \bibinfo {author} {\bibfnamefont {W}~\bibnamefont
  {Bialek}}} (\bibinfo {year} {2014}),\ \bibfield  {title} {\enquote {\bibinfo
  {title} {Inverse spin glass and related maximum entropy problems},}\
  }\href@noop {} {\bibfield  {journal} {\bibinfo  {journal} {Phys Rev Lett}\
  }\textbf {\bibinfo {volume} {113}}~(\bibinfo {number} {11}),\ \bibinfo
  {pages} {117204}}\BibitemShut {NoStop}%
\bibitem [{\citenamefont {Chai}\ \emph {et~al.}(2017)\citenamefont {Chai},
  \citenamefont {Khambhati}, \citenamefont {Ciric}, \citenamefont {Gur},
  \citenamefont {Gur}, \citenamefont {Satterthwaite},\ and\ \citenamefont
  {Bassett}}]{chai2017evolution}%
  \BibitemOpen
  \bibfield  {author} {\bibinfo {author} {\bibnamefont {Chai}, \bibfnamefont
  {L}}, \bibinfo {author} {\bibfnamefont {A~N}\ \bibnamefont {Khambhati}},
  \bibinfo {author} {\bibfnamefont {R}~\bibnamefont {Ciric}}, \bibinfo {author}
  {\bibfnamefont {R~C}\ \bibnamefont {Gur}}, \bibinfo {author} {\bibfnamefont
  {R~E}\ \bibnamefont {Gur}}, \bibinfo {author} {\bibfnamefont {T~D}\
  \bibnamefont {Satterthwaite}}, \ and\ \bibinfo {author} {\bibfnamefont {D~S}\
  \bibnamefont {Bassett}}} (\bibinfo {year} {2017}),\ \bibfield  {title}
  {\enquote {\bibinfo {title} {Evolution of brain network dynamics in
  neurodevelopment},}\ }\href@noop {} {\bibfield  {journal} {\bibinfo
  {journal} {Network Neuroscience}\ }\textbf {\bibinfo {volume} {In
  Press}}}\BibitemShut {NoStop}%
\bibitem [{\citenamefont {Chand}\ and\ \citenamefont
  {Dhamala}(2016)}]{chand2016salience}%
  \BibitemOpen
  \bibfield  {author} {\bibinfo {author} {\bibnamefont {Chand}, \bibfnamefont
  {G~B}}, \ and\ \bibinfo {author} {\bibfnamefont {M}~\bibnamefont {Dhamala}}}
  (\bibinfo {year} {2016}),\ \bibfield  {title} {\enquote {\bibinfo {title}
  {The salience network dynamics in perceptual decision-making},}\ }\href@noop
  {} {\bibfield  {journal} {\bibinfo  {journal} {Neuroimage}\ }\textbf
  {\bibinfo {volume} {134}},\ \bibinfo {pages} {85--93}}\BibitemShut {NoStop}%
\bibitem [{\citenamefont {Chand}\ \emph {et~al.}(2016)\citenamefont {Chand},
  \citenamefont {Lamichhane},\ and\ \citenamefont {Dhamala}}]{chand2016face}%
  \BibitemOpen
  \bibfield  {author} {\bibinfo {author} {\bibnamefont {Chand}, \bibfnamefont
  {G~B}}, \bibinfo {author} {\bibfnamefont {B}~\bibnamefont {Lamichhane}}, \
  and\ \bibinfo {author} {\bibfnamefont {M}~\bibnamefont {Dhamala}}} (\bibinfo
  {year} {2016}),\ \bibfield  {title} {\enquote {\bibinfo {title} {Face or
  house image perception: Beta and gamma bands of oscillations in brain
  networks carry out decision-making},}\ }\href@noop {} {\bibfield  {journal}
  {\bibinfo  {journal} {Brain Connect}\ }\textbf {\bibinfo {volume} {Epub ahead
  of print}}}\BibitemShut {NoStop}%
\bibitem [{\citenamefont {Chang}(2015)}]{chang2015towards}%
  \BibitemOpen
  \bibfield  {author} {\bibinfo {author} {\bibnamefont {Chang}, \bibfnamefont
  {E~F}}} (\bibinfo {year} {2015}),\ \bibfield  {title} {\enquote {\bibinfo
  {title} {Towards large-scale, human-based, mesoscopic neurotechnologies},}\
  }\href@noop {} {\bibfield  {journal} {\bibinfo  {journal} {Neuron}\ }\textbf
  {\bibinfo {volume} {86}}~(\bibinfo {number} {1}),\ \bibinfo {pages}
  {68--78}}\BibitemShut {NoStop}%
\bibitem [{\citenamefont {Chen}(2014)}]{Chen2014}%
  \BibitemOpen
  \bibfield  {author} {\bibinfo {author} {\bibnamefont {Chen}, \bibfnamefont
  {Guanrong}}} (\bibinfo {year} {2014}),\ \bibfield  {title} {\enquote
  {\bibinfo {title} {Pinning control and synchronization on complex dynamical
  networks},}\ }\href {\doibase 10.1007/s12555-014-9001-2} {\bibfield
  {journal} {\bibinfo  {journal} {International Journal of Control, Automation
  and Systems}\ }\textbf {\bibinfo {volume} {12}}~(\bibinfo {number} {2}),\
  \bibinfo {pages} {221--230}}\BibitemShut {NoStop}%
\bibitem [{\citenamefont {Chen}\ \emph {et~al.}(2014)\citenamefont {Chen},
  \citenamefont {Attiah}, \citenamefont {Baltuch}, \citenamefont {Smith},
  \citenamefont {Hamilton},\ and\ \citenamefont {Lucas}}]{chen2014harnessing}%
  \BibitemOpen
  \bibfield  {author} {\bibinfo {author} {\bibnamefont {Chen}, \bibfnamefont
  {H~I}}, \bibinfo {author} {\bibfnamefont {M}~\bibnamefont {Attiah}}, \bibinfo
  {author} {\bibfnamefont {G}~\bibnamefont {Baltuch}}, \bibinfo {author}
  {\bibfnamefont {D~H}\ \bibnamefont {Smith}}, \bibinfo {author} {\bibfnamefont
  {R~H}\ \bibnamefont {Hamilton}}, \ and\ \bibinfo {author} {\bibfnamefont
  {T~H}\ \bibnamefont {Lucas}}} (\bibinfo {year} {2014}),\ \bibfield  {title}
  {\enquote {\bibinfo {title} {Harnessing plasticity for the treatment of
  neurosurgical disorders: an overview},}\ }\href@noop {} {\bibfield  {journal}
  {\bibinfo  {journal} {World Neurosurg}\ }\textbf {\bibinfo {volume}
  {82}}~(\bibinfo {number} {5}),\ \bibinfo {pages} {648--659}}\BibitemShut
  {NoStop}%
\bibitem [{\citenamefont {Chiken}\ and\ \citenamefont
  {Nambu}(2014)}]{chiken2014disrupting}%
  \BibitemOpen
  \bibfield  {author} {\bibinfo {author} {\bibnamefont {Chiken}, \bibfnamefont
  {S}}, \ and\ \bibinfo {author} {\bibfnamefont {A}~\bibnamefont {Nambu}}}
  (\bibinfo {year} {2014}),\ \bibfield  {title} {\enquote {\bibinfo {title}
  {Disrupting neuronal transmission: mechanism of {DBS}?}}\ }\href@noop {}
  {\bibfield  {journal} {\bibinfo  {journal} {Front Syst Neurosci}\ }\textbf
  {\bibinfo {volume} {8}},\ \bibinfo {pages} {33}}\BibitemShut {NoStop}%
\bibitem [{\citenamefont {Ching}\ \emph
  {et~al.}(2012{\natexlab{a}})\citenamefont {Ching}, \citenamefont {Brown},\
  and\ \citenamefont {Kramer}}]{Ching2012}%
  \BibitemOpen
  \bibfield  {author} {\bibinfo {author} {\bibnamefont {Ching}, \bibfnamefont
  {S}}, \bibinfo {author} {\bibfnamefont {E~N}\ \bibnamefont {Brown}}, \ and\
  \bibinfo {author} {\bibfnamefont {M~A}\ \bibnamefont {Kramer}}} (\bibinfo
  {year} {2012}{\natexlab{a}}),\ \bibfield  {title} {\enquote {\bibinfo {title}
  {Distributed control in a mean-field cortical network model: implications for
  seizure suppression},}\ }\href@noop {} {\bibfield  {journal} {\bibinfo
  {journal} {Phys Rev E Stat Nonlin Soft Matter Phys}\ }\textbf {\bibinfo
  {volume} {86}}~(\bibinfo {number} {2 Pt 1}),\ \bibinfo {pages}
  {021920}}\BibitemShut {NoStop}%
\bibitem [{\citenamefont {Ching}\ \emph
  {et~al.}(2012{\natexlab{b}})\citenamefont {Ching}, \citenamefont {Purdon},
  \citenamefont {Vijayan}, \citenamefont {Kopell},\ and\ \citenamefont
  {Brown}}]{ching2012neurophysiological}%
  \BibitemOpen
  \bibfield  {author} {\bibinfo {author} {\bibnamefont {Ching}, \bibfnamefont
  {S}}, \bibinfo {author} {\bibfnamefont {P~L}\ \bibnamefont {Purdon}},
  \bibinfo {author} {\bibfnamefont {S}~\bibnamefont {Vijayan}}, \bibinfo
  {author} {\bibfnamefont {N~J}\ \bibnamefont {Kopell}}, \ and\ \bibinfo
  {author} {\bibfnamefont {E~N}\ \bibnamefont {Brown}}} (\bibinfo {year}
  {2012}{\natexlab{b}}),\ \bibfield  {title} {\enquote {\bibinfo {title} {A
  neurophysiological-metabolic model for burst suppression},}\ }\href@noop {}
  {\bibfield  {journal} {\bibinfo  {journal} {Proc Natl Acad Sci U S A}\
  }\textbf {\bibinfo {volume} {109}}~(\bibinfo {number} {8}),\ \bibinfo {pages}
  {3095--3100}}\BibitemShut {NoStop}%
\bibitem [{\citenamefont {Ching}\ \emph {et~al.}(2013)\citenamefont {Ching},
  \citenamefont {Liberman}, \citenamefont {Chemali}, \citenamefont {Westover},
  \citenamefont {Kenny}, \citenamefont {Solt}, \citenamefont {Purdon},\ and\
  \citenamefont {Brown}}]{doi:10.1097/ALN.0b013e31829d4ab4}%
  \BibitemOpen
  \bibfield  {author} {\bibinfo {author} {\bibnamefont {Ching}, \bibfnamefont
  {ShiNung}}, \bibinfo {author} {\bibfnamefont {Max~Y.}\ \bibnamefont
  {Liberman}}, \bibinfo {author} {\bibfnamefont {Jessica~J.}\ \bibnamefont
  {Chemali}}, \bibinfo {author} {\bibfnamefont {M.~Brandon}\ \bibnamefont
  {Westover}}, \bibinfo {author} {\bibfnamefont {Jonathan~D.}\ \bibnamefont
  {Kenny}}, \bibinfo {author} {\bibfnamefont {Ken}\ \bibnamefont {Solt}},
  \bibinfo {author} {\bibfnamefont {Patrick~L.}\ \bibnamefont {Purdon}}, \ and\
  \bibinfo {author} {\bibfnamefont {Emery~N.}\ \bibnamefont {Brown}}} (\bibinfo
  {year} {2013}),\ \bibfield  {title} {\enquote {\bibinfo {title} {Real-time
  closed-loop control in a rodent model of medically induced coma using burst
  suppression},}\ }\href {\doibase 10.1097/ALN.0b013e31829d4ab4} {\bibfield
  {journal} {\bibinfo  {journal} {Anesthesiology}\ }\textbf {\bibinfo {volume}
  {119}}~(\bibinfo {number} {4}),\ \bibinfo {pages} {848--860}}\BibitemShut
  {NoStop}%
\bibitem [{\citenamefont {Colizza}\ \emph {et~al.}(2006)\citenamefont
  {Colizza}, \citenamefont {Flammini}, \citenamefont {Serrano},\ and\
  \citenamefont {Vespignani}}]{colizza2006detecting}%
  \BibitemOpen
  \bibfield  {author} {\bibinfo {author} {\bibnamefont {Colizza}, \bibfnamefont
  {V}}, \bibinfo {author} {\bibfnamefont {A}~\bibnamefont {Flammini}}, \bibinfo
  {author} {\bibfnamefont {M~A}\ \bibnamefont {Serrano}}, \ and\ \bibinfo
  {author} {\bibfnamefont {A}~\bibnamefont {Vespignani}}} (\bibinfo {year}
  {2006}),\ \bibfield  {title} {\enquote {\bibinfo {title} {Detecting rich-club
  ordering in complex networks},}\ }\href@noop {} {\bibfield  {journal}
  {\bibinfo  {journal} {Nature Physics}\ }\textbf {\bibinfo {volume} {2}},\
  \bibinfo {pages} {110 -- 115}}\BibitemShut {NoStop}%
\bibitem [{\citenamefont {Corbetta}\ and\ \citenamefont
  {Shulman}(2002)}]{corbetta2002control}%
  \BibitemOpen
  \bibfield  {author} {\bibinfo {author} {\bibnamefont {Corbetta},
  \bibfnamefont {M}}, \ and\ \bibinfo {author} {\bibfnamefont {G~L}\
  \bibnamefont {Shulman}}} (\bibinfo {year} {2002}),\ \bibfield  {title}
  {\enquote {\bibinfo {title} {Control of goal-directed and stimulus-driven
  attention in the brain},}\ }\href@noop {} {\bibfield  {journal} {\bibinfo
  {journal} {Nat Rev Neurosci}\ }\textbf {\bibinfo {volume} {3}}~(\bibinfo
  {number} {3}),\ \bibinfo {pages} {201--215}}\BibitemShut {NoStop}%
\bibitem [{\citenamefont {Cornblath}\ \emph {et~al.}(2018)\citenamefont
  {Cornblath}, \citenamefont {Tang}, \citenamefont {Baum}, \citenamefont
  {Moore}, \citenamefont {Roalf}, \citenamefont {Gur}, \citenamefont {Gur},
  \citenamefont {Pasqualetti}, \citenamefont {Satterthwaite},\ and\
  \citenamefont {Bassett}}]{cornblath2018sex}%
  \BibitemOpen
  \bibfield  {author} {\bibinfo {author} {\bibnamefont {Cornblath},
  \bibfnamefont {E~J}}, \bibinfo {author} {\bibfnamefont {E}~\bibnamefont
  {Tang}}, \bibinfo {author} {\bibfnamefont {G~L}\ \bibnamefont {Baum}},
  \bibinfo {author} {\bibfnamefont {T~M}\ \bibnamefont {Moore}}, \bibinfo
  {author} {\bibfnamefont {D~R}\ \bibnamefont {Roalf}}, \bibinfo {author}
  {\bibfnamefont {R~C}\ \bibnamefont {Gur}}, \bibinfo {author} {\bibfnamefont
  {R~E}\ \bibnamefont {Gur}}, \bibinfo {author} {\bibfnamefont {F}~\bibnamefont
  {Pasqualetti}}, \bibinfo {author} {\bibfnamefont {T~D}\ \bibnamefont
  {Satterthwaite}}, \ and\ \bibinfo {author} {\bibfnamefont {D~S}\ \bibnamefont
  {Bassett}}} (\bibinfo {year} {2018}),\ \bibfield  {title} {\enquote {\bibinfo
  {title} {Sex differences in network controllability as a predictor of
  executive function in youth},}\ }\href@noop {} {\bibfield  {journal}
  {\bibinfo  {journal} {arxiv}\ }\textbf {\bibinfo {volume} {1801}},\ \bibinfo
  {pages} {04623}}\BibitemShut {NoStop}%
\bibitem [{\citenamefont {Cornelius}\ \emph {et~al.}(2013)\citenamefont
  {Cornelius}, \citenamefont {Kath},\ and\ \citenamefont
  {Motter}}]{Cornelius2013}%
  \BibitemOpen
  \bibfield  {author} {\bibinfo {author} {\bibnamefont {Cornelius},
  \bibfnamefont {S~P}}, \bibinfo {author} {\bibfnamefont {W~L}\ \bibnamefont
  {Kath}}, \ and\ \bibinfo {author} {\bibfnamefont {A~E}\ \bibnamefont
  {Motter}}} (\bibinfo {year} {2013}),\ \bibfield  {title} {\enquote {\bibinfo
  {title} {Realistic control of network dynamics},}\ }\href@noop {} {\bibfield
  {journal} {\bibinfo  {journal} {Nat Commun}\ }\textbf {\bibinfo {volume}
  {4}},\ \bibinfo {pages} {1942}}\BibitemShut {NoStop}%
\bibitem [{\citenamefont {Cowan}\ \emph {et~al.}(2012)\citenamefont {Cowan},
  \citenamefont {Chastain}, \citenamefont {Vilhena}, \citenamefont
  {Freudenberg},\ and\ \citenamefont
  {Bergstrom}}]{10.1371/journal.pone.0038398}%
  \BibitemOpen
  \bibfield  {author} {\bibinfo {author} {\bibnamefont {Cowan}, \bibfnamefont
  {Noah~J}}, \bibinfo {author} {\bibfnamefont {Erick~J.}\ \bibnamefont
  {Chastain}}, \bibinfo {author} {\bibfnamefont {Daril~A.}\ \bibnamefont
  {Vilhena}}, \bibinfo {author} {\bibfnamefont {James~S.}\ \bibnamefont
  {Freudenberg}}, \ and\ \bibinfo {author} {\bibfnamefont {Carl~T.}\
  \bibnamefont {Bergstrom}}} (\bibinfo {year} {2012}),\ \bibfield  {title}
  {\enquote {\bibinfo {title} {Nodal dynamics, not degree distributions,
  determine the structural controllability of complex networks},}\ }\href
  {\doibase 10.1371/journal.pone.0038398} {\bibfield  {journal} {\bibinfo
  {journal} {PLOS ONE}\ }\textbf {\bibinfo {volume} {7}}~(\bibinfo {number}
  {6}),\ \bibinfo {pages} {1--5}}\BibitemShut {NoStop}%
\bibitem [{\citenamefont {Crone}\ \emph {et~al.}(2016)\citenamefont {Crone},
  \citenamefont {Lutkenhoff}, \citenamefont {Bio}, \citenamefont {Laureys},\
  and\ \citenamefont {Monti}}]{crone2016testing}%
  \BibitemOpen
  \bibfield  {author} {\bibinfo {author} {\bibnamefont {Crone}, \bibfnamefont
  {J~S}}, \bibinfo {author} {\bibfnamefont {E~S}\ \bibnamefont {Lutkenhoff}},
  \bibinfo {author} {\bibfnamefont {B~J}\ \bibnamefont {Bio}}, \bibinfo
  {author} {\bibfnamefont {S}~\bibnamefont {Laureys}}, \ and\ \bibinfo {author}
  {\bibfnamefont {M~M}\ \bibnamefont {Monti}}} (\bibinfo {year} {2016}),\
  \bibfield  {title} {\enquote {\bibinfo {title} {Testing proposed neuronal
  models of effective connectivity within the cortico-basal
  ganglia-thalamo-cortical loop during loss of consciousness},}\ }\href@noop {}
  {\bibfield  {journal} {\bibinfo  {journal} {Cereb Cortex}\ }\textbf {\bibinfo
  {volume} {Apr 24}},\ \bibinfo {pages} {bhw112}}\BibitemShut {NoStop}%
\bibitem [{\citenamefont {Davison}\ \emph {et~al.}(2016)\citenamefont
  {Davison}, \citenamefont {Dey},\ and\ \citenamefont {Leonard}}]{lizdavison}%
  \BibitemOpen
  \bibfield  {author} {\bibinfo {author} {\bibnamefont {Davison}, \bibfnamefont
  {Elizabeth}}, \bibinfo {author} {\bibfnamefont {Biswadip}\ \bibnamefont
  {Dey}}, \ and\ \bibinfo {author} {\bibfnamefont {Naomi~Ehrich}\ \bibnamefont
  {Leonard}}} (\bibinfo {year} {2016}),\ \bibfield  {title} {\enquote {\bibinfo
  {title} {Synchronization bound for networks of nonlinear oscillators},}\ }in\
  \href@noop {} {\emph {\bibinfo {booktitle} {54th Annual Allerton Conference
  on Communication, Control and Computing}}}\BibitemShut {NoStop}%
\bibitem [{\citenamefont {Deco}\ and\ \citenamefont
  {Jirsa}(2012)}]{deco2012ongoing}%
  \BibitemOpen
  \bibfield  {author} {\bibinfo {author} {\bibnamefont {Deco}, \bibfnamefont
  {G}}, \ and\ \bibinfo {author} {\bibfnamefont {V~K}\ \bibnamefont {Jirsa}}}
  (\bibinfo {year} {2012}),\ \bibfield  {title} {\enquote {\bibinfo {title}
  {Ongoing cortical activity at rest: criticality, multistability, and ghost
  attractors},}\ }\href@noop {} {\bibfield  {journal} {\bibinfo  {journal} {J
  Neurosci}\ }\textbf {\bibinfo {volume} {32}}~(\bibinfo {number} {10}),\
  \bibinfo {pages} {3366--3375}}\BibitemShut {NoStop}%
\bibitem [{\citenamefont {DeVille}\ and\ \citenamefont
  {Lerman}(2015)}]{1303.3907}%
  \BibitemOpen
  \bibfield  {author} {\bibinfo {author} {\bibnamefont {DeVille}, \bibfnamefont
  {Lee}}, \ and\ \bibinfo {author} {\bibfnamefont {Eugene}\ \bibnamefont
  {Lerman}}} (\bibinfo {year} {2015}),\ \bibfield  {title} {\enquote {\bibinfo
  {title} {Modular dynamical systems on networks},}\ }\href {\doibase
  10.4171/JEMS/577} {\bibfield  {journal} {\bibinfo  {journal} {J. Eur. Math.
  Soc.}\ }\textbf {\bibinfo {volume} {17}}~(\bibinfo {number} {12}),\ \bibinfo
  {pages} {2977--3013}}\BibitemShut {NoStop}%
\bibitem [{\citenamefont {Duong}(2010)}]{duong2010diffusion}%
  \BibitemOpen
  \bibfield  {author} {\bibinfo {author} {\bibnamefont {Duong}, \bibfnamefont
  {T~Q}}} (\bibinfo {year} {2010}),\ \bibfield  {title} {\enquote {\bibinfo
  {title} {Diffusion tensor and perfusion {MRI} of non-human primates},}\
  }\href@noop {} {\bibfield  {journal} {\bibinfo  {journal} {Methods}\ }\textbf
  {\bibinfo {volume} {50}}~(\bibinfo {number} {3}),\ \bibinfo {pages}
  {125--135}}\BibitemShut {NoStop}%
\bibitem [{\citenamefont {Eisenreich}\ \emph {et~al.}(2016)\citenamefont
  {Eisenreich}, \citenamefont {Akaishi},\ and\ \citenamefont
  {Hayden}}]{Eisenreich077685}%
  \BibitemOpen
  \bibfield  {author} {\bibinfo {author} {\bibnamefont {Eisenreich},
  \bibfnamefont {Benjamin}}, \bibinfo {author} {\bibfnamefont {Rei}\
  \bibnamefont {Akaishi}}, \ and\ \bibinfo {author} {\bibfnamefont {Benjamin}\
  \bibnamefont {Hayden}}} (\bibinfo {year} {2016}),\ \bibfield  {title}
  {\enquote {\bibinfo {title} {Control without controllers: Towards a
  distributed neuroscience of executive control},}\ }\href {\doibase
  10.1101/077685} {\bibfield  {journal} {\bibinfo  {journal} {bioRxiv}\
  }10.1101/077685}\BibitemShut {NoStop}%
\bibitem [{\citenamefont {Eldar}\ \emph {et~al.}(2013)\citenamefont {Eldar},
  \citenamefont {Cohen},\ and\ \citenamefont {Niv}}]{eldar2013effects}%
  \BibitemOpen
  \bibfield  {author} {\bibinfo {author} {\bibnamefont {Eldar}, \bibfnamefont
  {E}}, \bibinfo {author} {\bibfnamefont {J~D}\ \bibnamefont {Cohen}}, \ and\
  \bibinfo {author} {\bibfnamefont {Y}~\bibnamefont {Niv}}} (\bibinfo {year}
  {2013}),\ \bibfield  {title} {\enquote {\bibinfo {title} {The effects of
  neural gain on attention and learning},}\ }\href@noop {} {\bibfield
  {journal} {\bibinfo  {journal} {Nat Neurosci}\ }\textbf {\bibinfo {volume}
  {16}}~(\bibinfo {number} {8}),\ \bibinfo {pages} {1146--1153}}\BibitemShut
  {NoStop}%
\bibitem [{\citenamefont {Fern\'{a}ndez~Gal\'{a}n}(2008)}]{galan2008how}%
  \BibitemOpen
  \bibfield  {author} {\bibinfo {author} {\bibnamefont
  {Fern\'{a}ndez~Gal\'{a}n}, \bibfnamefont {R}}} (\bibinfo {year} {2008}),\
  \bibfield  {title} {\enquote {\bibinfo {title} {On how network architecture
  determines the dominant patterns of spontaneous neural activity},}\
  }\href@noop {} {\bibfield  {journal} {\bibinfo  {journal} {PLoS One}\
  }\textbf {\bibinfo {volume} {3}}~(\bibinfo {number} {5}),\ \bibinfo {pages}
  {e2148}}\BibitemShut {NoStop}%
\bibitem [{\citenamefont {Fiedler}\ \emph {et~al.}(2013)\citenamefont
  {Fiedler}, \citenamefont {Mochizuki}, \citenamefont {Kurosawa},\ and\
  \citenamefont {Saito}}]{Fiedler2013}%
  \BibitemOpen
  \bibfield  {author} {\bibinfo {author} {\bibnamefont {Fiedler}, \bibfnamefont
  {Bernold}}, \bibinfo {author} {\bibfnamefont {Atsushi}\ \bibnamefont
  {Mochizuki}}, \bibinfo {author} {\bibfnamefont {Gen}\ \bibnamefont
  {Kurosawa}}, \ and\ \bibinfo {author} {\bibfnamefont {Daisuke}\ \bibnamefont
  {Saito}}} (\bibinfo {year} {2013}),\ \bibfield  {title} {\enquote {\bibinfo
  {title} {Dynamics and control at feedback vertex sets. i: Informative and
  determining nodes in regulatory networks},}\ }\href {\doibase
  10.1007/s10884-013-9312-7} {\bibfield  {journal} {\bibinfo  {journal}
  {Journal of Dynamics and Differential Equations}\ }\textbf {\bibinfo {volume}
  {25}}~(\bibinfo {number} {3}),\ \bibinfo {pages} {563--604}}\BibitemShut
  {NoStop}%
\bibitem [{\citenamefont {Fornito}\ and\ \citenamefont
  {Bullmore}(2015)}]{fornito2015connectomics}%
  \BibitemOpen
  \bibfield  {author} {\bibinfo {author} {\bibnamefont {Fornito}, \bibfnamefont
  {A}}, \ and\ \bibinfo {author} {\bibfnamefont {E~T}\ \bibnamefont
  {Bullmore}}} (\bibinfo {year} {2015}),\ \bibfield  {title} {\enquote
  {\bibinfo {title} {Connectomics: a new paradigm for understanding brain
  disease},}\ }\href@noop {} {\bibfield  {journal} {\bibinfo  {journal} {Eur
  Neuropsychopharmacol}\ }\textbf {\bibinfo {volume} {25}}~(\bibinfo {number}
  {5}),\ \bibinfo {pages} {733--748}}\BibitemShut {NoStop}%
\bibitem [{\citenamefont {Fraiman}\ \emph {et~al.}(2009)\citenamefont
  {Fraiman}, \citenamefont {Balenzuela}, \citenamefont {Foss},\ and\
  \citenamefont {Chialvo}}]{fraiman2009ising}%
  \BibitemOpen
  \bibfield  {author} {\bibinfo {author} {\bibnamefont {Fraiman}, \bibfnamefont
  {D}}, \bibinfo {author} {\bibfnamefont {P}~\bibnamefont {Balenzuela}},
  \bibinfo {author} {\bibfnamefont {J}~\bibnamefont {Foss}}, \ and\ \bibinfo
  {author} {\bibfnamefont {D~R}\ \bibnamefont {Chialvo}}} (\bibinfo {year}
  {2009}),\ \bibfield  {title} {\enquote {\bibinfo {title} {Ising-like dynamics
  in large-scale functional brain networks},}\ }\href@noop {} {\bibfield
  {journal} {\bibinfo  {journal} {Phys Rev E Stat Nonlin Soft Matter Phys}\
  }\textbf {\bibinfo {volume} {79}}~(\bibinfo {number} {6 Pt 1}),\ \bibinfo
  {pages} {061922}}\BibitemShut {NoStop}%
\bibitem [{\citenamefont {Freund}\ and\ \citenamefont
  {Schapire}(1999)}]{freund1999large}%
  \BibitemOpen
  \bibfield  {author} {\bibinfo {author} {\bibnamefont {Freund}, \bibfnamefont
  {Y}}, \ and\ \bibinfo {author} {\bibfnamefont {R~E}\ \bibnamefont
  {Schapire}}} (\bibinfo {year} {1999}),\ \bibfield  {title} {\enquote
  {\bibinfo {title} {Large margin classification using the perceptron
  algorithm},}\ }\href@noop {} {\bibfield  {journal} {\bibinfo  {journal}
  {Machine Learning}\ }\textbf {\bibinfo {volume} {37}}~(\bibinfo {number}
  {3}),\ \bibinfo {pages} {277--296}}\BibitemShut {NoStop}%
\bibitem [{\citenamefont {Gao}\ and\ \citenamefont
  {Ganguli}(2015)}]{gao2015simplicity}%
  \BibitemOpen
  \bibfield  {author} {\bibinfo {author} {\bibnamefont {Gao}, \bibfnamefont
  {P}}, \ and\ \bibinfo {author} {\bibfnamefont {S}~\bibnamefont {Ganguli}}}
  (\bibinfo {year} {2015}),\ \bibfield  {title} {\enquote {\bibinfo {title} {On
  simplicity and complexity in the brave new world of large-scale
  neuroscience},}\ }\href@noop {} {\bibfield  {journal} {\bibinfo  {journal}
  {Curr Opin Neurobiol}\ }\textbf {\bibinfo {volume} {32}},\ \bibinfo {pages}
  {148--155}}\BibitemShut {NoStop}%
\bibitem [{\citenamefont {Gardner}(1987)}]{0295-5075-4-4-016}%
  \BibitemOpen
  \bibfield  {author} {\bibinfo {author} {\bibnamefont {Gardner}, \bibfnamefont
  {E}}} (\bibinfo {year} {1987}),\ \bibfield  {title} {\enquote {\bibinfo
  {title} {Maximum storage capacity in neural networks},}\ }\href
  {http://stacks.iop.org/0295-5075/4/i=4/a=016} {\bibfield  {journal} {\bibinfo
   {journal} {EPL (Europhysics Letters)}\ }\textbf {\bibinfo {volume}
  {4}}~(\bibinfo {number} {4}),\ \bibinfo {pages} {481}}\BibitemShut {NoStop}%
\bibitem [{\citenamefont {Giusti}\ \emph {et~al.}(2016)\citenamefont {Giusti},
  \citenamefont {Ghrist},\ and\ \citenamefont {Bassett}}]{giusti2016twos}%
  \BibitemOpen
  \bibfield  {author} {\bibinfo {author} {\bibnamefont {Giusti}, \bibfnamefont
  {C}}, \bibinfo {author} {\bibfnamefont {R}~\bibnamefont {Ghrist}}, \ and\
  \bibinfo {author} {\bibfnamefont {D~S}\ \bibnamefont {Bassett}}} (\bibinfo
  {year} {2016}),\ \bibfield  {title} {\enquote {\bibinfo {title} {Two's
  company, three (or more) is a simplex: {A}lgebraic-topological tools for
  understanding higher-order structure in neural data},}\ }\href@noop {}
  {\bibfield  {journal} {\bibinfo  {journal} {J Comput Neurosci}\ }\textbf
  {\bibinfo {volume} {41}}~(\bibinfo {number} {1}),\ \bibinfo {pages}
  {1--14}}\BibitemShut {NoStop}%
\bibitem [{\citenamefont {Glaser}\ and\ \citenamefont
  {Kording}(2016)}]{glaser2016development}%
  \BibitemOpen
  \bibfield  {author} {\bibinfo {author} {\bibnamefont {Glaser}, \bibfnamefont
  {J~I}}, \ and\ \bibinfo {author} {\bibfnamefont {K~P}\ \bibnamefont
  {Kording}}} (\bibinfo {year} {2016}),\ \bibfield  {title} {\enquote {\bibinfo
  {title} {The development and analysis of integrated neuroscience data},}\
  }\href@noop {} {\bibfield  {journal} {\bibinfo  {journal} {Front Comput
  Neurosci}\ }\textbf {\bibinfo {volume} {10}},\ \bibinfo {pages}
  {11}}\BibitemShut {NoStop}%
\bibitem [{\citenamefont {Glasser}\ \emph {et~al.}(2016)\citenamefont
  {Glasser}, \citenamefont {Coalson}, \citenamefont {Robinson}, \citenamefont
  {Hacker}, \citenamefont {Harwell}, \citenamefont {Yacoub}, \citenamefont
  {Ugurbil}, \citenamefont {Andersson}, \citenamefont {Beckmann}, \citenamefont
  {Jenkinson}, \citenamefont {Smith},\ and\ \citenamefont
  {Van~Essen}}]{glasser2016multi}%
  \BibitemOpen
  \bibfield  {author} {\bibinfo {author} {\bibnamefont {Glasser}, \bibfnamefont
  {M~F}}, \bibinfo {author} {\bibfnamefont {T~S}\ \bibnamefont {Coalson}},
  \bibinfo {author} {\bibfnamefont {E~C}\ \bibnamefont {Robinson}}, \bibinfo
  {author} {\bibfnamefont {C~D}\ \bibnamefont {Hacker}}, \bibinfo {author}
  {\bibfnamefont {J}~\bibnamefont {Harwell}}, \bibinfo {author} {\bibfnamefont
  {E}~\bibnamefont {Yacoub}}, \bibinfo {author} {\bibfnamefont {K}~\bibnamefont
  {Ugurbil}}, \bibinfo {author} {\bibfnamefont {J}~\bibnamefont {Andersson}},
  \bibinfo {author} {\bibfnamefont {C~F}\ \bibnamefont {Beckmann}}, \bibinfo
  {author} {\bibfnamefont {M}~\bibnamefont {Jenkinson}}, \bibinfo {author}
  {\bibfnamefont {S~M}\ \bibnamefont {Smith}}, \ and\ \bibinfo {author}
  {\bibfnamefont {D~C}\ \bibnamefont {Van~Essen}}} (\bibinfo {year} {2016}),\
  \bibfield  {title} {\enquote {\bibinfo {title} {A multi-modal parcellation of
  human cerebral cortex},}\ }\href@noop {} {\bibfield  {journal} {\bibinfo
  {journal} {Nature}\ }\textbf {\bibinfo {volume} {536}}~(\bibinfo {number}
  {7615}),\ \bibinfo {pages} {171--178}}\BibitemShut {NoStop}%
\bibitem [{\citenamefont {Golos}\ \emph {et~al.}(2015)\citenamefont {Golos},
  \citenamefont {Jirsa},\ and\ \citenamefont
  {Dauce}}]{golos2015multistability}%
  \BibitemOpen
  \bibfield  {author} {\bibinfo {author} {\bibnamefont {Golos}, \bibfnamefont
  {M}}, \bibinfo {author} {\bibfnamefont {V}~\bibnamefont {Jirsa}}, \ and\
  \bibinfo {author} {\bibfnamefont {E}~\bibnamefont {Dauce}}} (\bibinfo {year}
  {2015}),\ \bibfield  {title} {\enquote {\bibinfo {title} {Multistability in
  large scale models of brain activity},}\ }\href@noop {} {\bibfield  {journal}
  {\bibinfo  {journal} {PLoS Comput Biol}\ }\textbf {\bibinfo {volume}
  {11}}~(\bibinfo {number} {12}),\ \bibinfo {pages} {e1004644}}\BibitemShut
  {NoStop}%
\bibitem [{\citenamefont {G\'omez-Garde\~nes}\ \emph
  {et~al.}(2007)\citenamefont {G\'omez-Garde\~nes}, \citenamefont {Moreno},\
  and\ \citenamefont {Arenas}}]{PhysRevLett.98.034101}%
  \BibitemOpen
  \bibfield  {author} {\bibinfo {author} {\bibnamefont {G\'omez-Garde\~nes},
  \bibfnamefont {Jes\'us}}, \bibinfo {author} {\bibfnamefont {Yamir}\
  \bibnamefont {Moreno}}, \ and\ \bibinfo {author} {\bibfnamefont {Alex}\
  \bibnamefont {Arenas}}} (\bibinfo {year} {2007}),\ \bibfield  {title}
  {\enquote {\bibinfo {title} {Paths to synchronization on complex networks},}\
  }\href {\doibase 10.1103/PhysRevLett.98.034101} {\bibfield  {journal}
  {\bibinfo  {journal} {Phys. Rev. Lett.}\ }\textbf {\bibinfo {volume} {98}},\
  \bibinfo {pages} {034101}}\BibitemShut {NoStop}%
\bibitem [{\citenamefont {G\'omez-Garde\~nes}\ \emph
  {et~al.}(2010)\citenamefont {G\'omez-Garde\~nes}, \citenamefont
  {Zamora-L\'opez}, \citenamefont {Moreno},\ and\ \citenamefont
  {Arenas}}]{10.1371/journal.pone.0012313}%
  \BibitemOpen
  \bibfield  {author} {\bibinfo {author} {\bibnamefont {G\'omez-Garde\~nes},
  \bibfnamefont {Jes\'us}}, \bibinfo {author} {\bibfnamefont {Gorka}\
  \bibnamefont {Zamora-L\'opez}}, \bibinfo {author} {\bibfnamefont {Yamir}\
  \bibnamefont {Moreno}}, \ and\ \bibinfo {author} {\bibfnamefont {Alex}\
  \bibnamefont {Arenas}}} (\bibinfo {year} {2010}),\ \bibfield  {title}
  {\enquote {\bibinfo {title} {From modular to centralized organization of
  synchronization in functional areas of the cat cerebral cortex},}\ }\href
  {\doibase 10.1371/journal.pone.0012313} {\bibfield  {journal} {\bibinfo
  {journal} {PLOS ONE}\ }\textbf {\bibinfo {volume} {5}}~(\bibinfo {number}
  {8}),\ \bibinfo {pages} {1--11}}\BibitemShut {NoStop}%
\bibitem [{\citenamefont {Grosenick}\ \emph {et~al.}(2016)\citenamefont
  {Grosenick}, \citenamefont {Marshel},\ and\ \citenamefont
  {Deisseroth}}]{ref1}%
  \BibitemOpen
  \bibfield  {author} {\bibinfo {author} {\bibnamefont {Grosenick},
  \bibfnamefont {Logan}}, \bibinfo {author} {\bibfnamefont {James~H}\
  \bibnamefont {Marshel}}, \ and\ \bibinfo {author} {\bibfnamefont {Karl}\
  \bibnamefont {Deisseroth}}} (\bibinfo {year} {2016}),\ \bibfield  {title}
  {\enquote {\bibinfo {title} {Closed-loop and activity-guided optogenetic
  control},}\ }\href {\doibase 10.1016/j.neuron.2015.03.034} {\bibfield
  {journal} {\bibinfo  {journal} {Neuron}\ }\textbf {\bibinfo {volume}
  {86}}~(\bibinfo {number} {1}),\ \bibinfo {pages} {106--139}}\BibitemShut
  {NoStop}%
\bibitem [{\citenamefont {Gu}\ \emph {et~al.}(2016)\citenamefont {Gu},
  \citenamefont {Betzel}, \citenamefont {Mattar}, \citenamefont {Cieslak},
  \citenamefont {Delio}, \citenamefont {Grafton}, \citenamefont {Pasqualetti},\
  and\ \citenamefont {Bassett}}]{1607.01706}%
  \BibitemOpen
  \bibfield  {author} {\bibinfo {author} {\bibnamefont {Gu}, \bibfnamefont
  {S}}, \bibinfo {author} {\bibfnamefont {R~F}\ \bibnamefont {Betzel}},
  \bibinfo {author} {\bibfnamefont {M~G}\ \bibnamefont {Mattar}}, \bibinfo
  {author} {\bibfnamefont {M}~\bibnamefont {Cieslak}}, \bibinfo {author}
  {\bibfnamefont {P~R}\ \bibnamefont {Delio}}, \bibinfo {author} {\bibfnamefont
  {S~T}\ \bibnamefont {Grafton}}, \bibinfo {author} {\bibfnamefont
  {F}~\bibnamefont {Pasqualetti}}, \ and\ \bibinfo {author} {\bibfnamefont
  {D~S}\ \bibnamefont {Bassett}}} (\bibinfo {year} {2016}),\ \bibfield  {title}
  {\enquote {\bibinfo {title} {Optimal trajectories of brain state
  transitions},}\ }\href@noop {} {\bibfield  {journal} {\bibinfo  {journal}
  {Neuroimage}\ }\textbf {\bibinfo {volume} {In Press}}~(\bibinfo {number}
  {arXiv:1607}),\ \bibinfo {pages} {01706}}\BibitemShut {NoStop}%
\bibitem [{\citenamefont {Gu}\ \emph {et~al.}(2015)\citenamefont {Gu},
  \citenamefont {Pasqualetti}, \citenamefont {Cieslak}, \citenamefont
  {Telesford}, \citenamefont {Alfred}, \citenamefont {Kahn}, \citenamefont
  {Medaglia}, \citenamefont {Vettel}, \citenamefont {Miller}, \citenamefont
  {Grafton} \emph {et~al.}}]{gu2015controllability}%
  \BibitemOpen
  \bibfield  {author} {\bibinfo {author} {\bibnamefont {Gu}, \bibfnamefont
  {Shi}}, \bibinfo {author} {\bibfnamefont {Fabio}\ \bibnamefont
  {Pasqualetti}}, \bibinfo {author} {\bibfnamefont {Matthew}\ \bibnamefont
  {Cieslak}}, \bibinfo {author} {\bibfnamefont {Qawi~K}\ \bibnamefont
  {Telesford}}, \bibinfo {author} {\bibfnamefont {B~Yu}\ \bibnamefont
  {Alfred}}, \bibinfo {author} {\bibfnamefont {Ari~E}\ \bibnamefont {Kahn}},
  \bibinfo {author} {\bibfnamefont {John~D}\ \bibnamefont {Medaglia}}, \bibinfo
  {author} {\bibfnamefont {Jean~M}\ \bibnamefont {Vettel}}, \bibinfo {author}
  {\bibfnamefont {Michael~B}\ \bibnamefont {Miller}}, \bibinfo {author}
  {\bibfnamefont {Scott~T}\ \bibnamefont {Grafton}},  \emph {et~al.}} (\bibinfo
  {year} {2015}),\ \bibfield  {title} {\enquote {\bibinfo {title}
  {Controllability of structural brain networks},}\ }\href@noop {} {\bibfield
  {journal} {\bibinfo  {journal} {Nature communications}\ }\textbf {\bibinfo
  {volume} {6}}}\BibitemShut {NoStop}%
\bibitem [{\citenamefont {Hagmann}\ \emph {et~al.}(2008)\citenamefont
  {Hagmann}, \citenamefont {Cammoun}, \citenamefont {Gigandet}, \citenamefont
  {Meuli}, \citenamefont {Honey}, \citenamefont {Wedeen},\ and\ \citenamefont
  {Sporns}}]{Hagmann2008}%
  \BibitemOpen
  \bibfield  {author} {\bibinfo {author} {\bibnamefont {Hagmann}, \bibfnamefont
  {P}}, \bibinfo {author} {\bibfnamefont {L}~\bibnamefont {Cammoun}}, \bibinfo
  {author} {\bibfnamefont {X}~\bibnamefont {Gigandet}}, \bibinfo {author}
  {\bibfnamefont {R}~\bibnamefont {Meuli}}, \bibinfo {author} {\bibfnamefont
  {C~J}\ \bibnamefont {Honey}}, \bibinfo {author} {\bibfnamefont {V~J}\
  \bibnamefont {Wedeen}}, \ and\ \bibinfo {author} {\bibfnamefont
  {O}~\bibnamefont {Sporns}}} (\bibinfo {year} {2008}),\ \bibfield  {title}
  {\enquote {\bibinfo {title} {Mapping the structural core of human cerebral
  cortex},}\ }\href@noop {} {\bibfield  {journal} {\bibinfo  {journal} {PLoS
  Biology}\ }\textbf {\bibinfo {volume} {6}}~(\bibinfo {number} {7}),\ \bibinfo
  {pages} {e159}}\BibitemShut {NoStop}%
\bibitem [{\citenamefont {Han}\ \emph {et~al.}(2015)\citenamefont {Han},
  \citenamefont {Preciado}, \citenamefont {Nowzari},\ and\ \citenamefont
  {Pappas}}]{victorconical}%
  \BibitemOpen
  \bibfield  {author} {\bibinfo {author} {\bibnamefont {Han}, \bibfnamefont
  {Shuo}}, \bibinfo {author} {\bibfnamefont {Victor~M.}\ \bibnamefont
  {Preciado}}, \bibinfo {author} {\bibfnamefont {Cameron}\ \bibnamefont
  {Nowzari}}, \ and\ \bibinfo {author} {\bibfnamefont {George~J.}\ \bibnamefont
  {Pappas}}} (\bibinfo {year} {2015}),\ \bibfield  {title} {\enquote {\bibinfo
  {title} {Data-driven network resource allocation for controlling spreading
  processes},}\ }\href@noop {} {\bibfield  {journal} {\bibinfo  {journal}
  {Network Science and Engineering, IEEE Transactions on}\ }\textbf {\bibinfo
  {volume} {2}}~(\bibinfo {number} {4}),\ \bibinfo {pages}
  {127--138}}\BibitemShut {NoStop}%
\bibitem [{\citenamefont {Haykin}\ and\ \citenamefont
  {Fuster}(2014)}]{6777299}%
  \BibitemOpen
  \bibfield  {author} {\bibinfo {author} {\bibnamefont {Haykin}, \bibfnamefont
  {S}}, \ and\ \bibinfo {author} {\bibfnamefont {J.~M.}\ \bibnamefont
  {Fuster}}} (\bibinfo {year} {2014}),\ \bibfield  {title} {\enquote {\bibinfo
  {title} {On cognitive dynamic systems: Cognitive neuroscience and engineering
  learning from each other},}\ }\href {\doibase 10.1109/JPROC.2014.2311211}
  {\bibfield  {journal} {\bibinfo  {journal} {Proceedings of the IEEE}\
  }\textbf {\bibinfo {volume} {102}}~(\bibinfo {number} {4}),\ \bibinfo {pages}
  {608--628}}\BibitemShut {NoStop}%
\bibitem [{\citenamefont {Heatherton}\ and\ \citenamefont
  {Wagner}(2011)}]{heatherton2011cognitive}%
  \BibitemOpen
  \bibfield  {author} {\bibinfo {author} {\bibnamefont {Heatherton},
  \bibfnamefont {T~F}}, \ and\ \bibinfo {author} {\bibfnamefont {D~D}\
  \bibnamefont {Wagner}}} (\bibinfo {year} {2011}),\ \bibfield  {title}
  {\enquote {\bibinfo {title} {Cognitive neuroscience of self-regulation
  failure},}\ }\href@noop {} {\bibfield  {journal} {\bibinfo  {journal} {Trends
  Cogn Sci}\ }\textbf {\bibinfo {volume} {15}}~(\bibinfo {number} {3}),\
  \bibinfo {pages} {132--139}}\BibitemShut {NoStop}%
\bibitem [{\citenamefont {Herculano-Houzel}\ \emph {et~al.}(2007)\citenamefont
  {Herculano-Houzel}, \citenamefont {Collins}, \citenamefont {Wong},\ and\
  \citenamefont {Kaas}}]{herculano2007cellular}%
  \BibitemOpen
  \bibfield  {author} {\bibinfo {author} {\bibnamefont {Herculano-Houzel},
  \bibfnamefont {S}}, \bibinfo {author} {\bibfnamefont {C~E}\ \bibnamefont
  {Collins}}, \bibinfo {author} {\bibfnamefont {P}~\bibnamefont {Wong}}, \ and\
  \bibinfo {author} {\bibfnamefont {J~H}\ \bibnamefont {Kaas}}} (\bibinfo
  {year} {2007}),\ \bibfield  {title} {\enquote {\bibinfo {title} {Cellular
  scaling rules for primate brains},}\ }\href@noop {} {\bibfield  {journal}
  {\bibinfo  {journal} {Proc Natl Acad Sci U S A}\ }\textbf {\bibinfo {volume}
  {104}}~(\bibinfo {number} {9}),\ \bibinfo {pages} {3562--3567}}\BibitemShut
  {NoStop}%
\bibitem [{\citenamefont {Holt}\ and\ \citenamefont {Netoff}(2014)}]{Holt2014}%
  \BibitemOpen
  \bibfield  {author} {\bibinfo {author} {\bibnamefont {Holt}, \bibfnamefont
  {Abbey~B}}, \ and\ \bibinfo {author} {\bibfnamefont {Theoden~I.}\
  \bibnamefont {Netoff}}} (\bibinfo {year} {2014}),\ \bibfield  {title}
  {\enquote {\bibinfo {title} {Origins and suppression of oscillations in a
  computational model of parkinson's disease},}\ }\href {\doibase
  10.1007/s10827-014-0523-7} {\bibfield  {journal} {\bibinfo  {journal}
  {Journal of Computational Neuroscience}\ }\textbf {\bibinfo {volume}
  {37}}~(\bibinfo {number} {3}),\ \bibinfo {pages} {505--521}}\BibitemShut
  {NoStop}%
\bibitem [{\citenamefont {Honey}\ \emph {et~al.}(2009)\citenamefont {Honey},
  \citenamefont {Sporns}, \citenamefont {Cammoun}, \citenamefont {Gigandet},
  \citenamefont {Thiran}, \citenamefont {Meuli},\ and\ \citenamefont
  {Hagmann}}]{honey2009predicting}%
  \BibitemOpen
  \bibfield  {author} {\bibinfo {author} {\bibnamefont {Honey}, \bibfnamefont
  {C~J}}, \bibinfo {author} {\bibfnamefont {O}~\bibnamefont {Sporns}}, \bibinfo
  {author} {\bibfnamefont {L}~\bibnamefont {Cammoun}}, \bibinfo {author}
  {\bibfnamefont {X}~\bibnamefont {Gigandet}}, \bibinfo {author} {\bibfnamefont
  {J~P}\ \bibnamefont {Thiran}}, \bibinfo {author} {\bibfnamefont
  {R}~\bibnamefont {Meuli}}, \ and\ \bibinfo {author} {\bibfnamefont
  {P}~\bibnamefont {Hagmann}}} (\bibinfo {year} {2009}),\ \bibfield  {title}
  {\enquote {\bibinfo {title} {Predicting human resting-state functional
  connectivity from structural connectivity},}\ }\href@noop {} {\bibfield
  {journal} {\bibinfo  {journal} {Proc Natl Acad Sci U S A}\ }\textbf {\bibinfo
  {volume} {106}}~(\bibinfo {number} {6}),\ \bibinfo {pages}
  {2035--2040}}\BibitemShut {NoStop}%
\bibitem [{\citenamefont {Hopfield}(1982)}]{Hopfield01041982}%
  \BibitemOpen
  \bibfield  {author} {\bibinfo {author} {\bibnamefont {Hopfield},
  \bibfnamefont {J~J}}} (\bibinfo {year} {1982}),\ \bibfield  {title} {\enquote
  {\bibinfo {title} {Neural networks and physical systems with emergent
  collective computational abilities},}\ }\href
  {http://www.pnas.org/content/79/8/2554.abstract} {\bibfield  {journal}
  {\bibinfo  {journal} {Proceedings of the National Academy of Sciences}\
  }\textbf {\bibinfo {volume} {79}}~(\bibinfo {number} {8}),\ \bibinfo {pages}
  {2554--2558}},\ \Eprint
  {http://arxiv.org/abs/http://www.pnas.org/content/79/8/2554.full.pdf}
  {http://www.pnas.org/content/79/8/2554.full.pdf} \BibitemShut {NoStop}%
\bibitem [{\citenamefont {Hopfinger}\ \emph {et~al.}(2017)\citenamefont
  {Hopfinger}, \citenamefont {Parsons},\ and\ \citenamefont
  {Fr\"{o}hlich}}]{flavio}%
  \BibitemOpen
  \bibfield  {author} {\bibinfo {author} {\bibnamefont {Hopfinger},
  \bibfnamefont {Joseph~B}}, \bibinfo {author} {\bibfnamefont {Jonathan}\
  \bibnamefont {Parsons}}, \ and\ \bibinfo {author} {\bibfnamefont {Flavio}\
  \bibnamefont {Fr\"{o}hlich}}} (\bibinfo {year} {2017}),\ \bibfield  {title}
  {\enquote {\bibinfo {title} {Differential effects of 10-hz and 40-hz
  transcranial alternating current stimulation (tacs) on endogenous versus
  exogenous attention},}\ }\href {\doibase 10.1080/17588928.2016.1194261}
  {\bibfield  {journal} {\bibinfo  {journal} {Cognitive Neuroscience}\ }\textbf
  {\bibinfo {volume} {8}}~(\bibinfo {number} {2}),\ \bibinfo {pages}
  {102--111}},\ \bibinfo {note} {pMID: 27297977}\BibitemShut {NoStop}%
\bibitem [{\citenamefont {Jirsa}\ \emph {et~al.}(2014)\citenamefont {Jirsa},
  \citenamefont {Stacey}, \citenamefont {Quilichini}, \citenamefont {Ivanov},\
  and\ \citenamefont {Bernard}}]{jirsa2014nature}%
  \BibitemOpen
  \bibfield  {author} {\bibinfo {author} {\bibnamefont {Jirsa}, \bibfnamefont
  {V~K}}, \bibinfo {author} {\bibfnamefont {W~C}\ \bibnamefont {Stacey}},
  \bibinfo {author} {\bibfnamefont {P~P}\ \bibnamefont {Quilichini}}, \bibinfo
  {author} {\bibfnamefont {A~I}\ \bibnamefont {Ivanov}}, \ and\ \bibinfo
  {author} {\bibfnamefont {C}~\bibnamefont {Bernard}}} (\bibinfo {year}
  {2014}),\ \bibfield  {title} {\enquote {\bibinfo {title} {On the nature of
  seizure dynamics},}\ }\href@noop {} {\bibfield  {journal} {\bibinfo
  {journal} {Brain}\ }\textbf {\bibinfo {volume} {137}}~(\bibinfo {number} {Pt
  8}),\ \bibinfo {pages} {2210--2230}}\BibitemShut {NoStop}%
\bibitem [{\citenamefont {Johnson}\ \emph {et~al.}(2013)\citenamefont
  {Johnson}, \citenamefont {Lim}, \citenamefont {Netoff}, \citenamefont
  {Connolly}, \citenamefont {Johnson}, \citenamefont {Roy}, \citenamefont
  {Holt}, \citenamefont {Lim}, \citenamefont {Carey}, \citenamefont {Vitek},\
  and\ \citenamefont {He}}]{johnson2013neuromodulation}%
  \BibitemOpen
  \bibfield  {author} {\bibinfo {author} {\bibnamefont {Johnson}, \bibfnamefont
  {M~D}}, \bibinfo {author} {\bibfnamefont {H~H}\ \bibnamefont {Lim}}, \bibinfo
  {author} {\bibfnamefont {T~I}\ \bibnamefont {Netoff}}, \bibinfo {author}
  {\bibfnamefont {A~T}\ \bibnamefont {Connolly}}, \bibinfo {author}
  {\bibfnamefont {N}~\bibnamefont {Johnson}}, \bibinfo {author} {\bibfnamefont
  {A}~\bibnamefont {Roy}}, \bibinfo {author} {\bibfnamefont {A}~\bibnamefont
  {Holt}}, \bibinfo {author} {\bibfnamefont {K~O}\ \bibnamefont {Lim}},
  \bibinfo {author} {\bibfnamefont {J~R}\ \bibnamefont {Carey}}, \bibinfo
  {author} {\bibfnamefont {J~L}\ \bibnamefont {Vitek}}, \ and\ \bibinfo
  {author} {\bibfnamefont {B}~\bibnamefont {He}}} (\bibinfo {year} {2013}),\
  \bibfield  {title} {\enquote {\bibinfo {title} {Neuromodulation for brain
  disorders: challenges and opportunities},}\ }\href@noop {} {\bibfield
  {journal} {\bibinfo  {journal} {IEEE Trans Biomed Eng}\ }\textbf {\bibinfo
  {volume} {60}}~(\bibinfo {number} {3}),\ \bibinfo {pages}
  {610--624}}\BibitemShut {NoStop}%
\bibitem [{\citenamefont {Kailath}(1980)}]{kailath1980linear}%
  \BibitemOpen
  \bibfield  {author} {\bibinfo {author} {\bibnamefont {Kailath}, \bibfnamefont
  {T}}} (\bibinfo {year} {1980}),\ \href
  {https://books.google.com/books?id=ggYqAQAAMAAJ} {\emph {\bibinfo {title}
  {Linear Systems}}},\ Information and System Sciences Series\ (\bibinfo
  {publisher} {Prentice-Hall})\BibitemShut {NoStop}%
\bibitem [{\citenamefont {Kaiser}(2011)}]{kaiser2011tutorial}%
  \BibitemOpen
  \bibfield  {author} {\bibinfo {author} {\bibnamefont {Kaiser}, \bibfnamefont
  {M}}} (\bibinfo {year} {2011}),\ \bibfield  {title} {\enquote {\bibinfo
  {title} {A tutorial in connectome analysis: topological and spatial features
  of brain networks},}\ }\href@noop {} {\bibfield  {journal} {\bibinfo
  {journal} {Neuroimage}\ }\textbf {\bibinfo {volume} {57}}~(\bibinfo {number}
  {3}),\ \bibinfo {pages} {892--907}}\BibitemShut {NoStop}%
\bibitem [{\citenamefont {Kalman}\ \emph {et~al.}(1963)\citenamefont {Kalman},
  \citenamefont {Ho},\ and\ \citenamefont {Narendra}}]{REK-YCH-SKN:63}%
  \BibitemOpen
  \bibfield  {author} {\bibinfo {author} {\bibnamefont {Kalman}, \bibfnamefont
  {R~E}}, \bibinfo {author} {\bibfnamefont {Y.~C.}\ \bibnamefont {Ho}}, \ and\
  \bibinfo {author} {\bibfnamefont {K.S.}\ \bibnamefont {Narendra}}} (\bibinfo
  {year} {1963}),\ \bibfield  {title} {\enquote {\bibinfo {title}
  {Controllability of linear dynamical systems},}\ }\href@noop {} {\bibfield
  {journal} {\bibinfo  {journal} {Contributions to Differential Equations}\
  }\textbf {\bibinfo {volume} {1}}~(\bibinfo {number} {2}),\ \bibinfo {pages}
  {189--213}}\BibitemShut {NoStop}%
\bibitem [{\citenamefont {Kedzior}\ \emph {et~al.}(2016)\citenamefont
  {Kedzior}, \citenamefont {Gierke}, \citenamefont {Gellersen},\ and\
  \citenamefont {Berlim}}]{kedzior2016cognitive}%
  \BibitemOpen
  \bibfield  {author} {\bibinfo {author} {\bibnamefont {Kedzior}, \bibfnamefont
  {K~K}}, \bibinfo {author} {\bibfnamefont {L}~\bibnamefont {Gierke}}, \bibinfo
  {author} {\bibfnamefont {H~M}\ \bibnamefont {Gellersen}}, \ and\ \bibinfo
  {author} {\bibfnamefont {M~T}\ \bibnamefont {Berlim}}} (\bibinfo {year}
  {2016}),\ \bibfield  {title} {\enquote {\bibinfo {title} {Cognitive
  functioning and deep transcranial magnetic stimulation (dtms) in major
  psychiatric disorders: A systematic review},}\ }\href@noop {} {\bibfield
  {journal} {\bibinfo  {journal} {J Psychiatr Res}\ }\textbf {\bibinfo {volume}
  {75}},\ \bibinfo {pages} {107--115}}\BibitemShut {NoStop}%
\bibitem [{\citenamefont {Khambhati}\ \emph {et~al.}(2015)\citenamefont
  {Khambhati}, \citenamefont {Davis}, \citenamefont {Oommen}, \citenamefont
  {Chen}, \citenamefont {Lucas}, \citenamefont {Litt},\ and\ \citenamefont
  {Bassett}}]{khambhati2015dynamic}%
  \BibitemOpen
  \bibfield  {author} {\bibinfo {author} {\bibnamefont {Khambhati},
  \bibfnamefont {A~N}}, \bibinfo {author} {\bibfnamefont {K~A}\ \bibnamefont
  {Davis}}, \bibinfo {author} {\bibfnamefont {B~S}\ \bibnamefont {Oommen}},
  \bibinfo {author} {\bibfnamefont {S~H}\ \bibnamefont {Chen}}, \bibinfo
  {author} {\bibfnamefont {T~H}\ \bibnamefont {Lucas}}, \bibinfo {author}
  {\bibfnamefont {B}~\bibnamefont {Litt}}, \ and\ \bibinfo {author}
  {\bibfnamefont {D~S}\ \bibnamefont {Bassett}}} (\bibinfo {year} {2015}),\
  \bibfield  {title} {\enquote {\bibinfo {title} {Dynamic network drivers of
  seizure generation, propagation and termination in human neocortical
  epilepsy},}\ }\href@noop {} {\bibfield  {journal} {\bibinfo  {journal} {PLoS
  Comput Biol}\ }\textbf {\bibinfo {volume} {11}}~(\bibinfo {number} {12}),\
  \bibinfo {pages} {e1004608}}\BibitemShut {NoStop}%
\bibitem [{\citenamefont {Khambhati}\ \emph {et~al.}(2016)\citenamefont
  {Khambhati}, \citenamefont {Davis}, \citenamefont {Lucas}, \citenamefont
  {Litt},\ and\ \citenamefont {Bassett}}]{Khambhati20161170}%
  \BibitemOpen
  \bibfield  {author} {\bibinfo {author} {\bibnamefont {Khambhati},
  \bibfnamefont {Ankit~N}}, \bibinfo {author} {\bibfnamefont {Kathryn~A.}\
  \bibnamefont {Davis}}, \bibinfo {author} {\bibfnamefont {Timothy~H.}\
  \bibnamefont {Lucas}}, \bibinfo {author} {\bibfnamefont {Brian}\ \bibnamefont
  {Litt}}, \ and\ \bibinfo {author} {\bibfnamefont {Danielle~S.}\ \bibnamefont
  {Bassett}}} (\bibinfo {year} {2016}),\ \bibfield  {title} {\enquote {\bibinfo
  {title} {Virtual cortical resection reveals push-pull network control
  preceding seizure evolution},}\ }\href {\doibase
  http://dx.doi.org/10.1016/j.neuron.2016.07.039} {\bibfield  {journal}
  {\bibinfo  {journal} {Neuron}\ }\textbf {\bibinfo {volume} {91}}~(\bibinfo
  {number} {5}),\ \bibinfo {pages} {1170 -- 1182}}\BibitemShut {NoStop}%
\bibitem [{\citenamefont {Kim}\ \emph {et~al.}(2018)\citenamefont {Kim},
  \citenamefont {Soffer}, \citenamefont {Kahn}, \citenamefont {Vettel},
  \citenamefont {Pasqualetti},\ and\ \citenamefont {Bassett}}]{kim2018role}%
  \BibitemOpen
  \bibfield  {author} {\bibinfo {author} {\bibnamefont {Kim}, \bibfnamefont
  {J~Z}}, \bibinfo {author} {\bibfnamefont {J~M}\ \bibnamefont {Soffer}},
  \bibinfo {author} {\bibfnamefont {A~E}\ \bibnamefont {Kahn}}, \bibinfo
  {author} {\bibfnamefont {J~M}\ \bibnamefont {Vettel}}, \bibinfo {author}
  {\bibfnamefont {F}~\bibnamefont {Pasqualetti}}, \ and\ \bibinfo {author}
  {\bibfnamefont {D~S}\ \bibnamefont {Bassett}}} (\bibinfo {year} {2018}),\
  \bibfield  {title} {\enquote {\bibinfo {title} {Role of graph architecture in
  controlling dynamical networks with applications to neural systems},}\
  }\href@noop {} {\bibfield  {journal} {\bibinfo  {journal} {Nature Physics}\
  }\textbf {\bibinfo {volume} {14}},\ \bibinfo {pages} {91--98}}\BibitemShut
  {NoStop}%
\bibitem [{\citenamefont {Kopell}\ \emph {et~al.}(2000)\citenamefont {Kopell},
  \citenamefont {Ermentrout}, \citenamefont {Whittington},\ and\ \citenamefont
  {Traub}}]{kopell2000gamma}%
  \BibitemOpen
  \bibfield  {author} {\bibinfo {author} {\bibnamefont {Kopell}, \bibfnamefont
  {N}}, \bibinfo {author} {\bibfnamefont {G~B}\ \bibnamefont {Ermentrout}},
  \bibinfo {author} {\bibfnamefont {M~A}\ \bibnamefont {Whittington}}, \ and\
  \bibinfo {author} {\bibfnamefont {R~D}\ \bibnamefont {Traub}}} (\bibinfo
  {year} {2000}),\ \bibfield  {title} {\enquote {\bibinfo {title} {Gamma
  rhythms and beta rhythms have different synchronization properties},}\
  }\href@noop {} {\bibfield  {journal} {\bibinfo  {journal} {Proc Natl Acad Sci
  U S A}\ }\textbf {\bibinfo {volume} {97}}~(\bibinfo {number} {4}),\ \bibinfo
  {pages} {1867--1872}}\BibitemShut {NoStop}%
\bibitem [{\citenamefont {Kopell}\ \emph {et~al.}(2014)\citenamefont {Kopell},
  \citenamefont {Gritton}, \citenamefont {Whittington},\ and\ \citenamefont
  {Kramer}}]{kopell2014beyond}%
  \BibitemOpen
  \bibfield  {author} {\bibinfo {author} {\bibnamefont {Kopell}, \bibfnamefont
  {N~J}}, \bibinfo {author} {\bibfnamefont {H~J}\ \bibnamefont {Gritton}},
  \bibinfo {author} {\bibfnamefont {M~A}\ \bibnamefont {Whittington}}, \ and\
  \bibinfo {author} {\bibfnamefont {M~A}\ \bibnamefont {Kramer}}} (\bibinfo
  {year} {2014}),\ \bibfield  {title} {\enquote {\bibinfo {title} {Beyond the
  connectome: the dynome},}\ }\href@noop {} {\bibfield  {journal} {\bibinfo
  {journal} {Neuron}\ }\textbf {\bibinfo {volume} {83}}~(\bibinfo {number}
  {6}),\ \bibinfo {pages} {1319--1328}}\BibitemShut {NoStop}%
\bibitem [{\citenamefont {Lee}\ \emph {et~al.}(2010)\citenamefont {Lee},
  \citenamefont {Durand}, \citenamefont {Gradinaru}, \citenamefont {Zhang},
  \citenamefont {Goshen}, \citenamefont {Kim}, \citenamefont {Fenno},
  \citenamefont {Ramakrishnan},\ and\ \citenamefont
  {Deisseroth}}]{lee2010global}%
  \BibitemOpen
  \bibfield  {author} {\bibinfo {author} {\bibnamefont {Lee}, \bibfnamefont
  {J~H}}, \bibinfo {author} {\bibfnamefont {R}~\bibnamefont {Durand}}, \bibinfo
  {author} {\bibfnamefont {V}~\bibnamefont {Gradinaru}}, \bibinfo {author}
  {\bibfnamefont {F}~\bibnamefont {Zhang}}, \bibinfo {author} {\bibfnamefont
  {I}~\bibnamefont {Goshen}}, \bibinfo {author} {\bibfnamefont {D~S}\
  \bibnamefont {Kim}}, \bibinfo {author} {\bibfnamefont {L~E}\ \bibnamefont
  {Fenno}}, \bibinfo {author} {\bibfnamefont {C}~\bibnamefont {Ramakrishnan}},
  \ and\ \bibinfo {author} {\bibfnamefont {K}~\bibnamefont {Deisseroth}}}
  (\bibinfo {year} {2010}),\ \bibfield  {title} {\enquote {\bibinfo {title}
  {Global and local {fMRI} signals driven by neurons defined optogenetically by
  type and wiring},}\ }\href@noop {} {\bibfield  {journal} {\bibinfo  {journal}
  {Nature}\ }\textbf {\bibinfo {volume} {465}}~(\bibinfo {number} {7299}),\
  \bibinfo {pages} {788--792}}\BibitemShut {NoStop}%
\bibitem [{\citenamefont {Li}\ \emph {et~al.}(2016)\citenamefont {Li},
  \citenamefont {Cornelius}, \citenamefont {Liu}, \citenamefont {Wang},\ and\
  \citenamefont {Barab{\'a}si}}]{1607.06168}%
  \BibitemOpen
  \bibfield  {author} {\bibinfo {author} {\bibnamefont {Li}, \bibfnamefont
  {Aming}}, \bibinfo {author} {\bibfnamefont {Sean~P.}\ \bibnamefont
  {Cornelius}}, \bibinfo {author} {\bibfnamefont {Yang-Yu}\ \bibnamefont
  {Liu}}, \bibinfo {author} {\bibfnamefont {Long}\ \bibnamefont {Wang}}, \ and\
  \bibinfo {author} {\bibfnamefont {Albert-L{\'a}szl{\'o}}\ \bibnamefont
  {Barab{\'a}si}}} (\bibinfo {year} {2016}),\ \href@noop {} {\enquote {\bibinfo
  {title} {The fundamental advantages of temporal networks},}\ }\Eprint
  {http://arxiv.org/abs/arXiv:1607.06168} {arXiv:1607.06168} \BibitemShut
  {NoStop}%
\bibitem [{\citenamefont {Liu}\ \emph {et~al.}(2011)\citenamefont {Liu},
  \citenamefont {Slotine},\ and\ \citenamefont
  {Barab{\'a}si}}]{YYL-JJS-ALB:11}%
  \BibitemOpen
  \bibfield  {author} {\bibinfo {author} {\bibnamefont {Liu}, \bibfnamefont
  {Y~Y}}, \bibinfo {author} {\bibfnamefont {J.~J.}\ \bibnamefont {Slotine}}, \
  and\ \bibinfo {author} {\bibfnamefont {A.~L.}\ \bibnamefont {Barab{\'a}si}}}
  (\bibinfo {year} {2011}),\ \bibfield  {title} {\enquote {\bibinfo {title}
  {Controllability of complex networks},}\ }\href@noop {} {\bibfield  {journal}
  {\bibinfo  {journal} {Nature}\ }\textbf {\bibinfo {volume} {473}}~(\bibinfo
  {number} {7346}),\ \bibinfo {pages} {167--173}}\BibitemShut {NoStop}%
\bibitem [{\citenamefont {Liu}\ and\ \citenamefont
  {Barab\'asi}(2016)}]{RevModPhys.88.035006}%
  \BibitemOpen
  \bibfield  {author} {\bibinfo {author} {\bibnamefont {Liu}, \bibfnamefont
  {Yang-Yu}}, \ and\ \bibinfo {author} {\bibfnamefont {Albert-L\'aszl\'o}\
  \bibnamefont {Barab\'asi}}} (\bibinfo {year} {2016}),\ \bibfield  {title}
  {\enquote {\bibinfo {title} {Control principles of complex systems},}\ }\href
  {\doibase 10.1103/RevModPhys.88.035006} {\bibfield  {journal} {\bibinfo
  {journal} {Rev. Mod. Phys.}\ }\textbf {\bibinfo {volume} {88}},\ \bibinfo
  {pages} {035006}}\BibitemShut {NoStop}%
\bibitem [{\citenamefont {Makris}\ \emph {et~al.}(1997)\citenamefont {Makris},
  \citenamefont {Worth}, \citenamefont {Sorensen}, \citenamefont
  {Papadimitriou}, \citenamefont {Wu}, \citenamefont {Reese}, \citenamefont
  {Wedeen}, \citenamefont {Davis}, \citenamefont {Stakes}, \citenamefont
  {Caviness}, \citenamefont {Kaplan}, \citenamefont {Rosen}, \citenamefont
  {Pandya},\ and\ \citenamefont {Kennedy}}]{makris1997morphometry}%
  \BibitemOpen
  \bibfield  {author} {\bibinfo {author} {\bibnamefont {Makris}, \bibfnamefont
  {N}}, \bibinfo {author} {\bibfnamefont {A~J}\ \bibnamefont {Worth}}, \bibinfo
  {author} {\bibfnamefont {A~G}\ \bibnamefont {Sorensen}}, \bibinfo {author}
  {\bibfnamefont {G~M}\ \bibnamefont {Papadimitriou}}, \bibinfo {author}
  {\bibfnamefont {O}~\bibnamefont {Wu}}, \bibinfo {author} {\bibfnamefont
  {T~G}\ \bibnamefont {Reese}}, \bibinfo {author} {\bibfnamefont {V~J}\
  \bibnamefont {Wedeen}}, \bibinfo {author} {\bibfnamefont {T~L}\ \bibnamefont
  {Davis}}, \bibinfo {author} {\bibfnamefont {J~W}\ \bibnamefont {Stakes}},
  \bibinfo {author} {\bibfnamefont {V~S}\ \bibnamefont {Caviness}}, \bibinfo
  {author} {\bibfnamefont {E}~\bibnamefont {Kaplan}}, \bibinfo {author}
  {\bibfnamefont {B~R}\ \bibnamefont {Rosen}}, \bibinfo {author} {\bibfnamefont
  {D~N}\ \bibnamefont {Pandya}}, \ and\ \bibinfo {author} {\bibfnamefont {D~N}\
  \bibnamefont {Kennedy}}} (\bibinfo {year} {1997}),\ \bibfield  {title}
  {\enquote {\bibinfo {title} {Morphometry of in vivo human white matter
  association pathways with diffusion-weighted magnetic resonance imaging},}\
  }\href@noop {} {\bibfield  {journal} {\bibinfo  {journal} {Ann Neurol}\
  }\textbf {\bibinfo {volume} {42}}~(\bibinfo {number} {6}),\ \bibinfo {pages}
  {951--962}}\BibitemShut {NoStop}%
\bibitem [{\citenamefont {Marblestone}\ \emph {et~al.}(2016)\citenamefont
  {Marblestone}, \citenamefont {Wayne},\ and\ \citenamefont
  {Kording}}]{marblestone2016toward}%
  \BibitemOpen
  \bibfield  {author} {\bibinfo {author} {\bibnamefont {Marblestone},
  \bibfnamefont {A~H}}, \bibinfo {author} {\bibfnamefont {G}~\bibnamefont
  {Wayne}}, \ and\ \bibinfo {author} {\bibfnamefont {K~P}\ \bibnamefont
  {Kording}}} (\bibinfo {year} {2016}),\ \bibfield  {title} {\enquote {\bibinfo
  {title} {Toward an integration of deep learning and neuroscience},}\
  }\href@noop {} {\bibfield  {journal} {\bibinfo  {journal} {Front Comput
  Neurosci}\ }\textbf {\bibinfo {volume} {10}},\ \bibinfo {pages}
  {94}}\BibitemShut {NoStop}%
\bibitem [{\citenamefont {Markov}\ \emph {et~al.}(2011)\citenamefont {Markov},
  \citenamefont {Misery}, \citenamefont {Falchier}, \citenamefont {Lamy},
  \citenamefont {Vezoli}, \citenamefont {Quilodran}, \citenamefont {Gariel},
  \citenamefont {Giroud}, \citenamefont {Ercsey-Ravasz}, \citenamefont {Pilaz},
  \citenamefont {Huissoud}, \citenamefont {Barone}, \citenamefont {Dehay},
  \citenamefont {Toroczkai}, \citenamefont {Van~Essen}, \citenamefont
  {Kennedy},\ and\ \citenamefont {Knoblauch}}]{markov2011weight}%
  \BibitemOpen
  \bibfield  {author} {\bibinfo {author} {\bibnamefont {Markov}, \bibfnamefont
  {N~T}}, \bibinfo {author} {\bibfnamefont {P}~\bibnamefont {Misery}}, \bibinfo
  {author} {\bibfnamefont {A}~\bibnamefont {Falchier}}, \bibinfo {author}
  {\bibfnamefont {C}~\bibnamefont {Lamy}}, \bibinfo {author} {\bibfnamefont
  {J}~\bibnamefont {Vezoli}}, \bibinfo {author} {\bibfnamefont {R}~\bibnamefont
  {Quilodran}}, \bibinfo {author} {\bibfnamefont {M~A}\ \bibnamefont {Gariel}},
  \bibinfo {author} {\bibfnamefont {P}~\bibnamefont {Giroud}}, \bibinfo
  {author} {\bibfnamefont {M}~\bibnamefont {Ercsey-Ravasz}}, \bibinfo {author}
  {\bibfnamefont {L~J}\ \bibnamefont {Pilaz}}, \bibinfo {author} {\bibfnamefont
  {C}~\bibnamefont {Huissoud}}, \bibinfo {author} {\bibfnamefont
  {P}~\bibnamefont {Barone}}, \bibinfo {author} {\bibfnamefont {C}~\bibnamefont
  {Dehay}}, \bibinfo {author} {\bibfnamefont {Z}~\bibnamefont {Toroczkai}},
  \bibinfo {author} {\bibfnamefont {D~C}\ \bibnamefont {Van~Essen}}, \bibinfo
  {author} {\bibfnamefont {H}~\bibnamefont {Kennedy}}, \ and\ \bibinfo {author}
  {\bibfnamefont {K}~\bibnamefont {Knoblauch}}} (\bibinfo {year} {2011}),\
  \bibfield  {title} {\enquote {\bibinfo {title} {Weight consistency specifies
  regularities of macaque cortical networks},}\ }\href@noop {} {\bibfield
  {journal} {\bibinfo  {journal} {Cereb Cortex}\ }\textbf {\bibinfo {volume}
  {21}}~(\bibinfo {number} {6}),\ \bibinfo {pages} {1254--1272}}\BibitemShut
  {NoStop}%
\bibitem [{\citenamefont {Marx}\ \emph {et~al.}(2004)\citenamefont {Marx},
  \citenamefont {Koenig},\ and\ \citenamefont {Georges}}]{BM-DK-DG:04}%
  \BibitemOpen
  \bibfield  {author} {\bibinfo {author} {\bibnamefont {Marx}, \bibfnamefont
  {B}}, \bibinfo {author} {\bibfnamefont {D.}~\bibnamefont {Koenig}}, \ and\
  \bibinfo {author} {\bibfnamefont {D.}~\bibnamefont {Georges}}} (\bibinfo
  {year} {2004}),\ \bibfield  {title} {\enquote {\bibinfo {title} {Optimal
  sensor and actuator location for descriptor systems using generalized
  {G}ramians and balanced realizations},}\ }in\ \href@noop {} {\emph {\bibinfo
  {booktitle} {{A}merican {C}ontrol {C}onference}}}\ (\bibinfo {address}
  {Boston, MA, USA})\ pp.\ \bibinfo {pages} {2729--2734}\BibitemShut {NoStop}%
\bibitem [{\citenamefont {McCulloch}\ and\ \citenamefont
  {Pitts}(1943)}]{mcculloch1943logical}%
  \BibitemOpen
  \bibfield  {author} {\bibinfo {author} {\bibnamefont {McCulloch},
  \bibfnamefont {W}}, \ and\ \bibinfo {author} {\bibfnamefont {W}~\bibnamefont
  {Pitts}}} (\bibinfo {year} {1943}),\ \bibfield  {title} {\enquote {\bibinfo
  {title} {A logical calculus of ideas immanent in nervous activity},}\
  }\href@noop {} {\bibfield  {journal} {\bibinfo  {journal} {Bulletin of
  Mathematical Biophysics}\ }\textbf {\bibinfo {volume} {5}}~(\bibinfo {number}
  {4}),\ \bibinfo {pages} {115--133}}\BibitemShut {NoStop}%
\bibitem [{\citenamefont {Medaglia}\ \emph {et~al.}(2016)\citenamefont
  {Medaglia}, \citenamefont {Gu}, \citenamefont {Pasqualetti}, \citenamefont
  {Ashare}, \citenamefont {Lerman}, \citenamefont {Kable},\ and\ \citenamefont
  {Bassett}}]{1606.09185}%
  \BibitemOpen
  \bibfield  {author} {\bibinfo {author} {\bibnamefont {Medaglia},
  \bibfnamefont {John~D}}, \bibinfo {author} {\bibfnamefont {Shi}\ \bibnamefont
  {Gu}}, \bibinfo {author} {\bibfnamefont {Fabio}\ \bibnamefont {Pasqualetti}},
  \bibinfo {author} {\bibfnamefont {Rebecca~L.}\ \bibnamefont {Ashare}},
  \bibinfo {author} {\bibfnamefont {Caryn}\ \bibnamefont {Lerman}}, \bibinfo
  {author} {\bibfnamefont {Joseph}\ \bibnamefont {Kable}}, \ and\ \bibinfo
  {author} {\bibfnamefont {Danielle~S.}\ \bibnamefont {Bassett}}} (\bibinfo
  {year} {2016}),\ \href@noop {} {\enquote {\bibinfo {title} {Cognitive control
  in the controllable connectome},}\ }\Eprint
  {http://arxiv.org/abs/arXiv:1606.09185} {arXiv:1606.09185} \BibitemShut
  {NoStop}%
\bibitem [{\citenamefont {Medler}(1998)}]{medler1998brief}%
  \BibitemOpen
  \bibfield  {author} {\bibinfo {author} {\bibnamefont {Medler}, \bibfnamefont
  {D~A}}} (\bibinfo {year} {1998}),\ \bibfield  {title} {\enquote {\bibinfo
  {title} {A brief history of connectionism},}\ }\href@noop {} {\bibfield
  {journal} {\bibinfo  {journal} {Neural Computing Surveys}\ }\textbf {\bibinfo
  {volume} {1}},\ \bibinfo {pages} {61--101}}\BibitemShut {NoStop}%
\bibitem [{\citenamefont {Menara}\ \emph {et~al.}(2017)\citenamefont {Menara},
  \citenamefont {Gu}, \citenamefont {Bassett},\ and\ \citenamefont
  {Pasqualetti}}]{menara2017structural}%
  \BibitemOpen
  \bibfield  {author} {\bibinfo {author} {\bibnamefont {Menara}, \bibfnamefont
  {T}}, \bibinfo {author} {\bibfnamefont {S}~\bibnamefont {Gu}}, \bibinfo
  {author} {\bibfnamefont {D~S}\ \bibnamefont {Bassett}}, \ and\ \bibinfo
  {author} {\bibfnamefont {F}~\bibnamefont {Pasqualetti}}} (\bibinfo {year}
  {2017}),\ \bibfield  {title} {\enquote {\bibinfo {title} {On structural
  controllability of symmetric (brain) networks},}\ }\href@noop {} {\bibfield
  {journal} {\bibinfo  {journal} {arXiv}\ }\textbf {\bibinfo {volume} {1706}},\
  \bibinfo {pages} {05120}}\BibitemShut {NoStop}%
\bibitem [{\citenamefont {Menck}\ \emph {et~al.}(2013)\citenamefont {Menck},
  \citenamefont {Heitzig}, \citenamefont {Marwan},\ and\ \citenamefont
  {Kurths}}]{Menck2013}%
  \BibitemOpen
  \bibfield  {author} {\bibinfo {author} {\bibnamefont {Menck}, \bibfnamefont
  {Peter~J}}, \bibinfo {author} {\bibfnamefont {Jobst}\ \bibnamefont
  {Heitzig}}, \bibinfo {author} {\bibfnamefont {Norbert}\ \bibnamefont
  {Marwan}}, \ and\ \bibinfo {author} {\bibfnamefont {J{\"u}rgen}\ \bibnamefont
  {Kurths}}} (\bibinfo {year} {2013}),\ \bibfield  {title} {\enquote {\bibinfo
  {title} {How basin stability complements the linear-stability paradigm},}\
  }\href {http://dx.doi.org/10.1038/nphys2516} {\bibfield  {journal} {\bibinfo
  {journal} {Nature Physics}\ }\textbf {\bibinfo {volume} {9}},\ \bibinfo
  {pages} {89}}\BibitemShut {NoStop}%
\bibitem [{\citenamefont {Morris}\ \emph {et~al.}(2008)\citenamefont {Morris},
  \citenamefont {Embleton},\ and\ \citenamefont {Parker}}]{MORRIS20081329}%
  \BibitemOpen
  \bibfield  {author} {\bibinfo {author} {\bibnamefont {Morris}, \bibfnamefont
  {David~M}}, \bibinfo {author} {\bibfnamefont {Karl~V.}\ \bibnamefont
  {Embleton}}, \ and\ \bibinfo {author} {\bibfnamefont {Geoff~J.M.}\
  \bibnamefont {Parker}}} (\bibinfo {year} {2008}),\ \bibfield  {title}
  {\enquote {\bibinfo {title} {Probabilistic fibre tracking: Differentiation of
  connections from chance events},}\ }\href {\doibase
  https://doi.org/10.1016/j.neuroimage.2008.06.012} {\bibfield  {journal}
  {\bibinfo  {journal} {NeuroImage}\ }\textbf {\bibinfo {volume}
  {42}}~(\bibinfo {number} {4}),\ \bibinfo {pages} {1329 -- 1339}}\BibitemShut
  {NoStop}%
\bibitem [{\citenamefont {Muldoon}\ \emph {et~al.}(2016)\citenamefont
  {Muldoon}, \citenamefont {Pasqualetti}, \citenamefont {Gu}, \citenamefont
  {Cieslak}, \citenamefont {Grafton}, \citenamefont {Vettel},\ and\
  \citenamefont {Bassett}}]{muldoon2016stimulation}%
  \BibitemOpen
  \bibfield  {author} {\bibinfo {author} {\bibnamefont {Muldoon}, \bibfnamefont
  {S~F}}, \bibinfo {author} {\bibfnamefont {F}~\bibnamefont {Pasqualetti}},
  \bibinfo {author} {\bibfnamefont {S}~\bibnamefont {Gu}}, \bibinfo {author}
  {\bibfnamefont {M}~\bibnamefont {Cieslak}}, \bibinfo {author} {\bibfnamefont
  {S~T}\ \bibnamefont {Grafton}}, \bibinfo {author} {\bibfnamefont {J~M}\
  \bibnamefont {Vettel}}, \ and\ \bibinfo {author} {\bibfnamefont {D~S}\
  \bibnamefont {Bassett}}} (\bibinfo {year} {2016}),\ \bibfield  {title}
  {\enquote {\bibinfo {title} {Stimulation-based control of dynamic brain
  networks},}\ }\href@noop {} {\bibfield  {journal} {\bibinfo  {journal} {PLoS
  Comput Biol}\ }\textbf {\bibinfo {volume} {12}}~(\bibinfo {number} {9}),\
  \bibinfo {pages} {e1005076}}\BibitemShut {NoStop}%
\bibitem [{\citenamefont {Nabi}\ and\ \citenamefont
  {Moehlis}(2011)}]{1741-2552-8-6-065008}%
  \BibitemOpen
  \bibfield  {author} {\bibinfo {author} {\bibnamefont {Nabi}, \bibfnamefont
  {Ali}}, \ and\ \bibinfo {author} {\bibfnamefont {Jeff}\ \bibnamefont
  {Moehlis}}} (\bibinfo {year} {2011}),\ \bibfield  {title} {\enquote {\bibinfo
  {title} {Single input optimal control for globally coupled neuron
  networks},}\ }\href {http://stacks.iop.org/1741-2552/8/i=6/a=065008}
  {\bibfield  {journal} {\bibinfo  {journal} {Journal of Neural Engineering}\
  }\textbf {\bibinfo {volume} {8}}~(\bibinfo {number} {6}),\ \bibinfo {pages}
  {065008}}\BibitemShut {NoStop}%
\bibitem [{\citenamefont {Nag}\ and\ \citenamefont
  {Thakor}(2016)}]{nag2016implantable}%
  \BibitemOpen
  \bibfield  {author} {\bibinfo {author} {\bibnamefont {Nag}, \bibfnamefont
  {S}}, \ and\ \bibinfo {author} {\bibfnamefont {N~V}\ \bibnamefont {Thakor}}}
  (\bibinfo {year} {2016}),\ \bibfield  {title} {\enquote {\bibinfo {title}
  {Implantable neurotechnologies: electrical stimulation and applications},}\
  }\href@noop {} {\bibfield  {journal} {\bibinfo  {journal} {Med Biol Eng
  Comput}\ }\textbf {\bibinfo {volume} {54}}~(\bibinfo {number} {1}),\ \bibinfo
  {pages} {63--76}}\BibitemShut {NoStop}%
\bibitem [{\citenamefont {Nelson}\ and\ \citenamefont
  {Turrigiano}(2008)}]{nelson2008strength}%
  \BibitemOpen
  \bibfield  {author} {\bibinfo {author} {\bibnamefont {Nelson}, \bibfnamefont
  {S~B}}, \ and\ \bibinfo {author} {\bibfnamefont {G~G}\ \bibnamefont
  {Turrigiano}}} (\bibinfo {year} {2008}),\ \bibfield  {title} {\enquote
  {\bibinfo {title} {Strength through diversity},}\ }\href@noop {} {\bibfield
  {journal} {\bibinfo  {journal} {Neuron}\ }\textbf {\bibinfo {volume}
  {60}}~(\bibinfo {number} {3}),\ \bibinfo {pages} {477--482}}\BibitemShut
  {NoStop}%
\bibitem [{\citenamefont {Nelson}\ and\ \citenamefont
  {Valakh}(2015)}]{nelson2015excitatory}%
  \BibitemOpen
  \bibfield  {author} {\bibinfo {author} {\bibnamefont {Nelson}, \bibfnamefont
  {S~B}}, \ and\ \bibinfo {author} {\bibfnamefont {V}~\bibnamefont {Valakh}}}
  (\bibinfo {year} {2015}),\ \bibfield  {title} {\enquote {\bibinfo {title}
  {Excitatory/inhibitory balance and circuit homeostasis in autism spectrum
  disorders},}\ }\href@noop {} {\bibfield  {journal} {\bibinfo  {journal}
  {Neuron}\ }\textbf {\bibinfo {volume} {87}}~(\bibinfo {number} {4}),\
  \bibinfo {pages} {684--698}}\BibitemShut {NoStop}%
\bibitem [{\citenamefont {Newman}(2010)}]{newman2010networks}%
  \BibitemOpen
  \bibfield  {author} {\bibinfo {author} {\bibnamefont {Newman}, \bibfnamefont
  {M~E~J}}} (\bibinfo {year} {2010}),\ \href@noop {} {\emph {\bibinfo {title}
  {Networks: An Introduction}}}\ (\bibinfo  {publisher} {Oxford University
  Press})\BibitemShut {NoStop}%
\bibitem [{\citenamefont {Nishikawa}\ and\ \citenamefont
  {Motter}(2010)}]{Nishikawa08062010}%
  \BibitemOpen
  \bibfield  {author} {\bibinfo {author} {\bibnamefont {Nishikawa},
  \bibfnamefont {Takashi}}, \ and\ \bibinfo {author} {\bibfnamefont
  {Adilson~E.}\ \bibnamefont {Motter}}} (\bibinfo {year} {2010}),\ \bibfield
  {title} {\enquote {\bibinfo {title} {Network synchronization landscape
  reveals compensatory structures, quantization, and the positive effect of
  negative interactions},}\ }\href {\doibase 10.1073/pnas.0912444107}
  {\bibfield  {journal} {\bibinfo  {journal} {Proceedings of the National
  Academy of Sciences}\ }\textbf {\bibinfo {volume} {107}}~(\bibinfo {number}
  {23}),\ \bibinfo {pages} {10342--10347}}\BibitemShut {NoStop}%
\bibitem [{\citenamefont {Okano}\ and\ \citenamefont
  {Mitra}(2015)}]{okano2015brain}%
  \BibitemOpen
  \bibfield  {author} {\bibinfo {author} {\bibnamefont {Okano}, \bibfnamefont
  {H}}, \ and\ \bibinfo {author} {\bibfnamefont {P}~\bibnamefont {Mitra}}}
  (\bibinfo {year} {2015}),\ \bibfield  {title} {\enquote {\bibinfo {title}
  {Brain-mapping projects using the common marmoset},}\ }\href@noop {}
  {\bibfield  {journal} {\bibinfo  {journal} {Neurosci Res}\ }\textbf {\bibinfo
  {volume} {93}},\ \bibinfo {pages} {3--7}}\BibitemShut {NoStop}%
\bibitem [{\citenamefont {Pasqualetti}\ \emph {et~al.}(2014)\citenamefont
  {Pasqualetti}, \citenamefont {Zampieri},\ and\ \citenamefont
  {Bullo}}]{pasqualetti2014controllability}%
  \BibitemOpen
  \bibfield  {author} {\bibinfo {author} {\bibnamefont {Pasqualetti},
  \bibfnamefont {Fabio}}, \bibinfo {author} {\bibfnamefont {Sandro}\
  \bibnamefont {Zampieri}}, \ and\ \bibinfo {author} {\bibfnamefont
  {Francesco}\ \bibnamefont {Bullo}}} (\bibinfo {year} {2014}),\ \bibfield
  {title} {\enquote {\bibinfo {title} {Controllability metrics, limitations and
  algorithms for complex networks},}\ }\href@noop {} {\bibfield  {journal}
  {\bibinfo  {journal} {Control of Network Systems, IEEE Transactions on}\
  }\textbf {\bibinfo {volume} {1}}~(\bibinfo {number} {1}),\ \bibinfo {pages}
  {40--52}}\BibitemShut {NoStop}%
\bibitem [{\citenamefont {Patil}\ and\ \citenamefont
  {Thakor}(2016)}]{patil2016implantable}%
  \BibitemOpen
  \bibfield  {author} {\bibinfo {author} {\bibnamefont {Patil}, \bibfnamefont
  {A~C}}, \ and\ \bibinfo {author} {\bibfnamefont {N~V}\ \bibnamefont
  {Thakor}}} (\bibinfo {year} {2016}),\ \bibfield  {title} {\enquote {\bibinfo
  {title} {Implantable neurotechnologies: a review of micro- and nanoelectrodes
  for neural recording},}\ }\href@noop {} {\bibfield  {journal} {\bibinfo
  {journal} {Med Biol Eng Comput}\ }\textbf {\bibinfo {volume} {54}}~(\bibinfo
  {number} {1}),\ \bibinfo {pages} {23--44}}\BibitemShut {NoStop}%
\bibitem [{\citenamefont {Pecora}\ and\ \citenamefont
  {Carroll}(1998)}]{PhysRevLett.80.2109}%
  \BibitemOpen
  \bibfield  {author} {\bibinfo {author} {\bibnamefont {Pecora}, \bibfnamefont
  {Louis~M}}, \ and\ \bibinfo {author} {\bibfnamefont {Thomas~L.}\ \bibnamefont
  {Carroll}}} (\bibinfo {year} {1998}),\ \bibfield  {title} {\enquote {\bibinfo
  {title} {Master stability functions for synchronized coupled systems},}\
  }\href {\doibase 10.1103/PhysRevLett.80.2109} {\bibfield  {journal} {\bibinfo
   {journal} {Phys. Rev. Lett.}\ }\textbf {\bibinfo {volume} {80}},\ \bibinfo
  {pages} {2109--2112}}\BibitemShut {NoStop}%
\bibitem [{\citenamefont {Pequito}\ \emph
  {et~al.}(2016{\natexlab{a}})\citenamefont {Pequito}, \citenamefont {Kar},\
  and\ \citenamefont {Aguiar}}]{PequitoJ1}%
  \BibitemOpen
  \bibfield  {author} {\bibinfo {author} {\bibnamefont {Pequito}, \bibfnamefont
  {S}}, \bibinfo {author} {\bibfnamefont {S.}~\bibnamefont {Kar}}, \ and\
  \bibinfo {author} {\bibfnamefont {A.~P.}\ \bibnamefont {Aguiar}}} (\bibinfo
  {year} {2016}{\natexlab{a}}),\ \bibfield  {title} {\enquote {\bibinfo {title}
  {A framework for structural input/output and control configuration selection
  of large-scale systems},}\ }\href {\doibase 10.1109/TAC.2015.2437525}
  {\bibfield  {journal} {\bibinfo  {journal} {IEEE Transactions on Automatic
  Control}\ }\textbf {\bibinfo {volume} {61}}~(\bibinfo {number} {2}),\
  \bibinfo {pages} {303--318}}\BibitemShut {NoStop}%
\bibitem [{\citenamefont {Pequito}\ \emph
  {et~al.}(2016{\natexlab{b}})\citenamefont {Pequito}, \citenamefont
  {Khambhati}, \citenamefont {Pappas}, \citenamefont {šiljak}, \citenamefont
  {Bassett},\ and\ \citenamefont {Litt}}]{7526572}%
  \BibitemOpen
  \bibfield  {author} {\bibinfo {author} {\bibnamefont {Pequito}, \bibfnamefont
  {S}}, \bibinfo {author} {\bibfnamefont {A.~N.}\ \bibnamefont {Khambhati}},
  \bibinfo {author} {\bibfnamefont {G.~J.}\ \bibnamefont {Pappas}}, \bibinfo
  {author} {\bibfnamefont {D.~D.}\ \bibnamefont {šiljak}}, \bibinfo {author}
  {\bibfnamefont {D.}~\bibnamefont {Bassett}}, \ and\ \bibinfo {author}
  {\bibfnamefont {B.}~\bibnamefont {Litt}}} (\bibinfo {year}
  {2016}{\natexlab{b}}),\ \bibfield  {title} {\enquote {\bibinfo {title}
  {Structural analysis and design of dynamic-flow networks: Implications in the
  brain dynamics},}\ }in\ \href {\doibase 10.1109/ACC.2016.7526572} {\emph
  {\bibinfo {booktitle} {2016 American Control Conference (ACC)}}},\ pp.\
  \bibinfo {pages} {5758--5764}\BibitemShut {NoStop}%
\bibitem [{\citenamefont {P{\'o}sfai}\ \emph {et~al.}(2013)\citenamefont
  {P{\'o}sfai}, \citenamefont {Liu}, \citenamefont {Slotine},\ and\
  \citenamefont {Barab{\'a}si}}]{Posfai2013}%
  \BibitemOpen
  \bibfield  {author} {\bibinfo {author} {\bibnamefont {P{\'o}sfai},
  \bibfnamefont {M{\'a}rton}}, \bibinfo {author} {\bibfnamefont {Yang-Yu}\
  \bibnamefont {Liu}}, \bibinfo {author} {\bibfnamefont {Jean-Jacques}\
  \bibnamefont {Slotine}}, \ and\ \bibinfo {author} {\bibfnamefont
  {Albert-L{\'a}szl{\'o}}\ \bibnamefont {Barab{\'a}si}}} (\bibinfo {year}
  {2013}),\ \bibfield  {title} {\enquote {\bibinfo {title} {Effect of
  correlations on network controllability},}\ }\href
  {http://dx.doi.org/10.1038/srep01067} {\bibfield  {journal} {\bibinfo
  {journal} {Scientific Reports}\ }\textbf {\bibinfo {volume} {3}},\ \bibinfo
  {pages} {1067}}\BibitemShut {NoStop}%
\bibitem [{\citenamefont {Power}\ \emph {et~al.}(2011)\citenamefont {Power},
  \citenamefont {Cohen}, \citenamefont {Nelson}, \citenamefont {Wig},
  \citenamefont {Barnes}, \citenamefont {Church}, \citenamefont {Vogel},
  \citenamefont {Laumann}, \citenamefont {Miezin}, \citenamefont {Schlaggar},\
  and\ \citenamefont {Petersen}}]{power2011functional}%
  \BibitemOpen
  \bibfield  {author} {\bibinfo {author} {\bibnamefont {Power}, \bibfnamefont
  {J~D}}, \bibinfo {author} {\bibfnamefont {A~L}\ \bibnamefont {Cohen}},
  \bibinfo {author} {\bibfnamefont {S~M}\ \bibnamefont {Nelson}}, \bibinfo
  {author} {\bibfnamefont {G~S}\ \bibnamefont {Wig}}, \bibinfo {author}
  {\bibfnamefont {K~A}\ \bibnamefont {Barnes}}, \bibinfo {author}
  {\bibfnamefont {J~A}\ \bibnamefont {Church}}, \bibinfo {author}
  {\bibfnamefont {A~C}\ \bibnamefont {Vogel}}, \bibinfo {author} {\bibfnamefont
  {T~O}\ \bibnamefont {Laumann}}, \bibinfo {author} {\bibfnamefont {F~M}\
  \bibnamefont {Miezin}}, \bibinfo {author} {\bibfnamefont {B~L}\ \bibnamefont
  {Schlaggar}}, \ and\ \bibinfo {author} {\bibfnamefont {S~E}\ \bibnamefont
  {Petersen}}} (\bibinfo {year} {2011}),\ \bibfield  {title} {\enquote
  {\bibinfo {title} {Functional network organization of the human brain},}\
  }\href@noop {} {\bibfield  {journal} {\bibinfo  {journal} {Neuron}\ }\textbf
  {\bibinfo {volume} {72}}~(\bibinfo {number} {4}),\ \bibinfo {pages}
  {665--678}}\BibitemShut {NoStop}%
\bibitem [{\citenamefont {Raichle}(2015)}]{raichle2015brains}%
  \BibitemOpen
  \bibfield  {author} {\bibinfo {author} {\bibnamefont {Raichle}, \bibfnamefont
  {M~E}}} (\bibinfo {year} {2015}),\ \bibfield  {title} {\enquote {\bibinfo
  {title} {The brain's default mode network},}\ }\href@noop {} {\bibfield
  {journal} {\bibinfo  {journal} {Annu Rev Neurosci}\ }\textbf {\bibinfo
  {volume} {38}},\ \bibinfo {pages} {433--447}}\BibitemShut {NoStop}%
\bibitem [{\citenamefont {Reinschke}(1988)}]{KJR:88}%
  \BibitemOpen
  \bibfield  {author} {\bibinfo {author} {\bibnamefont {Reinschke},
  \bibfnamefont {K~J}}} (\bibinfo {year} {1988}),\ \href@noop {} {\emph
  {\bibinfo {title} {Multivariable Control: A Graph-Theoretic Approach}}}\
  (\bibinfo  {publisher} {Springer})\BibitemShut {NoStop}%
\bibitem [{\citenamefont {Ritt}\ and\ \citenamefont {Ching}(2015)}]{7171915}%
  \BibitemOpen
  \bibfield  {author} {\bibinfo {author} {\bibnamefont {Ritt}, \bibfnamefont
  {J~T}}, \ and\ \bibinfo {author} {\bibfnamefont {S.}~\bibnamefont {Ching}}}
  (\bibinfo {year} {2015}),\ \bibfield  {title} {\enquote {\bibinfo {title}
  {Neurocontrol: Methods, models and technologies for manipulating dynamics in
  the brain},}\ }in\ \href {\doibase 10.1109/ACC.2015.7171915} {\emph {\bibinfo
  {booktitle} {2015 American Control Conference (ACC)}}},\ pp.\ \bibinfo
  {pages} {3765--3780}\BibitemShut {NoStop}%
\bibitem [{\citenamefont {Rosenblatt}(1958)}]{rosenblatt1958perceptron}%
  \BibitemOpen
  \bibfield  {author} {\bibinfo {author} {\bibnamefont {Rosenblatt},
  \bibfnamefont {F}}} (\bibinfo {year} {1958}),\ \bibfield  {title} {\enquote
  {\bibinfo {title} {The perceptron: A probabilistic model for information
  storage and organization in the brain},}\ }\href@noop {} {\bibfield
  {journal} {\bibinfo  {journal} {Psychological Review}\ }\textbf {\bibinfo
  {volume} {65}}~(\bibinfo {number} {6}),\ \bibinfo {pages}
  {386--408}}\BibitemShut {NoStop}%
\bibitem [{\citenamefont {Rumelhart}\ \emph {et~al.}(1986)\citenamefont
  {Rumelhart}, \citenamefont {McClelland},\ and\ \citenamefont {the PDP
  Research~Group}}]{rumelhart1986parallel}%
  \BibitemOpen
  \bibfield  {author} {\bibinfo {author} {\bibnamefont {Rumelhart},
  \bibfnamefont {D~E}}, \bibinfo {author} {\bibfnamefont {J~L}\ \bibnamefont
  {McClelland}}, \ and\ \bibinfo {author} {\bibnamefont {the PDP
  Research~Group}}} (\bibinfo {year} {1986}),\ \href@noop {} {\emph {\bibinfo
  {title} {Parallel Distributed Processing: Explorations in the Microstructure
  of Cognition}}},\ Vol.~\bibinfo {volume} {1}\ (\bibinfo  {publisher} {MIT
  Press})\BibitemShut {NoStop}%
\bibitem [{\citenamefont {Ruths}\ and\ \citenamefont
  {Ruths}(2014)}]{Ruths1373}%
  \BibitemOpen
  \bibfield  {author} {\bibinfo {author} {\bibnamefont {Ruths}, \bibfnamefont
  {Justin}}, \ and\ \bibinfo {author} {\bibfnamefont {Derek}\ \bibnamefont
  {Ruths}}} (\bibinfo {year} {2014}),\ \bibfield  {title} {\enquote {\bibinfo
  {title} {Control profiles of complex networks},}\ }\href {\doibase
  10.1126/science.1242063} {\bibfield  {journal} {\bibinfo  {journal}
  {Science}\ }\textbf {\bibinfo {volume} {343}}~(\bibinfo {number} {6177}),\
  \bibinfo {pages} {1373--1376}}\BibitemShut {NoStop}%
\bibitem [{\citenamefont {Sacré}\ \emph {et~al.}(2016)\citenamefont {Sacré},
  \citenamefont {Kerr}, \citenamefont {Subramanian}, \citenamefont {Kahn},
  \citenamefont {Gonzalez-Martinez}, \citenamefont {Johnson}, \citenamefont
  {Sarma},\ and\ \citenamefont {Gale}}]{7591459}%
  \BibitemOpen
  \bibfield  {author} {\bibinfo {author} {\bibnamefont {Sacré}, \bibfnamefont
  {P}}, \bibinfo {author} {\bibfnamefont {M.~S.~D.}\ \bibnamefont {Kerr}},
  \bibinfo {author} {\bibfnamefont {S.}~\bibnamefont {Subramanian}}, \bibinfo
  {author} {\bibfnamefont {K.}~\bibnamefont {Kahn}}, \bibinfo {author}
  {\bibfnamefont {J.}~\bibnamefont {Gonzalez-Martinez}}, \bibinfo {author}
  {\bibfnamefont {M.~A.}\ \bibnamefont {Johnson}}, \bibinfo {author}
  {\bibfnamefont {S.~V.}\ \bibnamefont {Sarma}}, \ and\ \bibinfo {author}
  {\bibfnamefont {J.~T.}\ \bibnamefont {Gale}}} (\bibinfo {year} {2016}),\
  \bibfield  {title} {\enquote {\bibinfo {title} {The precuneus may encode
  irrationality in human gambling},}\ }in\ \href {\doibase
  10.1109/EMBC.2016.7591459} {\emph {\bibinfo {booktitle} {2016 38th Annual
  International Conference of the IEEE Engineering in Medicine and Biology
  Society (EMBC)}}},\ pp.\ \bibinfo {pages} {3406--3409}\BibitemShut {NoStop}%
\bibitem [{\citenamefont {Santaniello}\ \emph {et~al.}(2015)\citenamefont
  {Santaniello}, \citenamefont {McCarthy}, \citenamefont {Montgomery},
  \citenamefont {Gale}, \citenamefont {Kopell},\ and\ \citenamefont
  {Sarma}}]{Santaniello10022015}%
  \BibitemOpen
  \bibfield  {author} {\bibinfo {author} {\bibnamefont {Santaniello},
  \bibfnamefont {Sabato}}, \bibinfo {author} {\bibfnamefont {Michelle~M.}\
  \bibnamefont {McCarthy}}, \bibinfo {author} {\bibfnamefont {Erwin~B.}\
  \bibnamefont {Montgomery}}, \bibinfo {author} {\bibfnamefont {John~T.}\
  \bibnamefont {Gale}}, \bibinfo {author} {\bibfnamefont {Nancy}\ \bibnamefont
  {Kopell}}, \ and\ \bibinfo {author} {\bibfnamefont {Sridevi~V.}\ \bibnamefont
  {Sarma}}} (\bibinfo {year} {2015}),\ \bibfield  {title} {\enquote {\bibinfo
  {title} {Therapeutic mechanisms of high-frequency stimulation in
  parkinson’s disease and neural restoration via loop-based reinforcement},}\
  }\href {\doibase 10.1073/pnas.1406549111} {\bibfield  {journal} {\bibinfo
  {journal} {Proceedings of the National Academy of Sciences}\ }\textbf
  {\bibinfo {volume} {112}}~(\bibinfo {number} {6}),\ \bibinfo {pages}
  {E586--E595}}\BibitemShut {NoStop}%
\bibitem [{\citenamefont {Schiff}(2012)}]{Schiff2012}%
  \BibitemOpen
  \bibfield  {author} {\bibinfo {author} {\bibnamefont {Schiff}, \bibfnamefont
  {S}}} (\bibinfo {year} {2012}),\ \href@noop {} {\emph {\bibinfo {title}
  {Neural Control Engineering: The Emerging Intersection between Control Theory
  and Neuroscience}}}\ (\bibinfo  {publisher} {MIT Press})\BibitemShut
  {NoStop}%
\bibitem [{\citenamefont {Scott}(1977)}]{scott1977neurophysics}%
  \BibitemOpen
  \bibfield  {author} {\bibinfo {author} {\bibnamefont {Scott}, \bibfnamefont
  {A}}} (\bibinfo {year} {1977}),\ \href@noop {} {\emph {\bibinfo {title}
  {Neurophysics}}}\ (\bibinfo  {publisher} {John Wiley \& Sons})\BibitemShut
  {NoStop}%
\bibitem [{\citenamefont {Shaker}\ and\ \citenamefont
  {Tahavori}(2012)}]{hrs-mt:12}%
  \BibitemOpen
  \bibfield  {author} {\bibinfo {author} {\bibnamefont {Shaker}, \bibfnamefont
  {Hamid~Reza}}, \ and\ \bibinfo {author} {\bibfnamefont {Maryamsadat}\
  \bibnamefont {Tahavori}}} (\bibinfo {year} {2012}),\ \bibfield  {title}
  {\enquote {\bibinfo {title} {Optimal sensor and actuator location for
  unstable systems},}\ }\href@noop {} {\bibinfo  {journal} {Journal of
  Vibration and Control}\ }\BibitemShut {NoStop}%
\bibitem [{\citenamefont {Sporns}\ \emph {et~al.}(2005)\citenamefont {Sporns},
  \citenamefont {Tononi},\ and\ \citenamefont {Kotter}}]{Sporns2005}%
  \BibitemOpen
\bibfield  {journal} {  }\bibfield  {author} {\bibinfo {author} {\bibnamefont
  {Sporns}, \bibfnamefont {O}}, \bibinfo {author} {\bibfnamefont
  {G}~\bibnamefont {Tononi}}, \ and\ \bibinfo {author} {\bibfnamefont
  {R}~\bibnamefont {Kotter}}} (\bibinfo {year} {2005}),\ \bibfield  {title}
  {\enquote {\bibinfo {title} {The human connectome: {A} structural description
  of the human brain},}\ }\href@noop {} {\bibfield  {journal} {\bibinfo
  {journal} {PLoS Comput Biol}\ }\textbf {\bibinfo {volume} {1}}~(\bibinfo
  {number} {4}),\ \bibinfo {pages} {e42}}\BibitemShut {NoStop}%
\bibitem [{\citenamefont {Stacey}\ and\ \citenamefont
  {Litt}(2008)}]{stacey2008technology}%
  \BibitemOpen
  \bibfield  {author} {\bibinfo {author} {\bibnamefont {Stacey}, \bibfnamefont
  {W~C}}, \ and\ \bibinfo {author} {\bibfnamefont {B}~\bibnamefont {Litt}}}
  (\bibinfo {year} {2008}),\ \bibfield  {title} {\enquote {\bibinfo {title}
  {Technology insight: neuroengineering and epilepsy-designing devices for
  seizure control},}\ }\href@noop {} {\bibfield  {journal} {\bibinfo  {journal}
  {Nat Clin Pract Neurol}\ }\textbf {\bibinfo {volume} {4}}~(\bibinfo {number}
  {4}),\ \bibinfo {pages} {190--201}}\BibitemShut {NoStop}%
\bibitem [{\citenamefont {Stam}(2004)}]{stam2004functional}%
  \BibitemOpen
  \bibfield  {author} {\bibinfo {author} {\bibnamefont {Stam}, \bibfnamefont
  {C~J}}} (\bibinfo {year} {2004}),\ \bibfield  {title} {\enquote {\bibinfo
  {title} {Functional connectivity patterns of human magnetoencephalographic
  recordings: a 'small-world' network?}}\ }\href@noop {} {\bibfield  {journal}
  {\bibinfo  {journal} {Neurosci Lett}\ }\textbf {\bibinfo {volume}
  {355}}~(\bibinfo {number} {1-2}),\ \bibinfo {pages} {25--28}}\BibitemShut
  {NoStop}%
\bibitem [{\citenamefont {Summers}\ and\ \citenamefont
  {Lygeros}(2014)}]{Summers20143784}%
  \BibitemOpen
  \bibfield  {author} {\bibinfo {author} {\bibnamefont {Summers}, \bibfnamefont
  {Tyler~H}}, \ and\ \bibinfo {author} {\bibfnamefont {John}\ \bibnamefont
  {Lygeros}}} (\bibinfo {year} {2014}),\ \bibfield  {title} {\enquote {\bibinfo
  {title} {Optimal sensor and actuator placement in complex dynamical
  networks},}\ }\href {\doibase
  http://dx.doi.org/10.3182/20140824-6-ZA-1003.00226} {\bibfield  {journal}
  {\bibinfo  {journal} {\{IFAC\} Proceedings Volumes}\ }\textbf {\bibinfo
  {volume} {47}}~(\bibinfo {number} {3}),\ \bibinfo {pages} {3784 -- 3789}},\
  \bibinfo {note} {19th \{IFAC\} World Congress}\BibitemShut {NoStop}%
\bibitem [{\citenamefont {Sun}\ and\ \citenamefont
  {Motter}(2013)}]{PhysRevLett.110.208701}%
  \BibitemOpen
  \bibfield  {author} {\bibinfo {author} {\bibnamefont {Sun}, \bibfnamefont
  {Jie}}, \ and\ \bibinfo {author} {\bibfnamefont {Adilson~E.}\ \bibnamefont
  {Motter}}} (\bibinfo {year} {2013}),\ \bibfield  {title} {\enquote {\bibinfo
  {title} {Controllability transition and nonlocality in network control},}\
  }\href {\doibase 10.1103/PhysRevLett.110.208701} {\bibfield  {journal}
  {\bibinfo  {journal} {Phys. Rev. Lett.}\ }\textbf {\bibinfo {volume} {110}},\
  \bibinfo {pages} {208701}}\BibitemShut {NoStop}%
\bibitem [{\citenamefont {Tang}\ \emph {et~al.}(2017)\citenamefont {Tang},
  \citenamefont {Giusti}, \citenamefont {Baum}, \citenamefont {Gu},
  \citenamefont {Pollock}, \citenamefont {Kahn}, \citenamefont {Roalf},
  \citenamefont {Moore}, \citenamefont {Ruparel}, \citenamefont {Gur},
  \citenamefont {Gur}, \citenamefont {Satterthwaite},\ and\ \citenamefont
  {Bassett}}]{Tang2017}%
  \BibitemOpen
  \bibfield  {author} {\bibinfo {author} {\bibnamefont {Tang}, \bibfnamefont
  {Evelyn}}, \bibinfo {author} {\bibfnamefont {Chad}\ \bibnamefont {Giusti}},
  \bibinfo {author} {\bibfnamefont {Graham~L.}\ \bibnamefont {Baum}}, \bibinfo
  {author} {\bibfnamefont {Shi}\ \bibnamefont {Gu}}, \bibinfo {author}
  {\bibfnamefont {Eli}\ \bibnamefont {Pollock}}, \bibinfo {author}
  {\bibfnamefont {Ari~E.}\ \bibnamefont {Kahn}}, \bibinfo {author}
  {\bibfnamefont {David~R.}\ \bibnamefont {Roalf}}, \bibinfo {author}
  {\bibfnamefont {Tyler~M.}\ \bibnamefont {Moore}}, \bibinfo {author}
  {\bibfnamefont {Kosha}\ \bibnamefont {Ruparel}}, \bibinfo {author}
  {\bibfnamefont {Ruben~C.}\ \bibnamefont {Gur}}, \bibinfo {author}
  {\bibfnamefont {Raquel~E.}\ \bibnamefont {Gur}}, \bibinfo {author}
  {\bibfnamefont {Theodore~D.}\ \bibnamefont {Satterthwaite}}, \ and\ \bibinfo
  {author} {\bibfnamefont {Danielle~S.}\ \bibnamefont {Bassett}}} (\bibinfo
  {year} {2017}),\ \bibfield  {title} {\enquote {\bibinfo {title}
  {Developmental increases in white matter network controllability support a
  growing diversity of brain dynamics},}\ }\href {\doibase
  10.1038/s41467-017-01254-4} {\bibfield  {journal} {\bibinfo  {journal}
  {Nature Communications}\ }\textbf {\bibinfo {volume} {8}}~(\bibinfo {number}
  {1}),\ \bibinfo {pages} {1252}}\BibitemShut {NoStop}%
\bibitem [{\citenamefont {Tang}\ \emph {et~al.}(2012)\citenamefont {Tang},
  \citenamefont {Gao}, \citenamefont {Zou},\ and\ \citenamefont
  {Kurths}}]{tang2012identifying}%
  \BibitemOpen
  \bibfield  {author} {\bibinfo {author} {\bibnamefont {Tang}, \bibfnamefont
  {Y}}, \bibinfo {author} {\bibfnamefont {H}~\bibnamefont {Gao}}, \bibinfo
  {author} {\bibfnamefont {W}~\bibnamefont {Zou}}, \ and\ \bibinfo {author}
  {\bibfnamefont {J}~\bibnamefont {Kurths}}} (\bibinfo {year} {2012}),\
  \bibfield  {title} {\enquote {\bibinfo {title} {Identifying controlling nodes
  in neuronal networks in different scales},}\ }\href@noop {} {\bibfield
  {journal} {\bibinfo  {journal} {PLoS One}\ }\textbf {\bibinfo {volume}
  {7}}~(\bibinfo {number} {7}),\ \bibinfo {pages} {e41375}}\BibitemShut
  {NoStop}%
\bibitem [{\citenamefont {Tass}\ \emph {et~al.}(1998)\citenamefont {Tass},
  \citenamefont {Rosenblum}, \citenamefont {Weule}, \citenamefont {Kurths},
  \citenamefont {Pikovsky}, \citenamefont {Volkmann}, \citenamefont
  {Schnitzler},\ and\ \citenamefont {Freund}}]{PhysRevLett.81.3291}%
  \BibitemOpen
  \bibfield  {author} {\bibinfo {author} {\bibnamefont {Tass}, \bibfnamefont
  {P}}, \bibinfo {author} {\bibfnamefont {M.~G.}\ \bibnamefont {Rosenblum}},
  \bibinfo {author} {\bibfnamefont {J.}~\bibnamefont {Weule}}, \bibinfo
  {author} {\bibfnamefont {J.}~\bibnamefont {Kurths}}, \bibinfo {author}
  {\bibfnamefont {A.}~\bibnamefont {Pikovsky}}, \bibinfo {author}
  {\bibfnamefont {J.}~\bibnamefont {Volkmann}}, \bibinfo {author}
  {\bibfnamefont {A.}~\bibnamefont {Schnitzler}}, \ and\ \bibinfo {author}
  {\bibfnamefont {H.-J.}\ \bibnamefont {Freund}}} (\bibinfo {year} {1998}),\
  \bibfield  {title} {\enquote {\bibinfo {title} {Detection of
  $\mathit{n}:\mathit{m}$ phase locking from noisy data: Application to
  magnetoencephalography},}\ }\href {\doibase 10.1103/PhysRevLett.81.3291}
  {\bibfield  {journal} {\bibinfo  {journal} {Phys. Rev. Lett.}\ }\textbf
  {\bibinfo {volume} {81}},\ \bibinfo {pages} {3291--3294}}\BibitemShut
  {NoStop}%
\bibitem [{\citenamefont {Taylor}\ \emph {et~al.}(2015)\citenamefont {Taylor},
  \citenamefont {Thomas}, \citenamefont {Sinha}, \citenamefont {Dauwels},
  \citenamefont {Kaiser}, \citenamefont {Thesen},\ and\ \citenamefont
  {Ruths}}]{taylorruths}%
  \BibitemOpen
  \bibfield  {author} {\bibinfo {author} {\bibnamefont {Taylor}, \bibfnamefont
  {Peter~N}}, \bibinfo {author} {\bibfnamefont {Jijju}\ \bibnamefont {Thomas}},
  \bibinfo {author} {\bibfnamefont {Nishant}\ \bibnamefont {Sinha}}, \bibinfo
  {author} {\bibfnamefont {Justin}\ \bibnamefont {Dauwels}}, \bibinfo {author}
  {\bibfnamefont {Marcus}\ \bibnamefont {Kaiser}}, \bibinfo {author}
  {\bibfnamefont {Thomas}\ \bibnamefont {Thesen}}, \ and\ \bibinfo {author}
  {\bibfnamefont {Justin}\ \bibnamefont {Ruths}}} (\bibinfo {year} {2015}),\
  \bibfield  {title} {\enquote {\bibinfo {title} {Optimal control based seizure
  abatement using patient derived connectivity},}\ }\href {\doibase
  10.3389/fnins.2015.00202} {\bibfield  {journal} {\bibinfo  {journal}
  {Frontiers in Neuroscience}\ }\textbf {\bibinfo {volume} {9}},\ \bibinfo
  {pages} {202}}\BibitemShut {NoStop}%
\bibitem [{\citenamefont {Varier}\ and\ \citenamefont
  {Kaiser}(2011)}]{varier2011neural}%
  \BibitemOpen
  \bibfield  {author} {\bibinfo {author} {\bibnamefont {Varier}, \bibfnamefont
  {S}}, \ and\ \bibinfo {author} {\bibfnamefont {M}~\bibnamefont {Kaiser}}}
  (\bibinfo {year} {2011}),\ \bibfield  {title} {\enquote {\bibinfo {title}
  {Neural development features: spatio-temporal development of the
  caenorhabditis elegans neuronal network},}\ }\href@noop {} {\bibfield
  {journal} {\bibinfo  {journal} {PLoS Comput Biol}\ }\textbf {\bibinfo
  {volume} {7}}~(\bibinfo {number} {1}),\ \bibinfo {pages}
  {e1001044}}\BibitemShut {NoStop}%
\bibitem [{\citenamefont {Wedeen}\ \emph {et~al.}(2012)\citenamefont {Wedeen},
  \citenamefont {Rosene}, \citenamefont {Wang}, \citenamefont {Dai},
  \citenamefont {Mortazavi}, \citenamefont {Hagmann}, \citenamefont {Kaas},\
  and\ \citenamefont {Tseng}}]{wedeen2012geometric}%
  \BibitemOpen
  \bibfield  {author} {\bibinfo {author} {\bibnamefont {Wedeen}, \bibfnamefont
  {V~J}}, \bibinfo {author} {\bibfnamefont {D~L}\ \bibnamefont {Rosene}},
  \bibinfo {author} {\bibfnamefont {R}~\bibnamefont {Wang}}, \bibinfo {author}
  {\bibfnamefont {G}~\bibnamefont {Dai}}, \bibinfo {author} {\bibfnamefont
  {F}~\bibnamefont {Mortazavi}}, \bibinfo {author} {\bibfnamefont
  {P}~\bibnamefont {Hagmann}}, \bibinfo {author} {\bibfnamefont {J~H}\
  \bibnamefont {Kaas}}, \ and\ \bibinfo {author} {\bibfnamefont {W~Y}\
  \bibnamefont {Tseng}}} (\bibinfo {year} {2012}),\ \bibfield  {title}
  {\enquote {\bibinfo {title} {The geometric structure of the brain fiber
  pathways},}\ }\href@noop {} {\bibfield  {journal} {\bibinfo  {journal}
  {Science}\ }\textbf {\bibinfo {volume} {335}}~(\bibinfo {number} {6076}),\
  \bibinfo {pages} {1628--1634}}\BibitemShut {NoStop}%
\bibitem [{\citenamefont {Wells}\ \emph {et~al.}(2015)\citenamefont {Wells},
  \citenamefont {Kath},\ and\ \citenamefont {Motter}}]{PhysRevX.5.031036}%
  \BibitemOpen
  \bibfield  {author} {\bibinfo {author} {\bibnamefont {Wells}, \bibfnamefont
  {Daniel~K}}, \bibinfo {author} {\bibfnamefont {William~L.}\ \bibnamefont
  {Kath}}, \ and\ \bibinfo {author} {\bibfnamefont {Adilson~E.}\ \bibnamefont
  {Motter}}} (\bibinfo {year} {2015}),\ \bibfield  {title} {\enquote {\bibinfo
  {title} {Control of stochastic and induced switching in biophysical
  networks},}\ }\href {\doibase 10.1103/PhysRevX.5.031036} {\bibfield
  {journal} {\bibinfo  {journal} {Phys. Rev. X}\ }\textbf {\bibinfo {volume}
  {5}},\ \bibinfo {pages} {031036}}\BibitemShut {NoStop}%
\bibitem [{\citenamefont {Whalen}\ \emph {et~al.}(2015)\citenamefont {Whalen},
  \citenamefont {Brennan}, \citenamefont {Sauer},\ and\ \citenamefont
  {Schiff}}]{PhysRevX.5.011005}%
  \BibitemOpen
  \bibfield  {author} {\bibinfo {author} {\bibnamefont {Whalen}, \bibfnamefont
  {Andrew~J}}, \bibinfo {author} {\bibfnamefont {Sean~N.}\ \bibnamefont
  {Brennan}}, \bibinfo {author} {\bibfnamefont {Timothy~D.}\ \bibnamefont
  {Sauer}}, \ and\ \bibinfo {author} {\bibfnamefont {Steven~J.}\ \bibnamefont
  {Schiff}}} (\bibinfo {year} {2015}),\ \bibfield  {title} {\enquote {\bibinfo
  {title} {Observability and controllability of nonlinear networks: The role of
  symmetry},}\ }\href {\doibase 10.1103/PhysRevX.5.011005} {\bibfield
  {journal} {\bibinfo  {journal} {Phys. Rev. X}\ }\textbf {\bibinfo {volume}
  {5}},\ \bibinfo {pages} {011005}}\BibitemShut {NoStop}%
\bibitem [{\citenamefont {Wilson}\ and\ \citenamefont
  {Moehlis}(2016)}]{10.1371/journal.pcbi.1004673}%
  \BibitemOpen
  \bibfield  {author} {\bibinfo {author} {\bibnamefont {Wilson}, \bibfnamefont
  {Dan}}, \ and\ \bibinfo {author} {\bibfnamefont {Jeff}\ \bibnamefont
  {Moehlis}}} (\bibinfo {year} {2016}),\ \bibfield  {title} {\enquote {\bibinfo
  {title} {Clustered desynchronization from high-frequency deep brain
  stimulation},}\ }\href {\doibase 10.1371/journal.pcbi.1004673} {\bibfield
  {journal} {\bibinfo  {journal} {PLoS Comput Biol}\ }\textbf {\bibinfo
  {volume} {11}}~(\bibinfo {number} {12}),\ \bibinfo {pages}
  {1--26}}\BibitemShut {NoStop}%
\bibitem [{\citenamefont {Womelsdorf}\ and\ \citenamefont
  {Everling}(2015)}]{womelsdorf2015long}%
  \BibitemOpen
  \bibfield  {author} {\bibinfo {author} {\bibnamefont {Womelsdorf},
  \bibfnamefont {T}}, \ and\ \bibinfo {author} {\bibfnamefont {S}~\bibnamefont
  {Everling}}} (\bibinfo {year} {2015}),\ \bibfield  {title} {\enquote
  {\bibinfo {title} {Long-range attention networks: Circuit motifs underlying
  endogenously controlled stimulus selection},}\ }\href@noop {} {\bibfield
  {journal} {\bibinfo  {journal} {Trends Neurosci}\ }\textbf {\bibinfo {volume}
  {38}}~(\bibinfo {number} {11}),\ \bibinfo {pages} {682--700}}\BibitemShut
  {NoStop}%
\bibitem [{\citenamefont {Wu-Yan}\ \emph {et~al.}(2018)\citenamefont {Wu-Yan},
  \citenamefont {Betzel}, \citenamefont {Tang}, \citenamefont {Gu},
  \citenamefont {Pasqualetti},\ and\ \citenamefont
  {Bassett}}]{wuyan2018benchmarking}%
  \BibitemOpen
  \bibfield  {author} {\bibinfo {author} {\bibnamefont {Wu-Yan}, \bibfnamefont
  {E}}, \bibinfo {author} {\bibfnamefont {R~F}\ \bibnamefont {Betzel}},
  \bibinfo {author} {\bibfnamefont {E}~\bibnamefont {Tang}}, \bibinfo {author}
  {\bibfnamefont {S}~\bibnamefont {Gu}}, \bibinfo {author} {\bibfnamefont
  {F}~\bibnamefont {Pasqualetti}}, \ and\ \bibinfo {author} {\bibfnamefont
  {D~S}\ \bibnamefont {Bassett}}} (\bibinfo {year} {2018}),\ \bibfield  {title}
  {\enquote {\bibinfo {title} {Benchmarking measures of network controllability
  on canonical graph models},}\ }\href@noop {} {\bibfield  {journal} {\bibinfo
  {journal} {Journal of Nonlinear Science}\ }\textbf {\bibinfo {volume} {In
  Press}}}\BibitemShut {NoStop}%
\bibitem [{\citenamefont {Yamada}(2009)}]{YamadaE14}%
  \BibitemOpen
  \bibfield  {author} {\bibinfo {author} {\bibnamefont {Yamada}, \bibfnamefont
  {Kei}}} (\bibinfo {year} {2009}),\ \bibfield  {title} {\enquote {\bibinfo
  {title} {Diffusion tensor tractography should be used with caution},}\ }\href
  {\doibase 10.1073/pnas.0812352106} {\bibfield  {journal} {\bibinfo  {journal}
  {Proceedings of the National Academy of Sciences}\ }\textbf {\bibinfo
  {volume} {106}}~(\bibinfo {number} {7}),\ \bibinfo {pages}
  {E14--E14}}\BibitemShut {NoStop}%
\bibitem [{\citenamefont {Yeo}\ \emph {et~al.}(2011)\citenamefont {Yeo},
  \citenamefont {Krienen}, \citenamefont {Sepulcre}, \citenamefont {Sabuncu},
  \citenamefont {Lashkari}, \citenamefont {Hollinshead}, \citenamefont
  {Roffman}, \citenamefont {Smoller}, \citenamefont {Zollei}, \citenamefont
  {Polimeni}, \citenamefont {Fischl}, \citenamefont {Liu},\ and\ \citenamefont
  {Buckner}}]{yeo2011organization}%
  \BibitemOpen
  \bibfield  {author} {\bibinfo {author} {\bibnamefont {Yeo}, \bibfnamefont
  {B~T}}, \bibinfo {author} {\bibfnamefont {F~M}\ \bibnamefont {Krienen}},
  \bibinfo {author} {\bibfnamefont {J}~\bibnamefont {Sepulcre}}, \bibinfo
  {author} {\bibfnamefont {M~R}\ \bibnamefont {Sabuncu}}, \bibinfo {author}
  {\bibfnamefont {D}~\bibnamefont {Lashkari}}, \bibinfo {author} {\bibfnamefont
  {M}~\bibnamefont {Hollinshead}}, \bibinfo {author} {\bibfnamefont {J~L}\
  \bibnamefont {Roffman}}, \bibinfo {author} {\bibfnamefont {J~W}\ \bibnamefont
  {Smoller}}, \bibinfo {author} {\bibfnamefont {L}~\bibnamefont {Zollei}},
  \bibinfo {author} {\bibfnamefont {J~R}\ \bibnamefont {Polimeni}}, \bibinfo
  {author} {\bibfnamefont {B}~\bibnamefont {Fischl}}, \bibinfo {author}
  {\bibfnamefont {H}~\bibnamefont {Liu}}, \ and\ \bibinfo {author}
  {\bibfnamefont {R~L}\ \bibnamefont {Buckner}}} (\bibinfo {year} {2011}),\
  \bibfield  {title} {\enquote {\bibinfo {title} {The organization of the human
  cerebral cortex estimated by intrinsic functional connectivity},}\
  }\href@noop {} {\bibfield  {journal} {\bibinfo  {journal} {J Neurophysiol}\
  }\textbf {\bibinfo {volume} {106}}~(\bibinfo {number} {3}),\ \bibinfo {pages}
  {1125--1165}}\BibitemShut {NoStop}%
\bibitem [{\citenamefont {Za{\~n}udo}\ and\ \citenamefont
  {Albert}(2015)}]{10.1371/journal.pcbi.1004193}%
  \BibitemOpen
  \bibfield  {author} {\bibinfo {author} {\bibnamefont {Za{\~n}udo},
  \bibfnamefont {Jorge G~T}}, \ and\ \bibinfo {author} {\bibfnamefont
  {R{\'e}ka}\ \bibnamefont {Albert}}} (\bibinfo {year} {2015}),\ \bibfield
  {title} {\enquote {\bibinfo {title} {Cell fate reprogramming by control of
  intracellular network dynamics},}\ }\href {\doibase
  10.1371/journal.pcbi.1004193} {\bibfield  {journal} {\bibinfo  {journal}
  {PLoS Comput Biol}\ }\textbf {\bibinfo {volume} {11}}~(\bibinfo {number}
  {4}),\ \bibinfo {pages} {1--24}}\BibitemShut {NoStop}%
\bibitem [{\citenamefont {Za{\~n}udo}\ \emph {et~al.}(2016)\citenamefont
  {Za{\~n}udo}, \citenamefont {Yang},\ and\ \citenamefont
  {Albert}}]{1605.08415}%
  \BibitemOpen
  \bibfield  {author} {\bibinfo {author} {\bibnamefont {Za{\~n}udo},
  \bibfnamefont {Jorge G~T}}, \bibinfo {author} {\bibfnamefont {Gang}\
  \bibnamefont {Yang}}, \ and\ \bibinfo {author} {\bibfnamefont {R{\'e}ka}\
  \bibnamefont {Albert}}} (\bibinfo {year} {2016}),\ \href@noop {} {\enquote
  {\bibinfo {title} {Structure-based control of complex networks with nonlinear
  dynamics},}\ }\Eprint {http://arxiv.org/abs/arXiv:1605.08415}
  {arXiv:1605.08415} \BibitemShut {NoStop}%
\end{thebibliography}%

\end{document}